# Optical and Radio Variability of the Blazar S4 0954+658

V.V. Vlasyuk,[1*] Yu.V. Sotnikova,[1,2] A.E. Volvach,[3] O.I. Spiridonova,[1] V.A. Stolyarov,[1,4]
A.G. Mikhailov,[1] Yu.A. Kovalev,[5] Y.Y. Kovalev,[6,5,7] M.L. Khabibullina,[1] M.A. Kharinov,[8]
L. Yang,[9] M.G. Mingaliev,[1,2,8] T.A. Semenova,[1] P.G. Zhekanis,[1] T.V. Mufakharov,[1,2]
R.Yu. Udovitskiy,[1] A.A. Kudryashova,[1] L.N. Volvach,[3] A.K. Erkenov,[1] A.S. Moskvitin,[1]
E.V. Emelianov,[1] T.A. Fatkhullin,[1] P.G. Tsybulev,[1] N.A. Nizhelsky,[1] G.V. Zhekanis,[1] E. V. Kravchenko[7,5]

[1] *Special Astrophysical Observatory of RAS, Nizhny Arkhyz, 369167, Russia*
[2] *Kazan (Volga Region) Federal University, Kazan 420008, Russia*
[3] *Crimean Astrophysical Observatory, Russian Academy of Sciences, Nauchny, 298409, Russia*
[4] *Astrophysics Group, Cavendish Laboratory, University of Cambridge, Cambridge, CB3 0HE, UK*
[5] *Astro Space Center, Lebedev Physical Institute, Russian Academy of Sciences, Moscow, 117997, Russia*
[6] *Max-Planck-Institut für Radioastronomie, Auf dem Hügel 69, Bonn, 53121, Germany*
[7] *Moscow Institute of Physics and Technology, Institutsky per. 9, Dolgoprudny, 141700, Russia*
[8] *Institute of Applied Astronomy, Russian Academy of Sciences, Kutuzova Embankment 10, St. Petersburg, 191187, Russia*
[9] *Department of Physics and Astronomy, Sun Yat-sen University, No 2 Daxue Road, Zhuhai, 519082, China*

**ABSTRACT**

We present an optical-to-radio study of the BL Lac object S4 0954+658 observations during 1998–2023. The measurements were obtained with the SAO RAS Zeiss-1000 1-m and AS-500/2 0.5-m telescopes in 2003–2023, with the RATAN-600 radio telescope at 1.25 (0.96, 1.1), 2.3, 4.7 (3.7, 3.9), 8.2 (7.7), 11.2, 22.3 (21.7) GHz in 1998–2023, with the IAA RAS RT-32 Zelenchukskaya and Badary telescopes at 5.05 and 8.63 GHz in 2020–2023, and with the RT-22 single-dish telescope of CrAO RAS at 36.8 GHz in 2009-2023. In this period the blazar had been showing extremely high broadband activity with the variability amplitude of flux densities up to 70–100% both in the optical and radio domains. In the period of 2014–2023 the blazar had been showing the historically highest activity in the radio wavelengths, and we detected multiple radio flares of varying amplitude and duration. The large flares last on average from 0.3 to 1 year at 22–36.8 GHz and slightly longer at 5–11.2 GHz. The optical flares are shorter and last 7–50 days. In the most active epoch of 2018–2023 the characteristic time scale $\tau$ of variation at 5–22 GHz is about 100 days and about 1000 days for the state with lower activity in 2009–2014. We found a general correlation between the optical, radio, and $\gamma$-ray flux variations, which suggests that we observe the same photon population from different emission regions. We estimated linear size of this region as 0.5-2 pc for different epochs. A broadband two components radio spectrum of S4 0954+658 jet was modelled by using both electrons and protons as emitting particles. It is shown that the synchrotron radio waves in this AGN may be generated by relativistic protons.

## 1 INTRODUCTION

Blazars are a specific type of active galactic nuclei (AGNs) known for emitting relativistic jets that align with the observer's line of sight (Urry & Padovani 1995). Blazars encompass two categories: BL Lacertae objects (BL Lacs) and flat spectrum radio quasars (FSRQs). When their optical/UV spectra are examined, BL Lacs display a continuous, featureless emission or weak narrow emission lines with an equivalent width (EW) of $\leqslant 5$Å (e.g., (Stickel et al. 1991; Marcha et al. 1996)), while FSRQs exhibit prominent broad emission lines. The flux, polarization, and spectra of blazars are highly variable across the electromagnetic spectrum from radio to $\gamma$-rays (e.g., (Gupta et al. 2017) and references therein). Blazars demonstrate variability on different timescales, ranging from minutes to several decades (Miller et al. 1989; Gopal-Krishna et al. 1993; Wagner & Witzel 1995; Gupta et al. 2004).

The emission of blazars, spanning the entire electromagnetic spectrum, is primarily dominated by nonthermal radiation originating from their relativistic jets. This broad-spectrum emission offers a valuable opportunity for investigating their spectral energy distribution (SED), which is commonly characterized by a distinctive double-hump structure (von Montigny et al. 1995; Fossati et al. 1998). In the frequency range from radio to soft X-rays, synchrotron emission serves as the dominant mechanism, while at higher energies such as hard X-rays and $\gamma$-rays, inverse Compton (IC) scattering is believed to be the primary process (Ulrich et al. 1997; Böttcher 2007). A classification system based on the peak synchrotron frequency $\nu_{\text{peak}}$ has categorized blazars into three subclasses: LSP, low synchrotron peaks with $\nu_{\text{peak}} \leqslant 10^{14}$ Hz; ISP, intermediate synchrotron peaks with $10^{14}$ Hz $< \nu_{\text{peak}} < 10^{15}$ Hz; and HSP, high synchrotron peaks with $\nu_{\text{peak}} \geqslant 10^{15}$ Hz (Abdo et al. 2010).

* E-mail: vvlas@sao.ru

When compared to the other ranges of the electromagnetic spectrum, the optical band appears relatively narrow. Nonetheless, it plays a crucial role in providing valuable information about nonthermal synchrotron emission and potential thermal emission from accretion discs. Typically, on short-term and long-term variability timescales, distinct spectral trends are observed. BL Lacertae objects exhibit a bluer-when-brighter (BWB) trend (Massaro et al. 1998; Villata et al. 2002; Vagnetti et al. 2003; Gu & Ai 2011; Gaur et al. 2012), while flat spectrum radio quasars tend to show a redder-when-brighter (RWB) trend (Ramírez et al. 2004; Gu et al. 2006; Osterman Meyer et al. 2008, 2009). However, it is worth noting that in some cases opposite trends have also been detected in certain blazars (Gu et al. 2006; Gaur et al. 2012; Isler et al. 2017). Recent studies have extensively investigated the optical variability of blazars across various timescales, utilizing observations from both space-based and ground-based telescopes. The findings reveal that long-term variability is primarily characterized by significant flux changes, occasionally accompanied by sudden flares and quasi-periodic oscillations (Bhatta et al. 2023).

The variable fluxes of blazars in the optical band exhibit a log-normal distribution on long timescales (Bhatta 2021) and a normal distribution on shorter timescales with most of the flux histograms following bimodal distribution, as have been reported in (Pininti et al. 2023) based on numerous light curves of blazars from the TESS (Transiting Exoplanet Survey Satellite) survey.

In order to find a possible link between the emission of blazars in different spectral ranges, Liodakis et al. (2018) studied radio, optical, and $\gamma$-ray light curves of 145 bright blazars, spanning up to 8 yr, to probe the flaring activity and interband correlations. Of these, 26 objects showed a $> 3\sigma$ correlation for at least one wavelength pair, as measured by the discrete correlation function. As it has been proved, the most common and strongest correlations are found between the optical and $\gamma$-ray bands, with fluctuations simultaneous within 30-day data resolution. The radio response is usually substantially delayed with respect to the other wavelengths with median time lags of about $100--160$ days.

The BL Lac-type blazar S4 0964+658 is one of the intriguing members of the blazar family, characterized by its unique emission properties and complex behavior which has sparked debates about its classification (Stickel et al. 1991; Hervet et al. 2016). It displays significant flux and polarization variability on both intraday and longer timescales (Hagen-Thorn et al. 2015; Morozova et al. 2016; Volvach et al. 2016; Liu et al. 2018). The time delay observed between flares in the optical and radio bands may find an explanation in the dynamic interplay of active processes originating from various regions in the central parts of the host galaxy. Moreover, the jet structure adds to the blazar enigmatic nature. Recent studies (Morozova et al. 2014; Jorstad et al. 2017; Wang et al. 2023) have provided additional insights, including the identification of jet components exhibiting superluminal motion, further elucidating the geometry and behavior of this blazar. Despite the ongoing debate, S4 0954+658 continues to captivate researchers, offering a unique window into the complex and dynamic phenomena occurring in active galactic nuclei and blazars.

In this paper we present the quasi-simultaneous observations of the blazar S4 0954+658 in the R band from the 1-meter Zeiss-1000 and 0.5-meter AS-500/2 optical reflectors of the Special Astrophysical Observatory of the Russian Academy of Sciences (SAO RAS) together with the radio observations at 1–22 GHz from RATAN-600, at 5.05 and 8.63 GHz from the RT-32 (IAA RAS[1]), and at 36.8 GHz from the RT-22 (CrAO RAS[2]). This collaboration's primary objective was to achieve dense sampling to identify potential correlations between the emissions in the two different bands. Such a vast database for this specific source provides an opportunity to investigate emission characteristics on both long and short time scales.

The paper is structured as follows. In Section 2 the results from the previous studies of the blazar are given, in Section 3 we describe the radio catalogues which were used as additional radio data. The photometric study with the optical telescopes of SAO RAS are presented in Section 4. Section 5 describes the radio observations with the RATAN-600, RT-32, and RT-22 instruments. We discuss the broad-band radio spectra and variability characteristics in Section 6. The flare characteristics in the optical and radio domains are presented in Section 7. In Section 8 we compute the structure functions for chosen observing epochs. The correlations between different bands are discussed in Section 9. In Section 10 we summarize the results and draw final conclusions.

## 2 THE BLAZAR S4 0954+658

The object S4 0954+658 at $z = 0.368$ (Lawrence et al. 1986), also known as QSO B0954+65, was identified as a radio source during the Jodrell-Bank 996 MHz Survey, and its optical counterpart was found by Cohen et al. (1977). Walsh et al. (1984) classified S4 0954+658 as a BL Lac object from the analysis of its spectrum, showing a smooth continuum with no detectable emission or absorption features. However, its classification remains ambiguous to this day. This source shows a one-sided radio jet with a polarized hotspot, and the polarization of the inner part of the jet indicates a longitudinal magnetic field (Kollgaard et al. 1992). Ghisellini et al. (2011) assigned this entity to the category of LBLs (low energy peaked BL Lacs, same as LSPs) based on its SED. But Hervet et al. (2016) classified S4 0954+658 as a flat spectrum radio quasar (FSRQ) due to the kinematic features of its radio jet. Nevertheless, in the literature it is mostly classified as a BL Lac object because of the small equivalent width of its emission lines in the spectrum (Stickel et al. 1991).

S4 0954+658 has been extensively studied due to its intricate variability as a blazar. The object was also studied in the

---

[1] Institute of Applied Astronomy, Russian Academy of Sciences
[2] Crimean Astrophysical Observatory, Russian Academy of Sciences

X-ray band (Perlman et al. 2006; Resconi et al. 2009), and it was one of the first detected extragalactic $\gamma$-ray sources (Thompson et al. 1995; Acero et al. 2015).

According to the Very Large Array (VLA) images, a curved jet is observed, extending south on arcsec scales (Perley 1982; Kollgaard et al. 1992). Very Long Baseline Interferometry (VLBI) maps revealed a core and jet components with remarkably strong polarization ($\sim$11%) and unusually high superluminal motion ($\beta_{\mathrm{app}}h = 7.4\pm0.7$ and $4.4\pm0.7$ for the K2 and K3 jet components respectively, (Gabuzda et al. 1992, 1994)), whose polarization is well aligned with the VLBI-structure axis, as is commonly seen in the BL Lacertae objects, indicating that the milliarcsecond jet is also curved (see Gabuzda & Cawthorne (1996) and references therein).

The multi-color photometric and polarization observations of S4 0954+658 carried out mainly in the Astronomical Institute of St. Petersburg State University and the Central Astronomical Observatory of the Russian Academy of Sciences in 2008–2012 were analyzed in Hagen-Thorn et al. (2015). The authors proved that there exist distinct individual variable components responsible for the activity; the power-law spectrum and high degree of polarization confirm that the emission is generated by synchrotron radiation. Modeling the observed dependencies between polarization and intensity were used to derive the parameters of both constant and variable components of the emission.

During an unprecedented bright optical flare in early 2015, a new bright polarized superluminal knot ejected from the VLBI core at 43 GHz during the peak of the flare was discovered by Morozova et al. (2016). This outburst was accompanied by very-high-energy (VHE > 150 GeV) and variable gamma-ray emission (Mirzoyan 2015; MAGIC Collaboration et al. 2018). A comprehensive analysis of observations from $\gamma$-rays to centimeter radio waves along with harmonic and structure analyses in Volvach et al. (2016) allowed the authors to derive the orbital ($T_{\mathrm{orb}} \approx 870$ yrs) and precessional ($T_{\mathrm{pr}} \approx 8700$ yrs) periods in the source rest frame for a model in which a binary super-massive black hole is present in this active galactic nucleus.

The blazar S4 0954+658 has a relatively small angular size the major and minor axes are $7.0 \times 4.4$ mas (Cassaro et al. 2002). Gabuzda et al. (2000) revealed a bent jet on both parsec and kiloparsec scales. Their study also unveiled significant intranight polarization variability (30–40%) of the radio core at 5 GHz. Furthermore, Kudryavtseva et al. (2010) made noteworthy observations of several moving components in the jet at 22 GHz with a mean velocity of $4.9 \pm 0.4\ c$.

During an analysis of the VLBA 43 GHz data spanning from May 2017 to May 2021, Wang et al. (2023) successfully identified three prominent jet components. Among these, two components exhibited clear superluminal motion, while one remained stationary in close proximity to the core. The angles between the moving jet components and the line of sight were measured as $5.5°$ and $6.9°$ with corresponding opening angles of $1.8°$ and $2.4°$. Notably, the trajectories of all jet components predominantly align along two distinct paths, corroborating the findings of a previous study conducted by Morozova et al. (2014).

Jorstad et al. (2017) estimated a curvature of the parsec-scale jet of this BL Lac object: about of 0.5 mas from the core, where the moving knots seem to decelerate.

Usually, the analysis of data acquired in different wave-bands over a long time period encounters limitations of good data sampling (Raiteri et al. 1999), which is an exception for S4 0954+658 due to its violent behaviour and intensive observing programs.

In a recent study conducted in 2021, Becerra González et al. (2021) identified a MgII emission line in S4 0954+658 at $z = 0.3694 \pm 0.0011$ with an equivalent width close to 5Å, which is often considered as the limit for classifying blazars as FSRQs. Therefore, it appears more reasonable to regard this $\gamma$-ray emitter as a transitional object.

## 3 LITERATURE DATA IN THE RADIO BAND

Additionally we used the radio data taken from the astrophysical CATalogs support System (CATS database,[3] Verkhodanov et al. 2005, 2009) to construct the broadband radio spectrum over the historical period of blazar observations. The data cover a time period of up to 40 years at the most representative frequency of $\sim$5 GHz, including the latest measurements presented in this paper.

The vast majority of the measurements is provided by several tens of radio catalogues that are summarized in Table 1, where we include the epochs, frequencies of observations, and corresponding literature references. The main data are represented by the NRAO VLA Sky Survey (NVSS, Condon et al. 1998), Faint Images of the Radio Sky at Twenty-cm Survey (FIRST, Becker et al. 1994), Westerbork Northern Sky Survey (WENSS, Rengelink et al. 1997), Green Bank 6-cm Survey (GB6, Gregory et al. 1996), Australia Telescope 20 GHz Survey (ATCA20, Murphy et al. 2010), Giant Metrewave Radio Telescope Sky Survey (TGSS) at 150 MHz (2015; Intema et al. 2017), VLA measurements (Healey et al. 2007), and others. The time period begins in 1974 with the Texas Survey of Radio Sources, covering $-35°\!.5 < \delta < 71°\!.5$ at 365 MHz (TXS, Douglas et al. 1996, Gregory et al. 1996). A significant contribution in the measurements was provided by the Multiyear Monitoring Program of Compact Radio Sources at 2.5 and 8.2 GHz (Lazio et al. 2001) and by the low-frequency GaLactic and Extragalactic All-sky Murchison Widefield Array (GLEAM) at 72–231 MHz (Hurley-Walker et al. 2017). The multi-frequency quasi-simultaneous observations with the RATAN-600 radio telescope are presented by the six-frequencies catalog from Mingaliev et al. (2017). The RATAN-600 measurements of the blazar are available in the electronic on-line catalog BLcat[4] (Mingaliev et al. 2014; Sotnikova et al. 2022a).

## 4 PHOTOMETRIC STUDY OF THE SOURCE WITH THE SAO RAS OPTICAL TELESCOPES

### 4.1 Optical observations

The optical study (mainly in the R band with episodic use of the BVI bands) was provided using the 1-meter (2003–June 2023) and 0.5-meter (2021–June 2023) optical reflectors of the Special Astrophysical Observatory of RAS. Below we analyze mainly the data collected in the R passband during the mentioned periods.

---

[3] https://www.sao.ru/cats
[4] https://www.sao.ru/blcat

**Table 1.** The list of catalogues with the radio measurements. The references in the last column are: [1] — Aller et al. (1985), [2] — Pauliny-Toth et al. (1978), [3] — Douglas et al. (1996), [4] — Hales et al. (1990), [5] — Owen et al. (1978), [6] — Lazio et al. (2001), [7] — Seielstad et al. (1983); Richards et al. (2011), [8] — Ulvestad et al. (1981), [9] — Johnston et al. (1995), [10] — Perley (1982), [11] — Riley et al. (1999), [12] — Zhang et al. (1997), [13] — Gregory et al. (1996), [14] — Gregory & Condon (1991), [15] — Becker et al. (1991), [16] — Reuter et al. (1997), Steppe et al. (1993), [17] — Teraesranta et al. (1998), [18] — Rengelink et al. (1997), [19] — Laurent-Muehleisen et al. (1997), [20] — Condon et al. (1998), Condon & Yin (2001), [21] — Wrobel et al. (1998), [22] — Robson et al. (2001), Jenness et al. (2010), [23] — Mingaliev et al. (2017), [24] — Angelakis et al. (2019), [25] — Massardi et al. (2011, 2008), [26] — Planck Collaboration et al. (2016), [27] — Intema et al. (2017), [28] — Hurley-Walker et al. (2017), [29] — Lacy et al. (2020).

| Catalog/Telescope | Epoch | Frequency, GHz | Reference |
|---|---|---|---|
| UMRAO | 1965–1984 | 4.8, 8, 14.5 | [1] |
| S4 | 1972, 1974 | 2.7, 10.7, 4.9 | [2] |
| TXS | 1974-1983 | 0.365 | [3] |
| 6CIII | 1976 | 0.151 | [4] |
| OPM78 | 1977 | 1.379, 4.585, 15.064, 22.185, 90 | [5] |
| GBIMO | 1979–1996 | 2.5, 8.2 | [6] |
| OVRO | 1979-1982, 2008-2009 | 10.8, 15 | [7] |
| SRCUl | 1979 | 1.48, 4.9 | [8] |
| RRF95 | 1979–1994 | 2.3, 8.4 | [9] |
| VLA4 | 1980–1981 | 4.885, 1.465 | [10] |
| 7CJR | 1984–1987 | 0.038, 0.151 | [11] |
| MIYUN | 1985–1993 | 0.232 | [12] |
| GB6 | 1986–1987 | 4.85 | [13] |
| 87GB | 1987 | 4.85 | [14] |
| 6CMN | 1987 | 4.85 | [15] |
| Mo30m | 1990–1994 | 90, 150, 142, 230 | [16] |
| SRCT | 1990–2000 | 22, 37, 87 | [17] |
| WENSS | 1991 | 0.325, 0.609 | [18] |
| RGB1 | 1992, 1994, 1995 | 5 | [19] |
| NVSS | 1993–1996 | 1.4 | [20] |
| JVAS | 1995–1997 | 8.4 | [21] |
| JCMT | 1997–2005 | 353 | [22] |
| RATAN-600 | 2005–2014 | 1.1, 2.3, 4.8, 7.7, 11.2, 21.7 | [23] |
| Effelsberg | 2006–2015 | 2.64–43 | [24] |
| ATCA | 2007 | 20 | [25] |
| PCCS2 | 2009–2013 | 30, 44, 70, 100, 143, 217, 353, 545, 8 57 | [26] |
| GMRT (TGSS) | 2010–2012 | 0.15 | [27] |
| GLEAM | 2013–2015 | 0.072–0.231 | [28] |
| VLASS | 2016–2019 | 2–4 | [29] |

The optical telescope Zeiss-1000 coupled with the CCD photometer in the Cassegrain focus provides a field of view of about $7'$ with the $0''\!.22$/pixel data sampling. The CCD camera was designed and manufactured in SAO RAS in 2002 based on the $2048 \times 2048$ back-illuminated E2V chip CCD 42-40 (Markelov et al. 2000). The pixel size of this popular chip is 13.5 microns.

This instrumental configuration has been almost unchangeable since the start of the observations: the CCD camera has demonstrated very stable operation with only few insignificant failures. The system readout noise is about 4 $e^-$, the gain is 2 $e^-$/ADU, and the dark current is about 0.01 ADU per second.

Regular observations of AGNs with a 0.5-meter Ritchey–Chretien telescope (further AS-500/2) manufactured by Astrosib (Novosibirsk, Russian Federation) started in January of 2021, while the observations aimed at search for exoplanets had started with a similar instrument at SAO RAS 1 year before. This instrument with a hyperbolic primary mirror is installed on a fast-tracking "10 Micron GM 4000" high-precision equatorial mount. Data acquisition is performed using a front-illuminated FLI Proline PL16801 4K×4K camera with 9-micron pixels. The FLI CCD camera with Peltier cooling (operation temperature about $-40°$) coupled with an Atlas focuser and a five-position motorized 50-mm filter wheel are controlled by an industrial PC, which supports all the functions of remote control and data acquisition under a Linux operating system (Valyavin et al. 2022).

The data for the blazar S4 0954+658 have been collected with the AS-500/2 in two instrumental configurations: from January 2021 to April 2022 the FLI camera had been installed in the primary focus (field of view of $1.5° \times 1.5°$ with the $1''\!.35$/pixel sampling), and then in the Cassegrain focus (field of view of $31' \times 31'$ with $0''\!.46$/pixel) since April 2022.

The latter variant provided a scale of $0''\!.92$/pixel with the $2 \times 2$ data binning, which is in good agreement with our typical $1''\!.5–2''$ seeing.

The readout noise and the gain were about 15 $e^-$ and 7 $e^-$/ADU respectively. The dark current for the FLI camera was about 1 ADU per second and was subtracted before other reduction steps.

Both CCD photometers are equipped by similar sets of filters, which are close to the standard broadband Johnson–Cousins filters given the sensitivity of both CCDs.

The typical integration time for the blazar observations was $300^s$ for Zeiss-1000 and $120^s$ for the AS-500/2 respectively.

During the period of object high activity, the integration time was less: up to $60^s$ for higher time cadence.

### 4.2 Data reduction

The standard reduction and photometrical techniques were applied to extract the magnitudes of the blazar and nearby reference stars: dark current subtraction, image flat-fielding, integration of the individual object signal within rings of increasing size, etc. The main details of data reduction were described earlier (Vlasyuk 1993).

Bias and dark frames were taken before and after an observing run each night along with twilight-sky flat-field images for preliminary data processing.

The stars from Raiteri et al. (1999) denoted from N2 to N9 were used for calibration of blazar magnitudes.

A comparison with the data obtained on the same nights with different telescopes showed general agreement; in the cases where variations have been detected, they are consistent with the usual variations of the source.

The typical photometrical accuracy of our data is better than $0^m.01$ and between $0^m.01$ and $0^m.02$ for individual frames from the Zeiss-1000 and AS-500/2 telescopes respectively. Under mediocre weather conditions, these values may be 2–3 times worse, as well as when the object is in the low brightness state ($R$ magnitude fainter $16^m$).

In order to provide a joint analysis of the optical and radio data, we averaged our measurements over individual nights and transformed the resulting values into fluxes according to the constant from Mead et al. (1990).

### 4.3 Long-term brightness curve

The long-term brightness curve of S4 0954+658 is shown in Fig.1 for a total of 670 nights over 20 yrs from February 2003 to June 2023. These data include 423 nights from the Zeiss-1000 and 306 nights from the AS-500/2 (January 2021–June 2023). For 60 nights data have been collected from both telescopes and we show the averaged values.

The average observing epochs (MJD and yyyy.mm.dd), number of observations for each night $N_{\text{obs}}$, and obtained fluxes $R$ with their errors $\sigma_R$ expressed in mJy are presented in Table A1. Below in this paper we use the fluxes in the R band, not corrected for the galactic extinction. The first few rows of this table are presented here in Table 2.

A detailed inspection of the data from Table A1 revealed the presence of some epochs with high $\sigma_R$, above 0.1 mJy. Sometimes $\sigma_R$ may exceed 1 mJy and even reaches a value as high as $\sim 2.6$ mJy for the epoch MJD = 59711. In combination with a significant amount of flux estimates, it can be a reliable indicator of an intraday variability event in the blazar. The more detailed analysis of these events are now in preparation. Let us just note that our total database of photometric observations in the R band includes about 2120 and 5200 individual images for the Zeiss-1000 and AS-500/2 telescopes respectively.

Our total set of observations can be divided into three time intervals for the following analysis.

**Table 2.** Averaged R-band measurements of S4 0954+658 in February 2003 – June 2023. Columns are as follows: (1) and (2) - observing epochs —MJD and date in yyyy.mm.dd format respectively, (3) - the number of observations $N_{\text{obs}}$, (4) — the flux densities in the $R$ band and their errors in mJy. A short fragment is presented; the full version is available in attachment.

| MJD epoch | yyyy.mm.dd | $N_{\text{obs}}$ | $R_{\text{flux}}$ and $\sigma_R$, mJy |
|:---:|:---:|:---:|:---:|
| 1 | 2 | 3 | 4 |
| 52678 | 2003.02.07 | 1 | $1.24 \pm 0.02$ |
| 52679 | 2003.02.08 | 3 | $1.39 \pm 0.03$ |
| 52699 | 2003.02.28 | 2 | $1.12 \pm 0.01$ |
| 52701 | 2003.03.02 | 4 | $1.56 \pm 0.09$ |
| 52702 | 2003.03.03 | 8 | $1.31 \pm 0.03$ |
| 52705 | 2003.03.06 | 10 | $2.02 \pm 0.06$ |

The first interval covers the period from February 2003 to December 2014 (MJD from 52600 to 57000), during which the blazar had been in a low state in the optics. According to our data, its flux did not exceed a level of 2 mJy. One should note the presence of seasonal gaps in our data, mainly caused by the location of the object at a high zenith distance during short summer nights. An analysis of the literature data showed that we have actually lost only one optical outburst of S4 0954+658 with a maximum flux of about 4.5 mJy in March–April 2011, which was detected and studied in (Morozova et al. 2014).

The second cycle spans the period of December 2014–June 2018 (MJD from 57000 to 58300), when S4 0954+658 underwent an outburst with $R \sim 14.4$ mJy in February 2015 (Morozova et al. 2016; Volvach et al. 2016) and less intensive post-outbursts with $R \sim 4.2$ mJy (MJD = 57187), 5.5 mJy (MJD = 57368), 3.1 mJy (MJD = 57830), and 5.4 mJy (MJD = 58227). During this interval the faintest R band flux was 0.44 mJy on MJD = 57844.

The third interval covers the observations between July 2018 and June 2023. The flux of S4 0954+658 ranged between 0.48 mJy (MJD = 59955) and 16.4 mJy (MJD = 59713) during this period. A careful analysis of the data from this cycle reveals a complex structure of the light curve, which may be approximated as a superposition of two processes.

The first one demonstrates quite slow flux variation with a characteristic time of about 50 days: one can count 5 episodes of blazar brightening over last 1000 days.

The second process presents by much faster flares. All these episodes with different intensity, being separated by short intervals of fainter flux, also consist of groups of faster events: between 5 and 9 occasions with a typical repeating time of about 7–20 days. One episode at MJD = 59670–59750, which includes the brightest optical state in our data at MJD = 59711, is shown in the top panel of Fig.2 with a grey strip.

Just to illustrate the fastest flux variation over our monitoring campaign, we present in Fig.2 (bottom) the S4 0954+658 flux variation during the night MJD = 59711. The data were taken using the AS-500/2 telescope with a $90^s$ cadence during a $4^h$ time interval. The brightness of the blazar changed from 10.5 to 16 mJy in 100 min and from 16 mJy to 13 mJy in 70 min. The steepest fragments of the light curve at epochs 0.81–0.83 and 0.84–0.88 demonstrated a curve slope of $\sim 6$ and ∼4 mJy/hour respectively. The following decline stage have a slope of ∼2 mJy/hour.

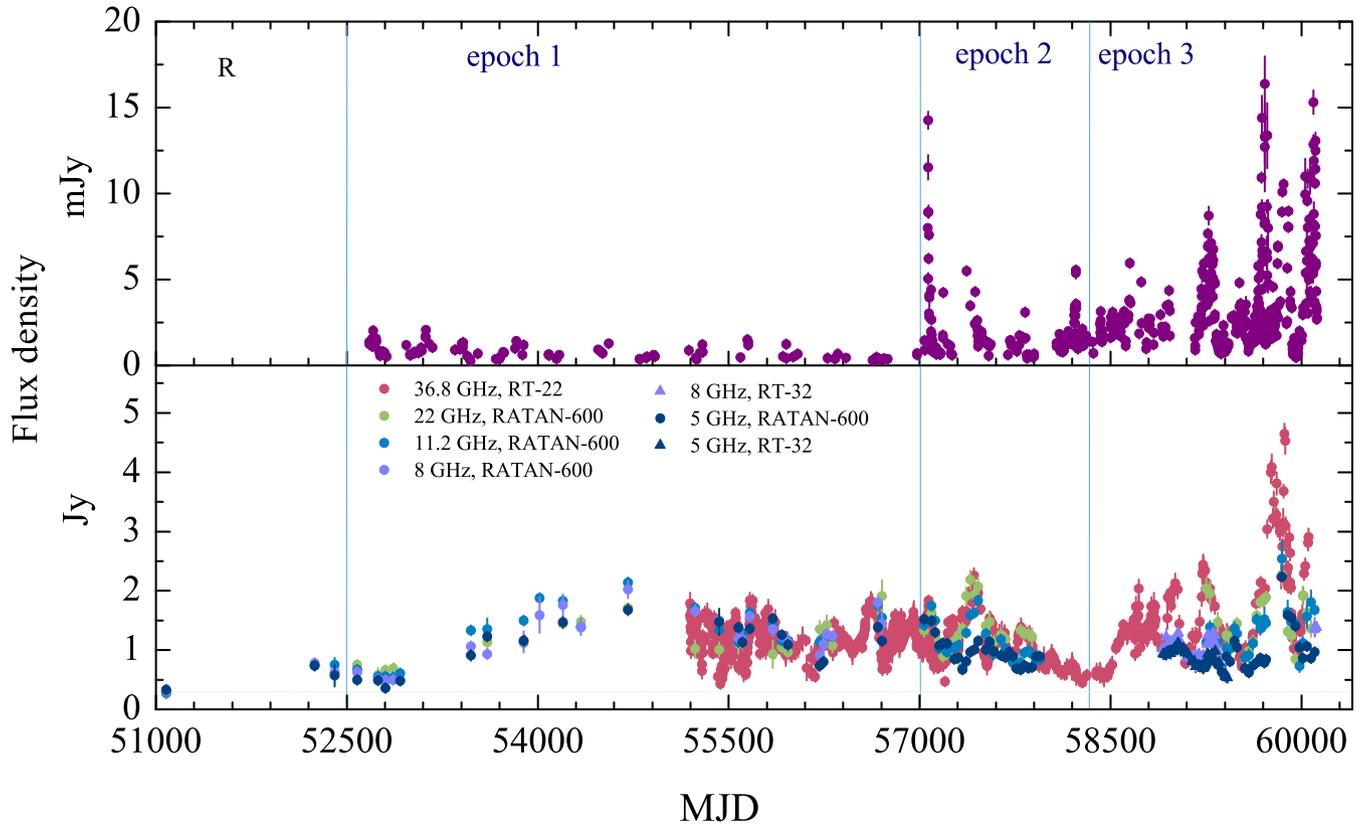

**Figure 1.** Top: optical light curve of S4 0954+658 over the time interval from Feb 2003 to Jun 2023. Bottom: the RATAN-600 (1998–2023) and RT-32 (2021–2023) flux densities at 5 (dark blue), 8 (lilac), 11 (turquoise), and 22 (green) GHz; the RT-22 36.8 GHz measurements in 2009–2023 (red). The blue lines indicate the observing epochs (1, 2, and 3) discussed in this paper, with different states in the optical band. The grey line indicates the lower level of radio flux densities of about 0.3 Jy at 5–22 GHz in 1998.

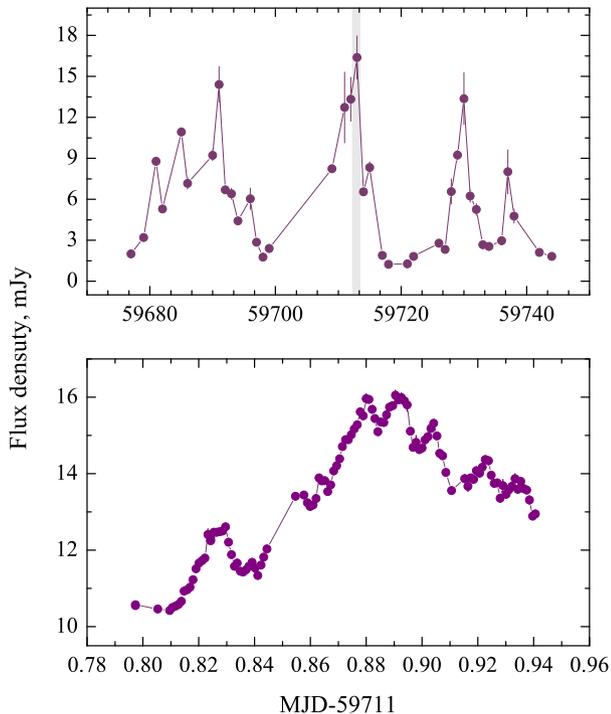

**Figure 2.** Top: a fragment of the optical light curve of S4 0954+658 over the time interval from Apr 2022 to Jun 2022 with the most prominent outbursts. The gray strip indicates the night of the strongest outburst. Bottom: an example of an intraday variability event (the outburst on the night of May 12/13, 2022). The cadence of the data is about $90^s$, the total time interval is about $4^h$.

## 5 RADIO OBSERVATIONS

### 5.1 RATAN-600

The radio flux densities were measured with the RATAN-600 telescope with a 600-m circular multi-element antenna operating in the transit mode (Parijskij (1993)). The observations provide measurements of 1–22 GHz broadband spectra simultaneously within 3–5 minutes when a source moves along the focal line where the receivers are located. The angular resolution (measured by the full width at half maximum parameter, FWHM) in this mode depends on the antenna elevation angle, and the resolution along declination $FWHM_{Dec}$ is three to five times worse than that along right ascension $FWHM_{RA}$. The angular resolution along RA and Dec calculated for the average angles is presented in Table 4 for the 6 observing frequencies. The detection limit for a RATAN-600 single sector is approximately 5 mJy at 4.7 GHz (integration time is about $3^s$) under good conditions and at the average antenna elevation angle (Dec $\sim 0°$).

The observations were carried out in 1998–2023 at the South sector and at "South+Flat" regime with two radiometric complexes at six frequencies: 0.96/1.25/1.1, 2.25, 3.95/4.7, 7.7/8.2, 11.2 and 21.7/22.3 GHz. The blazar was observed from 3 to 7 times for each epoch to improve the signal-to-noise (S/N) ratio. The measurements were processed using the automated data reduction system (Kovalev et al. 1999; Tsybulev 2011; Udovitskiy et al. 2016; Tsybulev et al. 2018) and the Flexible Astronomical Data Processing System (FADPS) standard package modules (Verkhodanov 1997) for the broadband RATAN-600 continuum radiometers. We used the following seven flux density secondary calibrators: 3C 48, 3C 147, 3C 286, NGC 7027, DR 21, 3C 295 and 3C 309.1. The flux density scales were calculated based on Baars et al. (1977) and Perley & Butler (2013, 2017). The measurements of the calibrators were corrected for the angular size and linear polarization according to the data from Ott et al. (1994) and Tabara & Inoue (1980).

The obtained flux densities with their errors $\sigma$, observing epochs averaged over $N_{obs}$ days (MJD and yyyy.mm.dd), and number of observations $N_{obs}$ are presented in Table A2. The fragment of this table is presented in Table 3. Since the observing frequencies of the RT-32 (5.05, 8.63 GHz), RT-22 (22.2 GHz), and RATAN-600 (4.7, 8.2, and 22.3 GHz) are very close, we further use the rounded values of the frequencies: 22, 8, 5, and 1 GHz for the ease of description and analysis.

The total flux density error includes the uncertainty of the RATAN-600 calibration curve, and the error of the antenna temperature measurement (Udovitskiy et al. 2016) is calculated by the equation:

$$\left(\frac{\sigma_S}{S_\nu}\right)^2 = \left(\frac{\sigma_c}{g_\nu(h)}\right)^2 + \left(\frac{\sigma_m}{T_{ant,\nu}}\right)^2, \quad (1)$$

where $\sigma_S$ is the total flux density standard error; $S_\nu$, the flux density at a frequency $\nu$; $\sigma_c$, the standard calibration curve error, which is about 1–2% and 2–5% at 4.7 and 8.2 GHz respectively; $g_\nu(h)$, the elevation angle calibration function; $\sigma_m$, the standard error of the antenna temperature measurement; and $T_{ant,\nu}$ is an antenna temperature. The systematic uncertainty of the absolute flux density scale (3–10% at 1–22 GHz) is not included in the total flux error. For S4 0954+658 the average flux density errors at 22, 11, 8, 5, 2, and 1 GHz are 10, 7, 7, 5, 17, and 22% respectively.

The RATAN-600 measurements during 1997–2023 at 22, 11, 8, and 5 GHz show continuous flux densities variations by a factor of five, from 0.5 up to 2.5 Jy.

### 5.2 RT-32

A weekly monitoring of radio flux densities of S4 0954+658 were carried out with two RT-32 radio telescopes of IAA RAS at the Zelenchukskaya and Badary observatories in the single-dish mode from 1 March 2020 to 24 July 2021. Alongside, in June–July 2023 several additional observations were made. All the measurements were performed at 5.05 and 8.63 GHz with a bandwidth of 900 MHz for both and with cooled receivers.

The observations were conducted in the drift scan mode. One scan lasted ~1 minute at 8.63 GHz and ~1.5 minutes at 5.05 GHz with 1 second registration time. To accumulate the signal, the scanning cycle was repeated a required number of times, forming a continuous observing set. About 30 to 50 scans were carried out for one object observation.

3C 48, 3C 147, 3C 295 and 3C 309.1 were used as reference sources. The flux density scales were calculated similarly to the RATAN-600 observations (Baars et al. 1977; Perley & Butler 2013).

The observed data were processed with the original program package CV (Kharinov & Yablokova 2012) and the Database of Radiometric Observations. The drift scans were filtered, rejected if spoiled considerably by weather or industrial noise, averaged and fitted with a Gaussian curve. Before averaging, the zero level of each scan was approximated by a parabola. The correction for pointing offsets was applied to the scans by the peak values of the Gaussian fits. The antenna temperature and its error were estimated from the Gaussian analysis of the averaged scan. The reference signal error of the noise generator is less than 1% and is also included into the result. The average flux density errors for the target source at 8.63 GHz and 5.05 GHz amount to 4% both.

**Table 3.** RATAN-600 measurements in 1998–2023 and the RT-32 data in 2020-2023: observing MJD epoch (Col. 1), date in yyyy.mm.dd format (Col. 2), the number of observations $N_{\rm obs}$ (Col. 3), the flux densities at 21.7/22.3, 11.2, 7.7/8.2/8.6 GHz, 4.7/5.1, 2.3, and 0.96/1.1/1.2 GHz and their errors in Jy (Cols. 4–9), the name of the telescope (Col. 10). The number of observations $N_{\rm obs}$ is indicated if information is available; for the RT-32 the number of observations is given for two frequencies. A short fragment is given here; the full version of Table is available in attachment.

| MJD epoch 1 | yyyy.mm.dd 2 | $N_{\rm obs}$ 3 | $S_{22}, \sigma$ 4 | $S_{11.2}, \sigma$ 5 | $S_8, \sigma$ 6 | $S_5, \sigma$ 7 | $S_{2.3}, \sigma$ 8 | $S_1, \sigma$ 9 | Telescope 10 |
|---|---|---|---|---|---|---|---|---|---|
| 57795 | 2017.02.10 | 5 | $1.33 \pm 0.10$ | $0.79 \pm 0.07$ | – | $0.67 \pm 0.03$ | – | – | RATAN-600 |
| 57823 | 2017.03.10 | 6 | $1.28 \pm 0.10$ | $0.95 \pm 0.03$ | – | $0.74 \pm 0.04$ | – | – | RATAN-600 |
| 57854 | 2017.04.10 | 5 | $1.27 \pm 0.10$ | $0.99 \pm 0.06$ | – | $0.70 \pm 0.04$ | – | – | RATAN-600 |
| 58909 | 2020.03.01 | 43 | – | – | $1.02 \pm 0.06$ | – | – | – | RT-32 |
| 58916 | 2020.03.08 | 62/42 | – | – | $0.95 \pm 0.04$ | $0.93 \pm 0.08$ | – | – | RT-32 |
| 58923 | 2020.03.15 | 88 | – | – | – | $0.93 \pm 0.04$ | – | – | RT-32 |
| 58937 | 2020.03.29 | 67/86 | – | – | $1.22 \pm 0.14$ | $0.92 \pm 0.03$ | – | – | RT-32 |

**Table 4.** RATAN-600 continuum radiometer parameters: the central frequency $f_0$, the bandwidth $\Delta f_0$, the detection limit for point sources per transit $\Delta F$. FWHM$_{\rm RA \times Dec}$ is the angular resolution along RA and Dec calculated for the average angles.

| $f_0$ GHz | $\Delta f_0$ GHz | $\Delta F$ mJy/beam | FWHM$_{\rm RA \times Dec.}$ |
|---|---|---|---|
| 22.3 | 2.5 | 50 | $0\rlap{.}''17 \times 1\rlap{.}'6$ |
| 11.2 | 1.4 | 15 | $0\rlap{.}''34 \times 3\rlap{.}'2$ |
| 8.2 | 1.0 | 10 | $0\rlap{.}''47 \times 4\rlap{.}'4$ |
| 4.7 | 0.6 | 8 | $0\rlap{.}''81 \times 7\rlap{.}'6$ |
| 2.25 | 0.08 | 40 | $1\rlap{.}''71 \times 15\rlap{.}'8$ |
| 1.25 | 0.08 | 200 | $3\rlap{.}''07 \times 27\rlap{.}'2$ |

**Table 5.** A fragment of Table A3: the RT-22 measurements in 2009–2023: MJD epoch (Col. 1), yyyy.mm.dd (Col. 2), the flux densities at 36.8 GHz and their errors in Jy (Col. 3).

| MJD epoch 1 | yyyy.mm.dd 2 | $S_{36.8}, \sigma$ 3 |
|---|---|---|
| 55196 | 2009.12.30 | $1.79 \pm 0.18$ |
| 55198 | 2010.01.01 | $1.59 \pm 0.22$ |
| 55199 | 2010.01.02 | $1.54 \pm 0.23$ |
| 55200 | 2010.01.03 | $1.39 \pm 0.21$ |
| 55202 | 2010.01.05 | $1.33 \pm 0.12$ |
| 55204 | 2010.01.07 | $0.99 \pm 0.13$ |

The obtained RT-32 flux densities with their errors $\sigma$, average observing epochs (MJD and yyyy.mm.dd), and number of observations $N_{\rm obs}$ are presented in Table A2 (the fragment is in Table 3).

### 5.3 RT-22

The observations at 36.8 GHz were carried out with the radio telescope RT-22 (CrAO, Simeiz) using modulation radiometers in the beam pattern modulation mode (Vol'vach et al. 2008). Before measuring an antenna temperature, the coordinates were refined by scanning in right ascension and declination. The antenna temperature of the source was determined as the difference between the responses of the radiometer at the two mentioned antenna positions, averaged over 30 seconds. Depending on the signal-to-noise ratio of the radiometer response, a series of 30–60 measurements was carried out, after which the average value of the signal and its root-mean-square error were estimated. The absorption of radiation in the Earth's atmosphere was taken into account by the "atmospheric cuts" method, in which the differences in antenna temperatures at fixed elevation angles were registered. The antenna temperatures corrected for the absorption of the radiation in the Earth's atmosphere were recalculated into spectral flux density according to observations of the calibration sources (Volvach et al. 2016; Sotnikova et al. 2023).

The median value of the flux density errors at 36.8 GHz is 10%. The obtained flux densities and their errors are presented in Table A3. The first few rows of this table is presented in Table 5.

## 6 RADIO PROPERTIES

### 6.1 Light curves

The light curves of the blazar at radio frequencies show continuous flux density variations from 1998 to 2023 (Fig. 1). The lower state of the source was observed in 1998, at that epoch the flux densities varied from 0.59 to 0.26 mJy at 22–0.96 GHz (Fig. 3). In the time period of 2003–2023, for which the optical measurements are available, we separate three epochs (1, 2, and 3) with different states of the blazar in the optical and radio bands (blue lines in Fig. 1). During epoch 1 (February 2003 – December 2014), the blazar showed flux density variations from 0.55 to 2.1 Jy at 11.2 GHz. The second epoch (epoch 2, December 2014 – July 2018) revealed a radio flare with a flux maximum of 1.99 Jy at 36.8 GHz in January 2016, which started later at 5–11 GHz (March 2016). During the rest of the time of epoch 2, the radio flux density was slowly decreasing. During epoch 3 (July 2018 – June 2023), five large radio flares were observed with differ-ent amplitudes (up to 4.65 Jy at 36.8 GHz) and duration. We have stopped the observing run of epoch 3 at the beginning of the next (sixth) flare, while in the optical band the flux is almost down.

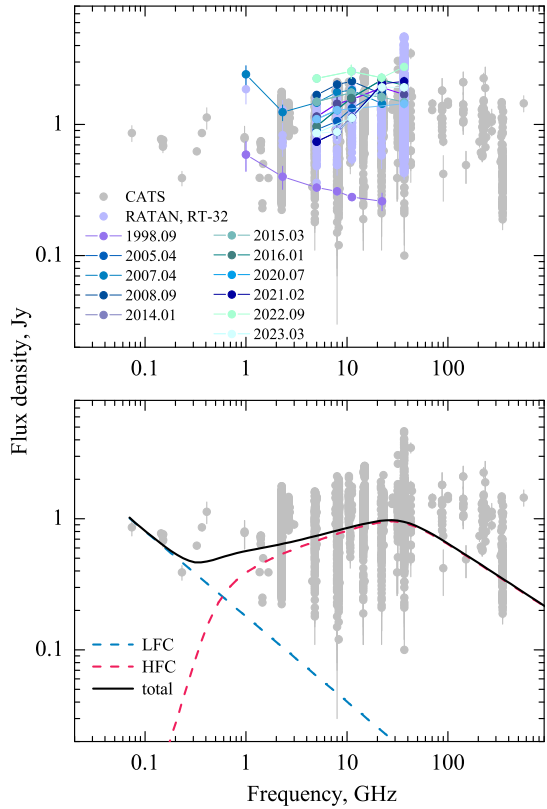

**Figure 3.** The broadband radio spectra of S4 0954+658. Top: the literature data from CATS are shown with the grey circles. The new RATAN-600, RT-32, and RT-22 data at 1–36.8 GHz are marked by lilac color. The spectra during the several flare epochs are shown by the same colors as in Fig. 1. The lower radio spectrum (blue circles) corresponds to the quietest state of the blazar in 1998. Bottom: the results of simulating the average ("total") spectrum via two model spectra components (LFC and HFC), similar to that for the blazar 1502+10 in Sotnikova et al. (2022b).

### 6.2 Broadband radio spectra

S4 0954+658 is an extremely variable AGN. Its broadband radio spectra (Figure 3) show a wide variety of shapes: from steep ($\alpha < -0.5$) to rising ($\alpha > 0$). During the 25 years of the RATAN-600 observations, the spectral index $\alpha_{11-22}$ was less then $-0.5$ only during three epochs: Feb 2010, Oct 2011, and Jan 2023. We defined the spectral index from the power-law $S_\nu \sim \nu^\alpha$ ($S_\nu$ is a flux density at the frequency $\nu$, and $\alpha$ is a spectral index). The average spectrum of S4 0954+658 has a peak at mm-waves in the observer's frame of reference.

According to the Hedgehog jet model (Kovalev et al. 2000), suggested by N. S. Kardashev in 1969, the longitudinal magnetic field $B$ of the radio jet depends on the distance $r$ as $B = B_0 \, (r/r_0)^{-2}$. In Figure 3 the fitting of the two components in this model to the observations is shown. The blue dotted line represents the fitting of the low frequency component (LFC) to the optically thin model synchrotron emission of nearly constant extended structures of the object up to kpc scales. The red dotted line represents the average spectrum of the model jet as the second component, the high frequency component (HFC). The black line (marked as "total") is the sum of the red and blue dotted lines. Such average model spectra could be observed if the continuous flux $dN(t)/dt$ of the emitted particles across the base of the jet is constant for a long time (about 10-20 years). The variability of $dN(t)/dt$ is converted in the model to the variability of the HFC and the total spectrum.

**Table 6.** Main physical parameters of the jet determined from fitting the flux densities in Figure 3 using equations (2)–(4) for two sorts of emitting particles: jet magnetic field $B$, brightness temperature $T_b$, angular diameter $\Theta$, and the relation of the magnetic energy density $W_H = B^2/8\pi$ to the energy density $W_E$ of emitting particles.

| Particles | $B$ Gauss | $T_b$ K | $\Theta$ mas | $W_H/W_E$ |
| --- | --- | --- | --- | --- |
| electrons | 20 | $5 \cdot 10^{10}$ | 0.08 | $W_H \gg W_E$ |
| protons | $4 \cdot 10^4$ | $9 \cdot 10^{13}$ | 0.002 | $W_H \gg W_E$ |

We obtained the following fitted physical parameters of the jet: the angle to the observer's line of sight $\vartheta \sim 1.5°$, the flux $S_m$ and the frequency $\nu_m$ of the jet spectrum maximum $S_m \sim 0.70$ Jy and $\nu_m \sim 60$ GHz, $\gamma \sim 2.0$, and supposed $\gamma_E \sim 300$.

The following physical parameters of the jet, neglecting the redshift, can be estimated using the fitted parameters (Sotnikova et al. 2022b):

$$B_\perp/M_{2e} \sim 0.82 \cdot 10^{-6} \cdot \nu_m \cdot \gamma_E^{-2}, \qquad (2)$$

$$T_b/M_{2e} \sim 1.5 \cdot 10^8 \cdot \gamma_E, \qquad (3)$$

$$\Theta \sim \lambda_m \left( \frac{2 S_m}{\pi k_B T_b} \right)^{1/2}. \qquad (4)$$

Here $\nu_m$ (Hz), $\lambda_m$ (m), and $S_m$ (Watt/(m$^2$Hz)) are the frequency, wavelength, and flux density of the spectrum maximum; the gamma-factor $\gamma_E = E/(Mc^2)$ of emitted particles is supposed to have the same value for electrons and protons; $B_\perp = B \sin \vartheta$, $M_{2e} = 1$ for electrons, and $M_{2e} = 1836$ for protons. $\Theta$ (rad), $k_B$ and $T_b$ (K) are the angular diameter of the emitting jet in the picture plane, the Boltzman constant, and the brightness temperature. The estimated parameters are shown in Table 6.

### 6.3 Variability estimates

In order to characterise the variability of the flux density, we have computed the variability and modulation indices and the fractional variability. The first and third characteristics take into account measurement uncertainties, while the modulation index and the fractional variability are less sensitive to outliers. The variability $V_S$ index was calculated using the formula from Aller et al. (1992):

$$V_S = \frac{(S_{\max} - \sigma_{S_{\max}}) - (S_{\min} + \sigma_{S_{\min}})}{(S_{\max} - \sigma_{S_{\max}}) + (S_{\min} + \sigma_{S_{\min}})} \qquad (5)$$

where $S_{\max}$ and $S_{\min}$ are the maximum and minimum flux densities over all epochs of observations; $\sigma_{S_{\max}}$ and $\sigma_{S_{\min}}$ are their errors. This formula prevents one from overestimating the variability when there are observations with large uncertainties in the data. We obtain a negative value of $V_S$ in the case where the flux error is greater than the observed scatter in the data.

The modulation index, defined as the standard deviation of flux density $\sigma_S$ divided by the mean flux density $\bar{S}$, was calculated as in Kraus et al. (2003):

$$M = \frac{\sigma_S}{\bar{S}}. \quad (6)$$

The fractional variability $F_S$ is defined as in Vaughan et al. (2003):

$$F_S = \sqrt{\frac{V^2 - \bar{\sigma}_{\text{err}}^2}{\bar{S}^2}} \quad (7)$$

where $V^2$ is the variance or dispersion of the process, $\bar{S}$ is the mean flux density, and $\sigma_{\text{err}}$ is the root mean square error. The uncertainty of $F_S$ is determined as:

$$\triangle F_S = \sqrt{\left(\sqrt{\frac{1}{2N}}\frac{\bar{\sigma}_{\text{err}}^2}{F_S * \bar{S}^2}\right)^2 + \left(\sqrt{\frac{\bar{\sigma}_{\text{err}}^2}{N}}\frac{1}{\bar{S}^2}\right)^2} \quad (8)$$

The values of $V_S$, $M$, and $F_S$ in the optical band and at four radio frequencies are given in Table 7. The number of observing epochs $N_{\text{obs}}$ and the timescale $t$ play a key role in searching for AGN variability. It is known that variability increases with the number of observations, which can be explained by the fact that peaks of variability are easily missed when sampling is very sparse (Tornikoski et al. 2000).

### 6.4 The contribution of refractive interstellar scintillation

The observed flux density variability could be caused by either intrinsic (related to the source's properties) or extrinsic factors. The extrinsic ones are caused by the interaction of source radio emission with an inhomogeneous propagation medium. The flickers caused by scattering on the inhomogeneities of the interplanetary medium have a characteristic timescale of the order of a second or less (Morgan et al. 2018), so they are smoothed out in RATAN-600 observations due to the relatively long scan time of 3–5 minutes during the source's transit. Similarly, for measurements at other radio telescopes, interplanetary flickering is also insignificant due to the typical continuum flux density measurement time being obviously longer than one second.

The scintillations caused by the propagation of radio waves in the interstellar medium can be diffractive and refractive. The diffractive scintillation typically requires an extremely compact source to be detected, and observations must be made in a narrow frequency band to resolve the fine-scale structures responsible for the diffractive scintillation (Narayan 1992). This is because the scintillation pattern can be highly frequency-dependent. The continuum radio measurements are carried out in a wide band (from hundreds of MHz to several GHz); therefore, this type of flickering can not be detected in the RATAN-600 and RT-32 data.

Let's estimate the possible contribution of refractive interstellar scintillations (RISS) to the variability at the RATAN-600 frequencies (1.2, 2.3, 4.7, 8.2, 11.2, and 22.3 GHz) for S4 0954+658. This source lies about 43° from the galactic plane. According to Walker (1998), the transition frequency defining the boundary between the modes of strong and weak scattering is approximately 8 GHz for the blazar. We will adopt this value of the transition frequency to estimate the modulation level of the flux density and its timescale.

We estimated the angular size as $\theta_s = 5 \times \theta_{\min}$, where $\theta_{\min} = 0.6\sqrt{S}/\nu$ is the minimum angular size for the stationary source of synchrotron radiation (Kellermann & Owen 1988), $S$ is the flux density (in Jy) at the observer's frequency (in GHz). We adopted $S = 1$ Jy, taking into account the shape of the S4 0954+658 average radio spectrum.

We followed the formulas from Walker (1998) to calculate the modulations of the flux density $m$ and their typical timescales, the results are presented in Table 8. We conclude that the modulation level is about 1%, which is significantly less than the average values of the obtained variability level. Therefore, the contribution of refractive scintillations to the total variability is insignificant.

## 7 CHARACTERISTICS OF THE FLARES

### 7.1 Optical flares

A detailed inspection of the brightness curve in the R band shows the presence of many flares with different temporal characteristics. Primarily, it is true for our data obtained during epochs 2 and 3.

The observed brightness curve for epoch 3 may be approximated as a combination of slow and fast flares. The first ones have typical variation time of about 250 days and amplitudes between 1 and 3 mJy. The MJD epochs of the maxima of slow flares are: 59250, 59510, 59710, 59870, and 60070. Their characteristic time scales are close to $50^d$.

The criterion of flare detection, flux exceeding the neighboring level 4–5 times, allowed us to identify 44 faster flares within the Jan 2021–Jun 2023 time interval. They have typical variation time between 7 and 20 days (Fig. 4) with timescales in the range of 2–5 days and amplitudes between 2 mJy (faintest slow flare with the maximum on MJD = 59510) and 20 mJy (most powerful flare with the maximum on MJD = 59710).

### 7.2 Radio flares

The radio light curve at 36.8 GHz demonstrates about 20 flares in 2009–2023 with the flux density maximum higher than 1.5 Jy. It is the best-sampled radio light curve, which allows us to investigate the behaviour of radio flares. The light curve ends in April 2023 by a sharp increasing of the flux density and, apparently, corresponds to the last optical flare which started in February 2023.

The total flux density variations in AGNs $\Delta S(t)$ can be well modeled by a superposition of flare components at high radio frequencies, more than 10 GHz (Valtaoja et al. 1999; Hovatta et al. 2009). The increase, decrease, and decay of each flare are modeled by the following relations:

$$\Delta S(t) = \begin{cases} \Delta S_{\max} \cdot e^{(t - t_{\max})/\tau_1}, & t < t_{\max} \\ \Delta S_{\max} \cdot e^{(t_{\max} - t)/\tau_2}, & t > t_{\max} \end{cases}$$

where $\Delta S_{\max}$ is the maximum amplitude of a radio flare (Jy),

**Table 7.** The values of $V_S$, $M$, and $F_S$ in the optical band and at radio frequencies of 5, 8, 11.2, 22, and 36.8 GHz.

| Frequency (GHz) | $V_S$ | $M$ | $F_S$ | $V_S$ | $M$ | $F_S$ | $V_S$ % | $M$ | $F_S$ | $V_S$ | $M$ | $F_S$ |
|---|---|---|---|---|---|---|---|---|---|---|---|---|
| | | all | | | epoch1 | | | epoch2 | | | epoch3 | |
| 5 | 73.1 | 30.5 | 29.8 | 61.4 | 36.8 | 35.1 | 35.0 | 23.4 | 22.5 | 58.1 | 28.2 | 27.6 |
| 8 | 70.1 | 30.6 | 29.1 | 53.5 | 35.7 | 33.4 | – | – | – | 35.9 | 20.5 | 19.0 |
| 11 | 77.1 | 33.0 | 31.9 | 56.0 | 33.6 | 32.6 | 40.7 | 28.0 | 26.8 | 46.9 | 30.2 | 27.6 |
| 22 | 74.9 | 31.8 | 30.1 | 40.2 | 31.3 | 28.2 | 37.4 | 28.1 | 26.2 | 37.5 | 27.0 | 24.8 |
| 36.8 | 80.6 | 38.3 | 37.0 | 56.2 | 21.3 | 18.6 | 64.0 | 33.9 | 32.0 | 73.3 | 49.1 | 48.2 |
| optical band | 96.4 | 93.2 | 92.7 | 76.8 | 50.0 | 49.7 | 93.7 | 101.6 | 101.0 | 93.6 | 78.5 | 77.9 |

**Table 8.** RISS modulation of the flux density $m$ and its typical timescale estimates $t$ for the RATAN-600 frequencies at a transitional frequency of 8 GHz.

| Frequency (GHz) | Source's size (mas) | $m$ (%) | $t$ days |
|---|---|---|---|
| 1.2 | 2.50 | 2.4 | 52.1 |
| 2.3 | 1.30 | 1.4 | 27.2 |
| 4.7 | 0.64 | 0.8 | 13.3 |
| 8.2 | 0.37 | 0.5 | 7.6 |
| 11.2 | 0.27 | 0.4 | 5.6 |
| 22.3 | 0.13 | 0.2 | 2.8 |

2002). The agreement between the appearance of VLBA components and the time period of radio flares is quite good. Indeed, the VLBA 43 GHz data[5] show the emergence of a new compact and highly polarized (polarization degree $p > 15\%$) structure in April 2015, August 2019, August 2020, and August 2022.

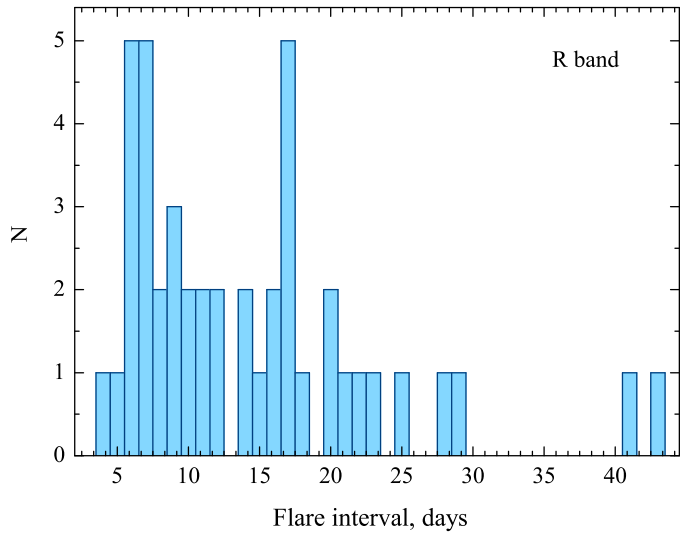

**Figure 4.** The distribution of time intervals between subsequent flares for epoch 3 in the R band.

$t_{\max}$ is the epoch of the flare maximum; $\tau_1$ and $\tau_2$ are the time of flare rising and falling (in days). We used the actual values of both $\tau_1$ and $\tau_2$ calculated from the light curves (Table 9), while in the original work (Valtaoja et al. 1999) the authors used $\tau_2 = 1.3\tau_1$. We revealed and modelled six stronger flares with amplitudes of about 2 Jy in the light curve at 36.8 GHz (Fig. 5). Their characteristics are presented in Table 9. It is also known that individual exponential flares correspond to the appearance of new VLBA components, which indicates that the flares obtained from the fits are related to actual jet physics (Lähteenmäki & Valtaoja 1999; Savolainen et al.

[5] https://www.bu.edu/blazars/VLBA_GLAST/0954.html

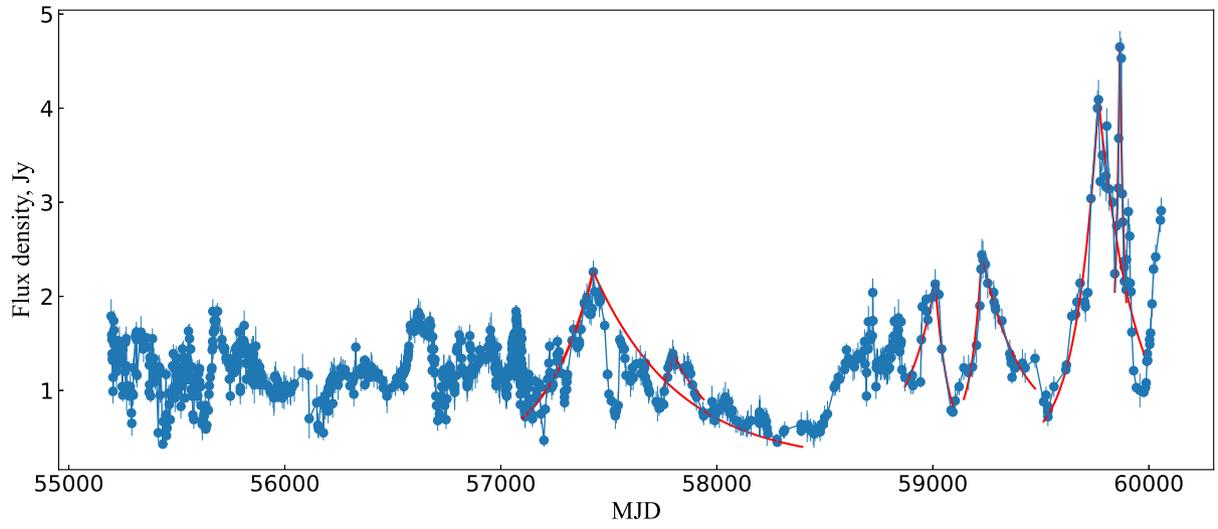

**Figure 5.** The light curve at 36.8 GHz approximated by the most bright six radio flares, with $S_{\max} > 2$ Jy, depicted by red lines.

**Table 9.** Main parameters of the large exponential flares at 36.8 GHz data.

| $t_{\max}$ MJD | $\Delta S_{\max}$ Jy | $\tau_1$ days | $\tau_2$ days | $\tau_2/\tau_1$ |
|---|---|---|---|---|
| 57428.19 | 2.035 | $227 \pm 10$ | $394 \pm 18$ | 1.7 |
| 57796.4* | 1.025 | $96 \pm 10$ | $223 \pm 25$ | 2.3 |
| 59010.87 | 1.745 | $143 \pm 17$ | $60 \pm 9$ | 0.4 |
| 59226.65 | 1.87 | $49 \pm 7$ | $173 \pm 10$ | 3.5 |
| 59766.85 | 3.73 | $102 \pm 5$ | $164 \pm 13$ | 1.6 |
| 59865.4 | 3.615 | $18 \pm 2$ | $22 \pm 2$ | 1.2 |

We have obtained different flare characteristics such as flare duration $\tau_1$, $\tau_2$ and $\tau_2/\tau_1$ relation (Table 9). It is an expected result because the light curve at 36.8 GHz has complex character with a great number of small flares. For example, we see a small flare at a decrease of a large flare with $t_{\max} = 57428.19$ MJD (it is denoted by asterisk in the Table 9). The large flare with $t_{\max} = 59865.4$ MJD happened during the decrease of the strongest flare ($t_{\max} = 59766.85$ MJD). These flares have approximately equal $\Delta S_{\max}$ of about 3.7 Jy, but their timescales differ by 5–8 times. Thus the light curve at the high frequency of 36.8 GHz reflects the features of non-stationary processes with timescales from weeks to months.

## 8 STRUCTURE FUNCTION ANALYSIS

The structure function (SF) is a method of searching for typical time scales and periodicities in non-stationary processes (Simonetti et al. 1985; Hughes et al. 1992). The SF analysis provides a method of quantifying time variability and gives information on the nature of the process that caused the variations. A characteristic time scale in a light curve, defined as the time interval between a maximum and an adjacent minimum or vice versa, is indicated by a maximum of the structure function, whereas periodicity in the light curve causes a minimum (Heidt & Wagner (1996)).

The structure function of the first order normalized to the variance of the signal $\sigma^2$ is usually determined as:

$$D_1(\tau) = \langle\{[f(t) - f(t+\tau)]\}^2\rangle, \quad (9)$$

where $f(t)$ is the signal at a time $t$, and $\tau$ is the lag. The slope of the power part of the curve is estimated as:

$$b = d\log D_1 / d\log \tau. \quad (10)$$

An ideal SF consists of two plateaus and a straight line with slope $b$ between them. The $X$-axis is the logarithm of the time lag, $\tau$, and $Y$-axis shows the logarithm of $D(\tau)$. One of the important characteristics of the structure function is the point where the SF reaches its higher plateau with an amplitude equal to $2\sigma^2$. This timescale gives the maximum timescale $T_{\max}$ of correlated signals or, equivalently, the minimum time scale of uncorrelated behavior. The slope between two plateaus determines the nature of the variable process. The light curve can be modelled as a combination of white (flicker) and red (shot) noises, and in this case the slope is between 0 and 1 (Hughes et al. 1992). For a single dominating outburst in the light curve, the slope is usually steeper than 1. If the slope $b \sim 2$, than there is a strong linear trend or a strong periodic oscillation (Hufnagel & Bregman 1992).

**Table 10.** Parameters of the SF for the R band and at 5, 11.2, 22, and 36.8 GHz.

| Frequency, GHz | $b$ | $log(\tau)$, days | $b$ | $log(\tau)$, days | $b$ | $log(\tau)$, days |
|---|---|---|---|---|---|---|
| | epoch 1 | | epoch 2 | | epoch 3 | |
| 5 | 2.2 | 3.0 | 1.8 | 2.3 | 2.7 | 2.2 |
| 11.2 | 2.3 | 3.0 | – | – | 2.0 | 2.1 |
| 22 | 3.3 | 2.7 | 1.7 | 2.2 | 2.9 | 2.0 |
| 36.8 | 1.3 | 1.9 | – | – | 1.1 | 2.3 |
| R band | 1.1 | 2.5 | 0.8 | 2.0 | 0.6 | 2.0 |

In this work, for the non-uniform and finite data series $f(i)$ with $i = 1, 2, ..., N$, obtained from the original temporally non-uniform observed data series by partitioning into intervals with the values averaged over the interval, the first-order structure function is calculated as:

$$D_1(k) = \frac{1}{N_1(k)} \sum_{i=1}^{N} w(i)w(i+k)[f(i+k) - f(i)]^2, \quad (11)$$

where $N_1(k) = \sum w(i)w(i+k)$ and the weight factor $w(i)$ is equal to 1 if the observations exist in the interval $i$ and is 0 if there are no observations. For $k = 1, 2, ..., L$, we construct our own set of intervals. The initial time interval $k = 1$ is chosen in such a way that it is equal to or slightly larger than the average time interval between observations (ignoring very long gaps in them). For radio frequencies an interval of about 60 days was taken as the initial time interval, at a frequency of 36.8 GHz it was 2 days, and 4 days for the optical data. The final interval $k = L$ is calculated based on the length of the time scale of the original sequence so that all values of the original series are divided into two intervals.

The error of the structure function is calculated from the error percentage of the original function in the interval:

$$\sigma_\% = \frac{1}{N-n-1}\sqrt{\left(\frac{\sigma_t}{f(t)}\right)^2 + \left(\frac{\sigma_{t+\tau}}{f(t+\tau)}\right)^2}. \quad (12)$$

We have computed the SF for the three epochs (1, 2, and 3) in the radio and optical bands (Fig. 6). To measure the slope $b$, we used all data points between the lower and upper plateaus. The SF parameters, $b$ and $lg(\tau_s)$, are given in Table 10. Due to the small number of data points, we could not construct the SF at 8 GHz. We also cannot determine the second plateau for epoch 2 at 11.2 GHz (large scattering of data points) and for epoch 2 at 36.8 GHz (it does not reach the value of the signal variance $\sigma^2$). We obtained $b \approx$ 2–3 for 5–22 GHz for the three epochs and $b \approx 1.0$ for 36.8 GHz and the R band. The characteristic time scale $\tau$ is about 100 days for epochs 2 and 3. For epoch 1, $lg(\tau)=3$, which corresponds to 1000 days at 5–22 GHz and the R band, and $lg(\tau)=2$, or $\tau$ is 100 days at 36.8 GHz.

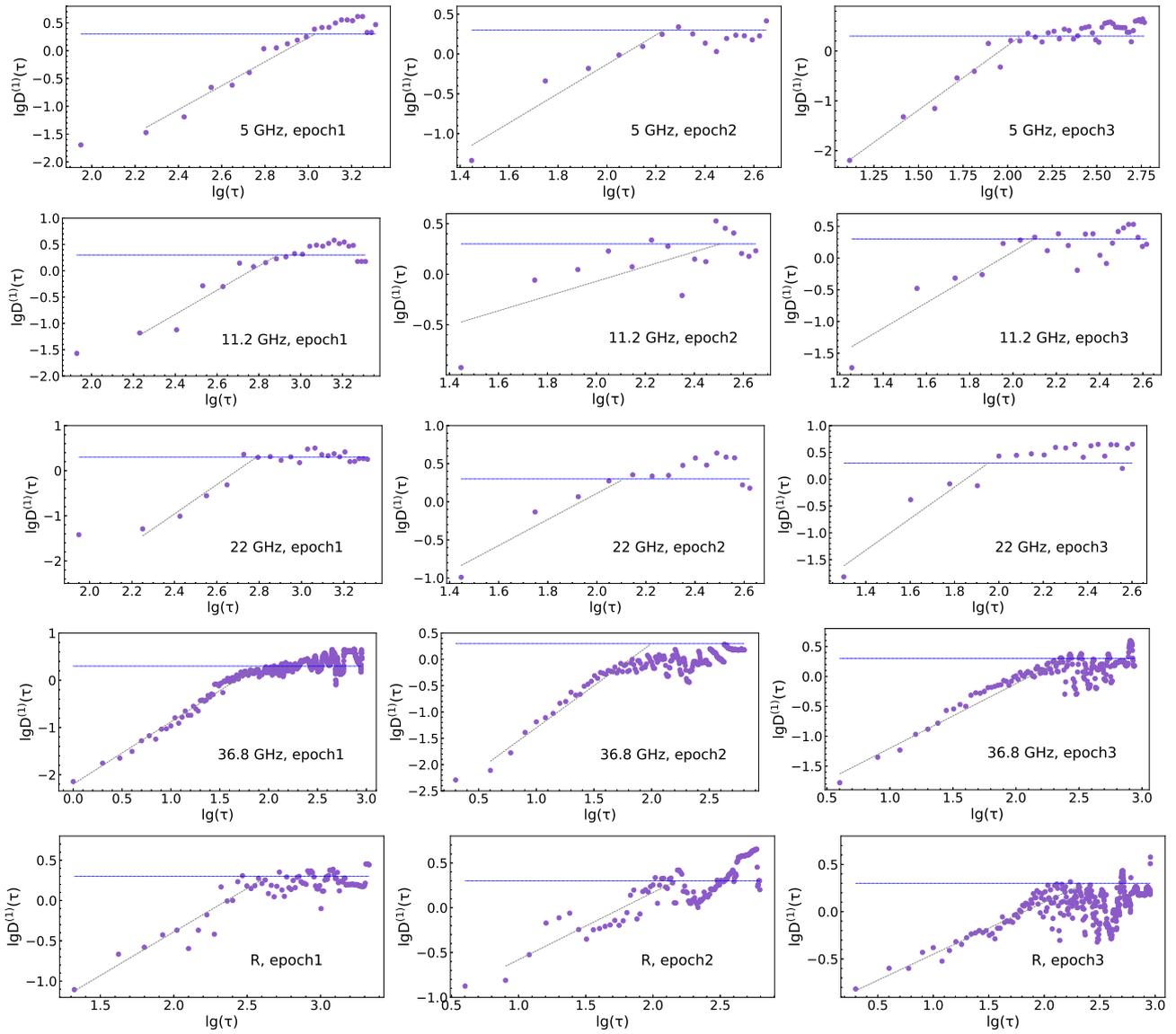

**Figure 6.** The SF for the total flux variations in the optical and radio bands for the three epochs.

## 9 RADIO-OPTICAL CORRELATION

We used the cross-correlation method to analyze time lags in the light curves at different frequencies. (Edelson & Krolik 1988). To calculate a discrete correlation function (DCF) of irregularly sampled time series, the software package described in Robertson et al. (2015) was used, and to find out the confidence levels of the cross-correlation function the Monte Carlo simulations were applied (Emmanoulopoulos et al. 2013).

### 9.1 Cross correlation function of irregularly sampled time series

Cross-correlation function (CCF) analysis is a common method to analyse multi-frequency light curves observed over the same period of time. The observations in most cases lead to irregularly sampled light curves. In order to determine time lags between frequencies, a discrete correlation function (DCF) for irregularly sampled time series is used (Edelson & Krolik 1988):

$$R_{12}(\tau) = \frac{1}{M}\sum_{ij} \frac{[f_1(i) - \mu_1][f_2(j) - \mu_2]}{\sqrt{(\sigma_1^2 - e_1^2)(\sigma_2^2 - e_2^2)}}, \quad (13)$$

where $\tau$ is a transform time (lag), $f_1(i)$ and $f_2(j)$ are the values of the original data series, for which $\Delta t_{ij} = t_j - t_i$ satisfy the condition $\tau - \Delta\tau/2 \leqslant \Delta t_{ij} < \tau + \Delta\tau/2$, $M$ is a number of data pairs that fall withing the particular interval $\Delta\tau$, $\mu_1$ and $\mu_2$ are the averages for two different data samples, $\sigma_1$ and $\sigma_2$ are their standard deviations, $e_1$ and $e_2$ are the uncertainties of the measurements.

The lag bin width $\Delta\tau$ is basically served as a resolution of the DCF. Its choice is a trade-off between the resolution itself and the stability of DCF calculation. For strongly irregular time series with big gaps that we have in the case of blazar observations, the optimal bin width is around 30 days.

To calculate the DCF of irregularly sampled data we used the software package *pydcf*[6] (Robertson et al. 2015). The package was converted into a *Python 3* module, and was used in a *Jupyter Notebook* script to calculate the DCFs for an ensemble of simulated light curve pairs (see 9.2).

### 9.2 A confidence level calculation for the cross correlation function

In the CCF/DCF analysis an estimation of the confidence levels is usually done by the Monte-Carlo simulations. In this process a large number of the artificial light curve pairs is generated, having the same statistical properties (power spectrum density, probability density function) as in the observed data sets. Finding the cross correlation functions for the ensemble of the paired artificial light curves shows the probability of getting a given CCF value purely by coincidence.

In this paper we used the software tool `DELightcurveSimulation`[7] (Emmanoulopoulos et al. 2013) to model the artificial light curves. Our data for all frequency bands including RATAN-600 and third-party data were pre-processed in advance, which included the interpolation of the irregularly sampled light curves with a step of one day using the Steffen spline interpolation (Steffen 1990). A `Jupyter Notebook` script was written for the artificial light curve modelling, and for the DCF calculation only the observed dates were chosen. Simulation of 5000–10000 light curves containing up to 10000 samples for each frequency band can be done relatively fast using a personal computer.

### 9.3 Cross-correlation analysis of the S4 J0954+658 light curves

In order to find the time lags between the optical and radio light curves, cross–correlation functions were calculated using the approach described in the sections above for the full time interval 2003–2023 and for the chosen epochs 1, 2, and 3. The cross-correlation functions between the mean-subtracted light curves at different frequencies (optical, 5 GHz, and 36.8 GHz) for epochs 1 to 3 and for the entire observing period 2003–2023 are shown in Figs 7–10. The resulting time lags with the maximum DCF values for epochs 1 to 3 are listed in Table 11. We consider the case of a significance level of $2\sigma$ and greater.

Comparison of different light curves is not quite easy due to different data sampling in the radio and optical ranges and the presence of gaps in the measurements. Therefore we used 30-day time lag intervals for the analysis of the light curves over the entire period under study as well as for epochs 1–3 at 5 GHz, while for the optical and 36.8 GHz data over epochs 1–3, 10-day time lag intervals were more acceptable, as these epochs have better cadence, up to 3–5 days.

Analysis of the cross-correlation over the total time interval allow us to make some preliminary conclusions. The most evident result is obtained for the cross-correlation between the radio light curves (top panel in Fig. 7). As we can see, the maximum peak with a significance of about $1.75\sigma$ is located at a time lag of $60^d \pm 5^d$. The cross-correlation over individual epochs (Fig. 8) gives peaks close to the same position (of course, we should exclude the highest peak at a lag of $-400^d$ for epoch 2 as non-physical). For epoch 3 with the best data sampling, the significance of the peak is about $3\sigma$. Thus, we can conclude that there exists a delay of about 60 days between the 5 GHz and 36.8 GHz light curves, and an increase of the flux density at higher frequencies precedes that at the lower ones.

Analysis of the cross-correlation between the radio and optical data is more complex due to the presence of many fast optical flares, as it was noted earlier in 7.1. Here we take into consideration only preceding optical flares, therefore only the right parts of the CCFs with positive lags are analyzed. Evidently, the local maxima at $\sim -200^d$ and $\sim -230^d$ (the center and bottom panels in Fig. 7, 5 GHz, and 36.8 GHz) correspond to correlation of radio flares with much earlier optical flares not related to their true counterparts. Thus, the real correlations between the radio and optical events estimated from the total ranges may be presented by wide humps in the range between $0^d$ and $200^d$ with CCF values of about 0.4 for both ranges: 5 GHz and 36.8 GHz.

We can expect that the correlation between the optical and radio light curves should be more evident for the second and third epochs, where our data are presented with the best cadence and in the absence of serious gaps. These reasons allowed us to increase the signal-to-noise ratio in the resulting CCF (confidence exceeds $2\sigma$ at CCF values of about 0.5 and higher, see Fig.9 and Fig.10), but the inter-range delay may still be varying from 0 to 200 days.

---

[6] https://github.com/astronomerdamo/pydcf
[7] https://github.com/samconnolly/DELightcurveSimulation/

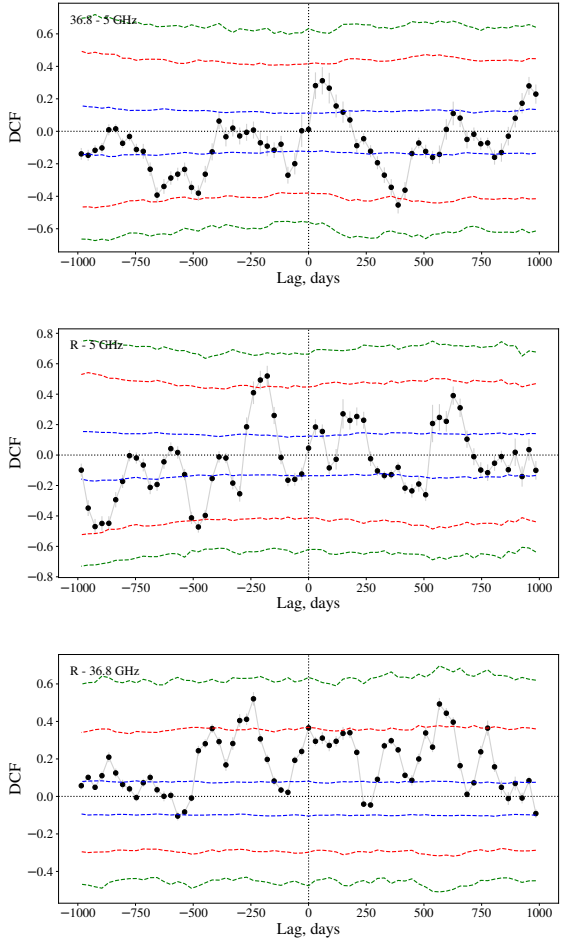

Figure 7. Cross-correlation functions over the entire observing period with a time lag resolution of 30 days. Top panel: between the 36.8 GHz and 5 GHz light curves, center panel: between the optical-range and 5 GHz data, bottom panel: between the optical-range and 36.8 GHz data. The confidence intervals of 1, 2, and $3\sigma$ are shown with the dashed lines (blue, red and green respectively).

Due to the confusion between neighboring fast flares, it is difficult to localize the matching flares in different spectral ranges. Rather, we can identify slow flares in the radio range with groups of fast optical flares with a total duration of about $200^d$, as it was noted in 7.1.

We suggest that the data for epoch 1 show a time lag between the optical and radio 36.8 GHz ranges of about $70^d$, and the data for epoch 2 give local maxima at delays of $5^d$, $200^d$, and $400^d$. Epoch 3, which includes many powerful short flares in the R band, produces two wide humps: the first one is around the zero time lag with CCF values of about 0.5 (the mean position is $30 \pm 20^d$), and the second one is located at a lag of about $160 \pm 30^d$.

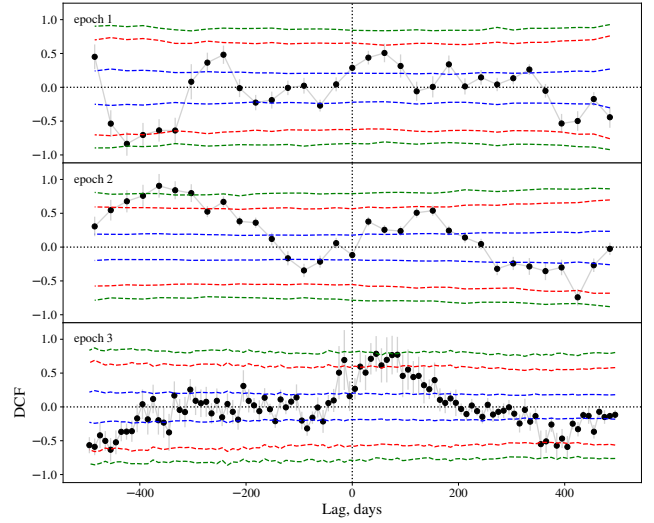

Figure 8. Cross-correlation functions between the 36.8 GHz and 5 GHz light curves with a time lag resolution of 30 days for epochs 1 (top) and 2 (middle) and 10 days for epoch 3 (bottom). The confidence intervals of 1, 2, and $3\sigma$ are shown with the dashed lines, as on Fig.7.

### 9.4 Correlation of optical flares with $\gamma$ photons

In this subsection we will try to find some connection between the set of optical flares and the $\gamma$-ray flux in range 0.1–100 GeV of S4 0954+658 over the last 1000 days (period MJD=59150 – 60150). The data from the Fermi satellite Large Area Telescope (LAT) (Atwood et al. (2009)) with the shortest 3-day binning can be taken from the Fermi LAT Light Curve Repository (Abdollahi et al. (2023)). The binning of the Fermi light curve is well suited to the typical interval in our night-averaged optical data.

For clarity, we plotted both light curves for the R band and $\gamma$-flux in one graph (Fig.11). The original Fermi data, expressed in photons per second per square centimeter, were multiplied by a factor of 30000. Even visual inspection of the optical and $\gamma$-ray fluxes shows generally good agreement with the exception of only a few fast optical flares on MJD $\sim$ 59200–59260, 59500–59550, 59650–59680 and during a quite long interval between MJD = 59800–59900. Within these epochs $\gamma$-emission was in the quiescent state or demonstrated faint events (below 1 in an adopted units or 3-5*$10^{-8}$ photons per second per square centimeter).

The inverse situation is not true: we could not find any $\gamma$-ray flare without an optical counterpart within a 1–2 day interval.

We studied the gamma–optical correlation with the DCF calculated in the manner as described above. The result is shown in Fig. 12, where the optical and $\gamma$-ray data have preliminary been averaged over $5^d$ intervals and the DCF calculated over the same bins within the MJD range between 59150 and 60150. The main peak with $r \sim 0.68$ indicates a fairly strong correlation at an almost zero-time delay (the peak position is $2.5 \pm 2.5$). This is in good agreement with the hypothesis about the common origin of optical photons and hard $\gamma$-rays.

**Table 11.** Time lags with the maximum DCF values for the light curve pairs 36.8 GHz–5 GHz, R band–5 GHz, and R band–36.8 GHz.

| Frequency pair | $\tau$, days | DCF level | $\tau$, days | DCF level | $\tau$, days | DCF level |
|---|---|---|---|---|---|---|
| | epoch 1 | | epoch 2 | | epoch 3 | |
| R-band - 5 GHz | $275 \pm 25$ | 0.21 | $250 \pm 30$ | 0.60 | $210 \pm 20$ | 0.25 |
| 36.8 GHz - 5 GHz | $60 \pm 15$ | 0.51 | $130 \pm 20$ | 0.50 | $55 \pm 10$ | 0.79 |
| R band - 36.8 GHz | $45 \pm 5$ | 0.79 | $380 \pm 10$ | 0.70 | $25 \pm 5$ | 0.55 |

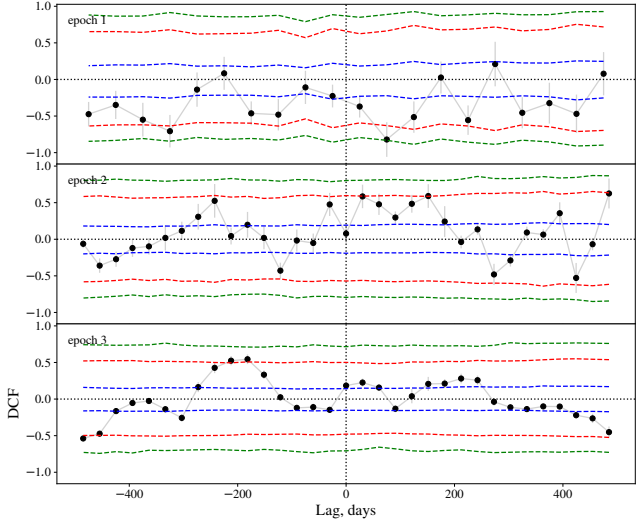

**Figure 9.** Cross-correlation functions between the R band and 5 GHz light curves with a time lag resolution of 50 days for epoch 1 (top) and 30 days for epochs 2 and 3 (bottom). The confidence intervals of 1, 2, and $3\sigma$ are shown with the dashed lines, as on Fig.7.

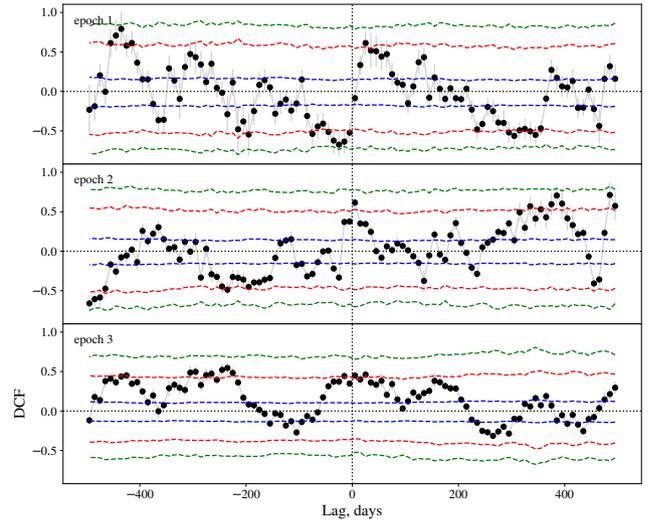

**Figure 10.** Cross-correlation functions between the R band and 36.8 GHz light curves with a time lag resolution of 10 days for the epochs from 1 (top) to 3 (bottom). The confidence intervals of 1, 2, and $3\sigma$ are shown with the dashed lines, as on Fig.7.

## 10  SUMMARY AND DISCUSSION

During a long time period of 2003–2023, S4 0954+658 has been showing extremely high broadband activity with amplitude of flux density variation up to 70–100% in the optical and radio domains. The high optical activity is accompanied by a series of bright radio flares, the amplitude and frequency of which have strongly increased in 2022–2023. During the high radio state period in S4 0954+658, we detect multiple radio flares of various amplitude and duration. The large radio flares last on average from 0.3 to 1 year at 22–36.7 GHz and slightly longer at 5–11.2 GHz. The optical flares consist of 2 types: the slower low-amplitude ones with characteristic time of about 50 days and the faster ones which are shorter and last 7–20 days.

The time structure of non-stationary emission shows that the characteristic time scale $\tau$ of the variable process at 5–22 GHz is about 100 days for epochs 2 and 3 and about 1000 days for the relatively low-active state during epoch 1.

The optical and radio 36.8 GHz variations correlate at de-lays in the interval of 25–200 days. This indicates co-spatiality of the radio and optical emission region, we can observe the same photon population from different emission area (Larionov et al. 2020). The linear size of this region can be estimated as

$$R < c \cdot t_{\text{obs}} \cdot \delta/(1+z), \qquad (14)$$

where $c$ is the speed of light, $t_{\text{obs}}$ is the time delay, $\delta$ is the Doppler factor, which lies in the range of $6.1 < \delta < 35$ (Kishore et al. 2023) for the source. The upper limit of the linear size for $t_{\text{obs}} = 100$ is 0.4 pc, the lower limit is 2 pc.

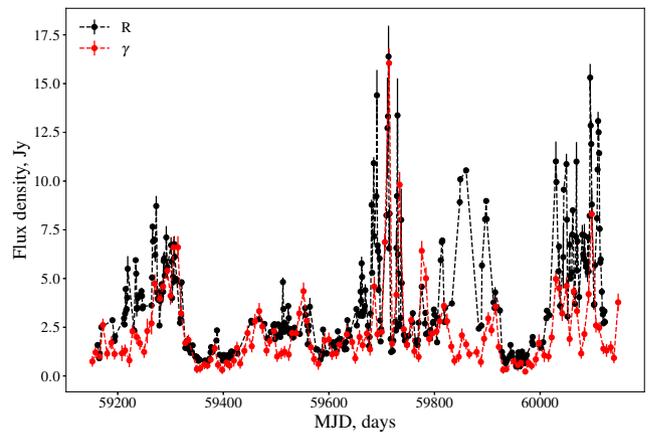

**Figure 11.** Optical and $\gamma$ fluxes of S4 0965+658 over period MJD=59150–60150.

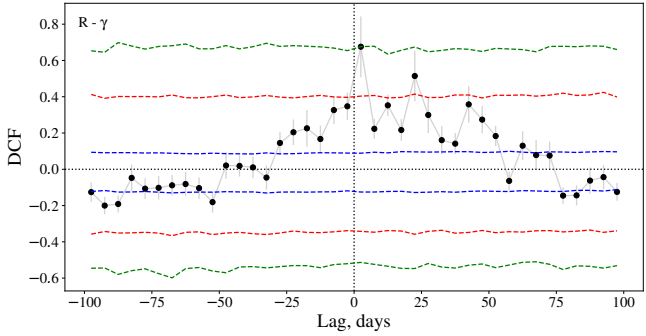

**Figure 12.** Cross-correlation between optical (R band) and $\gamma$ fluxes of S4 0965+658 over period MJD=59150–60150. All data were binned with $5^d$ window.

The DCF between the $\gamma$-ray and optical (R band) light curves over the entire period shows a main peak compatible with a zero-time lag ($2^d.5$).

Since the detection of ultra-high energy neutrino events (IceCube Collaboration et al. 2018), AGNs have become intriguing candidates for high-energy neutrino sources and effective proton accelerators (Kovalev et al. 2020b, 2022). The sources of these neutrinos are still unknown, but it turned out that the sky areas where the ultra-high-energy neutrinos come from are statistically associated with VLBI-bright quasar positions, and the time of their arrival coincide with powerful flares of synchrotron emission in the compact jets of these objects (Plavin et al. 2020, 2021). We have considered the broadband radio spectra of S4 0954+658 using both electrons and protons as emitting particles and showed that the spectra can be well explained by each particle type. The estimated parameters $B_\perp$ and $\Theta$ are in good agreement with those which are generally accepted in the current theories. The estimated brightness temperature $T_b$ for S4 0954+658 does not exceed the known limit of $\sim 10^{12}$ K for relativistic electrons (Kellermann & Pauliny-Toth 1969; Readhead 1994) and $\sim 10^{15}$ K for relativistic protons (Kardashev 2000; Kovalev et al. 2022). The discovered excess of the $10^{12}$ K limit in the RadioAstron space experiment (Kovalev et al. 2020a) for many AGNs could be explained both by the Doppler boosting through the bulk motion of emitting plasma (for a part of such objects) and by relativistic proton emission (for all such sources).

## ACKNOWLEDGEMENTS

The reported study was funded by the Ministry of Science and Higher Education of the Russian Federation under contract No. 075-15-2022-1227. YYK was supported by the M2FINDERS project, which has received funding from the European Research Council (ERC) under the European Union's Horizon 2020 Research and Innovation Programme (grant agreement No. 101018682). LY is supported by the National Natural Science Foundation of China (NSFC) grants No. 12205388. The observations were carried out with the RATAN-600 scientific facility, the Zeiss-1000 and AS-500/2 optical reflectors of SAO RAS, and the RT-22 of CrAO RAS. VAE and VLN are grateful to the staff of the Radio Astronomy Department of CrAO RAS for their participation in the observations. VVV and SOI would like to thank N. S. Kardashev and V. S. Bychkova for their involvement into blazar studies and valuable discussions on the topic of blazar variability. The observations at 5.05 and 8.63 GH was performed with the Badary and Zelenchukskaya radio RT-32 telescopes operated by the Shared Research Facility Center for the Quasar VLBI Network of IAA RAS (https://iaaras.ru/cu-center/). This research has made use of the NASA/IPAC Extragalactic Database (NED), which is operated by the Jet Propulsion Laboratory, California Institute of Technology, under contract with the National Aeronautics and Space Administration; the CATS database, available on the Special Astrophysical Observatory website; the SIMBAD database, operated at CDS, Strasbourg, France. This research has made use of the VizieR catalogue access tool, CDS, Strasbourg, France. The development of the Fermi LAT Light Curve Repository has been funded in part through the Fermi Guest Investigator Program (NASA Research Announcements NNH19ZDA001N and NNH20ZDA001N). This study makes use of the VLBA data from the VLBA-BU Blazar Monitoring Program (BEAM-ME and VLBA-BU-BLAZAR; http://www.bu.edu/blazars/BEAM-ME.html), funded by NASA through the Fermi Guest Investigator Program. The VLBA is an instrument of the National Radio Astronomy Observatory. The National Radio Astronomy Observatory is a facility of the National Science Foundation operated by Associated Universities, Inc.

# APPENDIX A: OPTICAL AND RADIO MEASUREMENTS

Table A1: The R-band measurements of S4 0954+658 in 2003–2023: epoch MJD (Col. 1), epoch yyyy.mm.dd (Col. 2), number of observations $N_{obs}$ (Col. 3), the flux densities in $R$ filter and their errors, mJy (Col. 4).

| MJD epoch | yyyy.mm.dd | $N_{obs}$ | $R_{flux}$, mJy, $\sigma_R$, mJy |
| 1 | 2 | 3 | 4 |
|---|---|---|---|
| 52678 | 2003.02.07 | 1  | $1.24 \pm 0.02$ |
| 52679 | 2003.02.08 | 3  | $1.39 \pm 0.03$ |
| 52699 | 2003.02.28 | 2  | $1.12 \pm 0.01$ |
| 52701 | 2003.03.02 | 4  | $1.56 \pm 0.09$ |
| 52702 | 2003.03.03 | 8  | $1.31 \pm 0.03$ |
| 52705 | 2003.03.06 | 10 | $2.02 \pm 0.06$ |
| 52727 | 2003.03.28 | 2  | $1.45 \pm 0.01$ |
| 52730 | 2003.03.31 | 1  | $1.58 \pm 0.05$ |
| 52731 | 2003.04.01 | 3  | $1.41 \pm 0.02$ |
| 52753 | 2003.04.23 | 2  | $0.86 \pm 0.01$ |
| 52758 | 2003.04.28 | 3  | $0.79 \pm 0.01$ |
| 52761 | 2003.05.01 | 11 | $0.68 \pm 0.03$ |
| 52762 | 2003.05.02 | 10 | $0.58 \pm 0.01$ |
| 52763 | 2003.05.03 | 5  | $0.60 \pm 0.01$ |
| 52782 | 2003.05.22 | 4  | $0.72 \pm 0.03$ |
| 52783 | 2003.05.23 | 4  | $0.78 \pm 0.02$ |
| 52785 | 2003.05.25 | 3  | $0.81 \pm 0.01$ |
| 52796 | 2003.06.05 | 12 | $0.75 \pm 0.04$ |
| 52812 | 2003.06.26 | 1  | $0.53 \pm 0.03$ |
| 52968 | 2003.11.24 | 3  | $1.19 \pm 0.01$ |
| 52996 | 2003.12.22 | 3  | $0.61 \pm 0.01$ |
| 53002 | 2003.12.28 | 3  | $0.66 \pm 0.01$ |
| 53003 | 2003.12.29 | 4  | $0.63 \pm 0.01$ |
| 53032 | 2004.01.27 | 3  | $0.72 \pm 0.01$ |
| 53034 | 2004.01.29 | 3  | $0.81 \pm 0.01$ |
| 53062 | 2004.02.26 | 3  | $0.75 \pm 0.01$ |
| 53083 | 2004.03.18 | 4  | $1.03 \pm 0.01$ |
| 53089 | 2004.03.24 | 3  | $0.86 \pm 0.01$ |
| 53117 | 2004.04.21 | 1  | $1.69 \pm 0.04$ |
| 53118 | 2004.04.22 | 3  | $1.64 \pm 0.01$ |
| 53122 | 2004.04.26 | 3  | $2.07 \pm 0.01$ |
| 53146 | 2004.05.20 | 3  | $1.27 \pm 0.01$ |
| 53174 | 2004.06.17 | 3  | $1.05 \pm 0.02$ |
| 53348 | 2004.12.08 | 6  | $0.91 \pm 0.02$ |
| 53387 | 2005.01.16 | 2  | $0.96 \pm 0.01$ |
| 53388 | 2005.01.17 | 3  | $0.99 \pm 0.01$ |
| 53404 | 2005.02.02 | 2  | $1.28 \pm 0.01$ |
| 53406 | 2005.02.04 | 3  | $1.18 \pm 0.01$ |
| 53407 | 2005.02.05 | 3  | $1.00 \pm 0.01$ |
| 53411 | 2005.02.09 | 3  | $1.06 \pm 0.01$ |
| 53412 | 2005.02.10 | 3  | $1.34 \pm 0.01$ |
| 53444 | 2005.03.14 | 3  | $0.55 \pm 0.01$ |
| 53466 | 2005.04.05 | 3  | $0.37 \pm 0.01$ |
| 53471 | 2005.04.10 | 2  | $0.26 \pm 0.01$ |
| 53529 | 2005.06.07 | 3  | $0.70 \pm 0.01$ |
| 53670 | 2005.10.26 | 3  | $0.37 \pm 0.02$ |
| 53698 | 2005.11.23 | 3  | $0.38 \pm 0.01$ |
| 53705 | 2005.11.30 | 3  | $0.51 \pm 0.01$ |
| 53730 | 2005.12.25 | 5  | $0.78 \pm 0.01$ |
| 53735 | 2005.12.30 | 8  | $0.75 \pm 0.01$ |
| 53822 | 2006.03.27 | 5  | $1.00 \pm 0.01$ |
| 53830 | 2006.04.04 | 5  | $1.42 \pm 0.01$ |
| 53881 | 2006.05.25 | 4  | $0.63 \pm 0.01$ |
| 53888 | 2006.06.01 | 3  | $1.20 \pm 0.01$ |
| 54086 | 2006.12.16 | 5  | $0.64 \pm 0.01$ |
| 54087 | 2006.12.16 | 5  | $0.64 \pm 0.01$ |
| 54088 | 2006.12.17 | 5  | $0.59 \pm 0.01$ |
| 54145 | 2007.02.13 | 6  | $0.40 \pm 0.01$ |
| 54150 | 2007.02.19 | 5  | $0.41 \pm 0.01$ |
| 54170 | 2007.03.10 | 5  | $0.62 \pm 0.01$ |
| 54473 | 2008.01.07 | 5  | $0.93 \pm 0.01$ |
| 54500 | 2008.02.04 | 6  | $0.79 \pm 0.01$ |

Table A1: continued.

| MJD epoch | yyyy.mm.dd | $N_{obs}$ | $R_{flux}$, mJy, $\sigma_R$, mJy |
|---|---|---|---|
| 1 | 2 | 3 | 4 |
| 54509 | 2008.02.12 | 3 | $0.72 \pm 0.01$ |
| 54557 | 2008.03.31 | 3 | $1.28 \pm 0.01$ |
| 54798 | 2008.11.26 | 5 | $0.38 \pm 0.01$ |
| 54805 | 2008.12.03 | 5 | $0.37 \pm 0.01$ |
| 54853 | 2009.01.21 | 8 | $0.45 \pm 0.01$ |
| 54857 | 2009.01.24 | 4 | $0.51 \pm 0.01$ |
| 54912 | 2009.03.21 | 17 | $0.62 \pm 0.01$ |
| 54920 | 2009.03.29 | 5 | $0.53 \pm 0.01$ |
| 55189 | 2009.12.23 | 3 | $0.89 \pm 0.01$ |
| 55241 | 2010.02.13 | 5 | $0.46 \pm 0.01$ |
| 55247 | 2010.02.19 | 5 | $0.36 \pm 0.01$ |
| 55274 | 2010.03.18 | 3 | $0.73 \pm 0.01$ |
| 55293 | 2010.04.06 | 5 | $1.22 \pm 0.01$ |
| 55297 | 2010.04.10 | 5 | $0.78 \pm 0.01$ |
| 55592 | 2011.01.30 | 4 | $0.47 \pm 0.01$ |
| 55646 | 2011.03.30 | 2 | $1.50 \pm 0.02$ |
| 55652 | 2011.03.31 | 5 | $1.39 \pm 0.01$ |
| 55655 | 2011.04.03 | 5 | $1.20 \pm 0.01$ |
| 55925 | 2011.12.28 | 5 | $0.53 \pm 0.01$ |
| 55951 | 2012.01.24 | 5 | $1.25 \pm 0.02$ |
| 55953 | 2012.01.26 | 6 | $0.42 \pm 0.01$ |
| 56006 | 2012.03.19 | 3 | $0.51 \pm 0.01$ |
| 56043 | 2012.04.25 | 3 | $0.67 \pm 0.01$ |
| 56278 | 2012.12.16 | 3 | $0.44 \pm 0.01$ |
| 56335 | 2013.02.11 | 3 | $0.39 \pm 0.01$ |
| 56338 | 2013.02.14 | 16 | $0.44 \pm 0.01$ |
| 56358 | 2013.03.06 | 8 | $0.67 \pm 0.01$ |
| 56423 | 2013.05.10 | 9 | $0.45 \pm 0.01$ |
| 56631 | 2013.12.04 | 3 | $0.32 \pm 0.01$ |
| 56652 | 2013.12.25 | 3 | $0.35 \pm 0.01$ |
| 56659 | 2014.01.01 | 12 | $0.45 \pm 0.01$ |
| 56662 | 2014.01.04 | 3 | $0.48 \pm 0.01$ |
| 56691 | 2014.02.02 | 3 | $0.34 \pm 0.01$ |
| 56692 | 2014.02.03 | 3 | $0.35 \pm 0.01$ |
| 56713 | 2014.02.24 | 5 | $0.38 \pm 0.01$ |
| 56717 | 2014.02.28 | 3 | $0.40 \pm 0.01$ |
| 56722 | 2014.03.05 | 3 | $0.40 \pm 0.01$ |
| 56748 | 2014.03.31 | 6 | $0.38 \pm 0.01$ |
| 56981 | 2014.11.19 | 3 | $0.71 \pm 0.01$ |
| 56982 | 2014.11.20 | 4 | $0.63 \pm 0.01$ |
| 57038 | 2015.01.15 | 3 | $1.46 \pm 0.01$ |
| 57043 | 2015.01.20 | 3 | $0.90 \pm 0.01$ |
| 57064 | 2015.02.10 | 2 | $7.99 \pm 0.02$ |
| 57066 | 2015.02.12 | 4 | $14.26 \pm 0.50$ |
| 57067 | 2015.02.13 | 7 | $11.52 \pm 0.72$ |
| 57068 | 2015.02.14 | 12 | $5.04 \pm 0.33$ |
| 57069 | 2015.02.15 | 3 | $8.90 \pm 0.31$ |
| 57071 | 2015.02.17 | 10 | $8.94 \pm 0.38$ |
| 57072 | 2015.02.18 | 9 | $6.21 \pm 0.25$ |
| 57073 | 2015.02.19 | 1 | $7.59 \pm 0.01$ |
| 57075 | 2015.02.21 | 7 | $3.03 \pm 0.25$ |
| 57076 | 2015.02.22 | 9 | $2.82 \pm 0.20$ |
| 57077 | 2015.02.23 | 10 | $3.95 \pm 0.56$ |
| 57078 | 2015.02.24 | 26 | $4.10 \pm 0.23$ |
| 57089 | 2015.03.07 | 18 | $4.40 \pm 0.18$ |
| 57090 | 2015.03.08 | 21 | $1.98 \pm 0.04$ |
| 57091 | 2015.03.09 | 3 | $1.59 \pm 0.02$ |
| 57092 | 2015.03.10 | 9 | $1.61 \pm 0.02$ |
| 57093 | 2015.03.11 | 7 | $2.66 \pm 0.05$ |
| 57095 | 2015.03.13 | 3 | $1.31 \pm 0.01$ |
| 57096 | 2015.03.14 | 3 | $1.78 \pm 0.03$ |
| 57097 | 2015.03.15 | 11 | $1.10 \pm 0.02$ |
| 57100 | 2015.03.18 | 3 | $0.84 \pm 0.01$ |
| 57101 | 2015.03.19 | 2 | $0.83 \pm 0.01$ |

Table A1: continued.

| MJD epoch | yyyy.mm.dd | $N_{obs}$ | $R_{flux}$, mJy, $\sigma_R$, mJy |
|---|---|---|---|
| 1 | 2 | 3 | 4 |
| 57103 | 2015.03.21 | 5 | $1.43 \pm 0.06$ |
| 57104 | 2015.03.22 | 4 | $0.89 \pm 0.03$ |
| 57105 | 2015.03.23 | 2 | $0.94 \pm 0.01$ |
| 57106 | 2015.03.24 | 13 | $0.90 \pm 0.03$ |
| 57127 | 2015.04.14 | 2 | $1.04 \pm 0.02$ |
| 57129 | 2015.04.16 | 9 | $0.77 \pm 0.01$ |
| 57133 | 2015.04.20 | 2 | $0.66 \pm 0.01$ |
| 57135 | 2015.04.22 | 1 | $0.96 \pm 0.01$ |
| 57136 | 2015.04.23 | 1 | $0.88 \pm 0.01$ |
| 57149 | 2015.05.06 | 1 | $0.71 \pm 0.01$ |
| 57158 | 2015.05.15 | 3 | $0.70 \pm 0.01$ |
| 57160 | 2015.05.17 | 3 | $0.74 \pm 0.01$ |
| 57161 | 2015.05.18 | 3 | $0.65 \pm 0.01$ |
| 57162 | 2015.05.18 | 1 | $0.77 \pm 0.01$ |
| 57163 | 2015.05.20 | 2 | $0.70 \pm 0.01$ |
| 57182 | 2015.06.08 | 2 | $1.62 \pm 0.01$ |
| 57183 | 2015.06.09 | 3 | $1.77 \pm 0.02$ |
| 57187 | 2015.06.13 | 3 | $4.24 \pm 0.03$ |
| 57223 | 2015.07.19 | 2 | $1.25 \pm 0.04$ |
| 57242 | 2015.08.07 | 2 | $1.15 \pm 0.01$ |
| 57245 | 2015.08.10 | 2 | $0.72 \pm 0.01$ |
| 57250 | 2015.08.15 | 2 | $0.65 \pm 0.01$ |
| 57251 | 2015.08.16 | 2 | $0.66 \pm 0.01$ |
| 57258 | 2015.08.23 | 2 | $0.62 \pm 0.01$ |
| 57368 | 2015.12.11 | 3 | $5.50 \pm 0.05$ |
| 57399 | 2016.01.11 | 9 | $3.47 \pm 0.13$ |
| 57434 | 2016.02.15 | 20 | $2.44 \pm 0.13$ |
| 57437 | 2016.02.18 | 17 | $4.28 \pm 0.18$ |
| 57452 | 2016.03.04 | 15 | $2.63 \pm 0.06$ |
| 57454 | 2016.03.06 | 7 | $2.17 \pm 0.18$ |
| 57455 | 2016.03.07 | 3 | $1.74 \pm 0.03$ |
| 57459 | 2016.03.11 | 3 | $1.96 \pm 0.05$ |
| 57460 | 2016.03.12 | 4 | $1.12 \pm 0.02$ |
| 57464 | 2016.03.16 | 2 | $1.41 \pm 0.03$ |
| 57483 | 2016.03.17 | 1 | $1.45 \pm 0.01$ |
| 57484 | 2016.04.05 | 3 | $1.98 \pm 0.01$ |
| 57490 | 2016.04.11 | 3 | $1.87 \pm 0.01$ |
| 57491 | 2016.04.12 | 3 | $1.59 \pm 0.01$ |
| 57525 | 2016.05.16 | 3 | $1.28 \pm 0.01$ |
| 57526 | 2016.05.17 | 4 | $1.36 \pm 0.01$ |
| 57528 | 2016.05.19 | 3 | $1.38 \pm 0.02$ |
| 57543 | 2016.06.03 | 3 | $0.56 \pm 0.01$ |
| 57550 | 2016.06.10 | 3 | $1.11 \pm 0.02$ |
| 57558 | 2016.06.18 | 2 | $1.25 \pm 0.01$ |
| 57695 | 2016.11.02 | 2 | $0.63 \pm 0.02$ |
| 57710 | 2016.11.17 | 2 | $1.44 \pm 0.01$ |
| 57712 | 2016.11.19 | 3 | $1.43 \pm 0.01$ |
| 57713 | 2016.11.20 | 3 | $1.47 \pm 0.01$ |
| 57719 | 2016.11.26 | 2 | $1.28 \pm 0.02$ |
| 57720 | 2016.11.27 | 2 | $1.20 \pm 0.01$ |
| 57749 | 2016.12.26 | 3 | $1.06 \pm 0.01$ |
| 57764 | 2017.01.10 | 3 | $0.83 \pm 0.01$ |
| 57784 | 2017.01.30 | 3 | $1.74 \pm 0.02$ |
| 57785 | 2017.01.31 | 2 | $1.22 \pm 0.01$ |
| 57811 | 2017.02.26 | 3 | $1.64 \pm 0.03$ |
| 57813 | 2017.02.28 | 2 | $1.61 \pm 0.01$ |
| 57830 | 2017.03.17 | 4 | $3.09 \pm 0.02$ |
| 57833 | 2017.03.20 | 1 | $1.59 \pm 0.01$ |
| 57843 | 2017.03.30 | 3 | $0.63 \pm 0.01$ |
| 57844 | 2017.03.31 | 3 | $0.44 \pm 0.01$ |
| 57860 | 2017.04.16 | 3 | $0.54 \pm 0.01$ |
| 57861 | 2017.04.17 | 3 | $0.63 \pm 0.01$ |
| 57862 | 2017.04.18 | 3 | $0.57 \pm 0.01$ |
| 57886 | 2017.05.12 | 3 | $0.51 \pm 0.01$ |

Table A1: continued.

| MJD epoch | yyyy.mm.dd | $N_{obs}$ | $R_{flux}$, mJy, $\sigma_R$, mJy |
|---|---|---|---|
| 1 | 2 | 3 | 4 |
| 57899 | 2017.05.25 | 1 | $0.67 \pm 0.01$ |
| 57900 | 2017.05.26 | 2 | $0.52 \pm 0.01$ |
| 57901 | 2017.05.27 | 2 | $0.52 \pm 0.01$ |
| 58075 | 2017.11.17 | 2 | $1.11 \pm 0.01$ |
| 58077 | 2017.11.19 | 2 | $1.30 \pm 0.01$ |
| 58079 | 2017.11.20 | 2 | $1.15 \pm 0.02$ |
| 58096 | 2017.12.08 | 2 | $1.12 \pm 0.02$ |
| 58100 | 2017.12.12 | 2 | $1.83 \pm 0.01$ |
| 58107 | 2017.12.19 | 2 | $0.84 \pm 0.01$ |
| 58116 | 2017.12.28 | 2 | $1.13 \pm 0.01$ |
| 58121 | 2018.01.02 | 4 | $1.33 \pm 0.04$ |
| 58124 | 2018.01.05 | 2 | $0.99 \pm 0.01$ |
| 58125 | 2018.01.06 | 2 | $0.79 \pm 0.01$ |
| 58134 | 2018.01.15 | 1 | $0.89 \pm 0.01$ |
| 58155 | 2018.02.05 | 2 | $1.72 \pm 0.01$ |
| 58156 | 2018.02.06 | 2 | $1.36 \pm 0.01$ |
| 58163 | 2018.02.13 | 3 | $1.24 \pm 0.02$ |
| 58169 | 2018.02.19 | 2 | $1.99 \pm 0.01$ |
| 58177 | 2018.02.27 | 2 | $1.57 \pm 0.01$ |
| 58183 | 2018.03.05 | 2 | $1.79 \pm 0.01$ |
| 58190 | 2018.03.12 | 2 | $1.29 \pm 0.01$ |
| 58194 | 2018.03.16 | 2 | $1.29 \pm 0.01$ |
| 58199 | 2018.03.21 | 1 | $1.48 \pm 0.01$ |
| 58203 | 2018.03.25 | 2 | $1.60 \pm 0.03$ |
| 58212 | 2018.04.03 | 2 | $1.19 \pm 0.03$ |
| 58213 | 2018.04.04 | 2 | $1.23 \pm 0.01$ |
| 58214 | 2018.04.05 | 2 | $1.74 \pm 0.01$ |
| 58215 | 2018.04.06 | 2 | $2.51 \pm 0.04$ |
| 58216 | 2018.04.07 | 2 | $2.93 \pm 0.01$ |
| 58217 | 2018.04.08 | 2 | $3.27 \pm 0.02$ |
| 58220 | 2018.04.11 | 2 | $2.45 \pm 0.01$ |
| 58221 | 2018.04.12 | 13 | $3.48 \pm 0.36$ |
| 58222 | 2018.04.13 | 2 | $3.36 \pm 0.08$ |
| 58224 | 2018.04.15 | 1 | $2.90 \pm 0.01$ |
| 58225 | 2018.04.16 | 1 | $1.89 \pm 0.01$ |
| 58226 | 2018.04.17 | 1 | $5.55 \pm 0.01$ |
| 58227 | 2018.04.18 | 2 | $5.37 \pm 0.32$ |
| 58230 | 2018.04.21 | 1 | $3.34 \pm 0.01$ |
| 58231 | 2018.04.22 | 1 | $3.58 \pm 0.01$ |
| 58238 | 2018.04.29 | 2 | $1.66 \pm 0.01$ |
| 58240 | 2018.05.01 | 2 | $1.33 \pm 0.01$ |
| 58243 | 2018.05.04 | 2 | $0.97 \pm 0.01$ |
| 58255 | 2018.05.15 | 1 | $1.08 \pm 0.01$ |
| 58256 | 2018.05.17 | 2 | $1.42 \pm 0.02$ |
| 58278 | 2018.06.08 | 2 | $2.11 \pm 0.06$ |
| 58283 | 2018.06.13 | 2 | $1.28 \pm 0.01$ |
| 58288 | 2018.06.18 | 3 | $1.16 \pm 0.01$ |
| 58308 | 2018.07.08 | 3 | $1.72 \pm 0.01$ |
| 58309 | 2018.07.09 | 1 | $1.75 \pm 0.01$ |
| 58310 | 2018.07.10 | 3 | $1.37 \pm 0.01$ |
| 58335 | 2018.08.04 | 4 | $1.43 \pm 0.02$ |
| 58360 | 2018.08.29 | 2 | $1.37 \pm 0.01$ |
| 58366 | 2018.09.04 | 2 | $0.72 \pm 0.01$ |
| 58418 | 2018.10.26 | 2 | $2.42 \pm 0.02$ |
| 58419 | 2018.10.27 | 5 | $1.91 \pm 0.01$ |
| 58422 | 2018.10.30 | 2 | $2.63 \pm 0.03$ |
| 58423 | 2018.10.31 | 2 | $3.15 \pm 0.03$ |
| 58424 | 2018.11.01 | 2 | $1.63 \pm 0.01$ |
| 58425 | 2018.11.02 | 3 | $2.16 \pm 0.01$ |
| 58426 | 2018.11.03 | 2 | $1.80 \pm 0.01$ |
| 58427 | 2018.11.04 | 2 | $1.72 \pm 0.01$ |
| 58428 | 2018.11.05 | 1 | $1.57 \pm 0.01$ |
| 58430 | 2018.11.07 | 5 | $1.87 \pm 0.03$ |
| 58431 | 2018.11.08 | 2 | $1.44 \pm 0.01$ |

Table A1: continued.

| MJD epoch | yyyy.mm.dd | $N_{obs}$ | $R_{flux}$, mJy, $\sigma_R$, mJy |
|---|---|---|---|
| 1 | 2 | 3 | 4 |
| 58432 | 2018.11.09 | 2 | 1.50 ± 0.01 |
| 58433 | 2018.11.10 | 1 | 1.39 ± 0.01 |
| 58434 | 2018.11.11 | 1 | 2.04 ± 0.01 |
| 58435 | 2018.11.12 | 3 | 1.83 ± 0.01 |
| 58501 | 2019.01.17 | 4 | 2.40 ± 0.08 |
| 58502 | 2019.01.18 | 2 | 1.63 ± 0.01 |
| 58503 | 2019.01.19 | 2 | 2.20 ± 0.01 |
| 58504 | 2019.01.20 | 24 | 2.82 ± 0.05 |
| 58517 | 2019.02.02 | 4 | 3.10 ± 0.01 |
| 58518 | 2019.02.03 | 2 | 2.75 ± 0.01 |
| 58530 | 2019.02.15 | 10 | 2.30 ± 0.10 |
| 58531 | 2019.02.16 | 3 | 1.78 ± 0.01 |
| 58532 | 2019.02.17 | 3 | 2.16 ± 0.01 |
| 58533 | 2019.02.18 | 1 | 1.92 ± 0.01 |
| 58534 | 2019.02.19 | 1 | 1.90 ± 0.01 |
| 58549 | 2019.03.06 | 2 | 2.42 ± 0.04 |
| 58550 | 2019.03.07 | 2 | 2.60 ± 0.01 |
| 58568 | 2019.03.25 | 2 | 2.79 ± 0.01 |
| 58573 | 2019.03.30 | 2 | 2.56 ± 0.01 |
| 58578 | 2019.04.04 | 2 | 1.72 ± 0.01 |
| 58580 | 2019.04.06 | 2 | 2.33 ± 0.01 |
| 58582 | 2019.04.08 | 2 | 1.98 ± 0.01 |
| 58608 | 2019.05.04 | 2 | 3.15 ± 0.03 |
| 58609 | 2019.05.05 | 10 | 2.54 ± 0.02 |
| 58610 | 2019.05.06 | 9 | 2.45 ± 0.02 |
| 58611 | 2019.05.07 | 10 | 2.52 ± 0.01 |
| 58619 | 2019.05.15 | 2 | 1.43 ± 0.06 |
| 58620 | 2019.05.16 | 2 | 1.39 ± 0.01 |
| 58621 | 2019.05.17 | 1 | 1.15 ± 0.01 |
| 58639 | 2019.06.04 | 2 | 2.69 ± 0.01 |
| 58643 | 2019.06.08 | 2 | 2.94 ± 0.01 |
| 58648 | 2019.06.13 | 3 | 3.80 ± 0.03 |
| 58649 | 2019.06.14 | 4 | 3.75 ± 0.02 |
| 58651 | 2019.06.16 | 3 | 5.96 ± 0.01 |
| 58657 | 2019.06.22 | 1 | 3.65 ± 0.01 |
| 58727 | 2019.08.31 | 1 | 1.85 ± 0.01 |
| 58743 | 2019.09.16 | 2 | 4.86 ± 0.06 |
| 58744 | 2019.09.17 | 3 | 2.46 ± 0.06 |
| 58781 | 2019.10.24 | 3 | 0.97 ± 0.02 |
| 58782 | 2019.10.25 | 3 | 1.34 ± 0.01 |
| 58800 | 2019.11.12 | 3 | 2.73 ± 0.02 |
| 58801 | 2019.11.13 | 2 | 2.18 ± 0.01 |
| 58802 | 2019.11.15 | 2 | 2.39 ± 0.03 |
| 58804 | 2019.11.16 | 2 | 2.27 ± 0.05 |
| 58810 | 2019.11.22 | 1 | 2.33 ± 0.01 |
| 58812 | 2019.11.24 | 1 | 2.71 ± 0.01 |
| 58839 | 2019.12.21 | 1 | 1.22 ± 0.01 |
| 58841 | 2019.12.23 | 2 | 1.90 ± 0.01 |
| 58894 | 2020.02.14 | 2 | 2.96 ± 0.01 |
| 58905 | 2020.02.25 | 2 | 2.26 ± 0.01 |
| 58907 | 2020.02.27 | 2 | 1.91 ± 0.01 |
| 58915 | 2020.20.03 | 4 | 1.70 ± 0.05 |
| 58930 | 2020.20.03 | 5 | 2.24 ± 0.02 |
| 58931 | 2020.20.03 | 2 | 2.15 ± 0.01 |
| 58942 | 2020.04.02 | 3 | 3.47 ± 0.03 |
| 58958 | 2020.20.04 | 2 | 3.52 ± 0.03 |
| 58959 | 2020.20.04 | 3 | 4.36 ± 0.01 |
| 58966 | 2020.04.26 | 2 | 3.19 ± 0.03 |
| 58968 | 2020.04.28 | 2 | 1.74 ± 0.01 |
| 59162 | 2020.11.08 | 2 | 1.59 ± 0.01 |
| 59165 | 2020.11.11 | 4 | 0.91 ± 0.01 |
| 59166 | 2020.11.12 | 2 | 1.15 ± 0.01 |
| 59169 | 2020.11.15 | 2 | 2.50 ± 0.02 |
| 59190 | 2020.12.06 | 2 | 2.86 ± 0.01 |

Table A1: continued.

| MJD epoch | yyyy.mm.dd | $N_{obs}$ | $R_{flux}$, mJy, $\sigma_R$, mJy |
|---|---|---|---|
| 1 | 2 | 3 | 4 |
| 59194 | 2020.12.10 | 3 | $1.80 \pm 0.01$ |
| 59195 | 2020.12.11 | 2 | $2.03 \pm 0.02$ |
| 59212 | 2020.12.27 | 4 | $2.96 \pm 0.02$ |
| 59213 | 2020.12.28 | 3 | $2.64 \pm 0.05$ |
| 59214 | 2020.12.29 | 2 | $3.20 \pm 0.02$ |
| 59215 | 2020.12.30 | 2 | $4.46 \pm 0.01$ |
| 59216 | 2021.01.01 | 6 | $3.38 \pm 0.39$ |
| 59217 | 2021.01.02 | 19 | $3.43 \pm 0.06$ |
| 59218 | 2021.01.03 | 33 | $4.02 \pm 0.25$ |
| 59219 | 2021.01.04 | 12 | $5.50 \pm 0.65$ |
| 59231 | 2021.01.16 | 5 | $2.34 \pm 0.06$ |
| 59234 | 2021.01.19 | 2 | $5.94 \pm 0.02$ |
| 59235 | 2021.01.20 | 21 | $5.25 \pm 0.35$ |
| 59236 | 2021.01.21 | 21 | $3.45 \pm 0.40$ |
| 59237 | 2021.01.22 | 27 | $3.50 \pm 0.20$ |
| 59238 | 2021.01.23 | 34 | $4.00 \pm 0.30$ |
| 59240 | 2021.01.25 | 1 | $4.13 \pm 0.01$ |
| 59245 | 2021.01.30 | 5 | $4.35 \pm 0.18$ |
| 59246 | 2021.01.31 | 12 | $4.10 \pm 0.17$ |
| 59247 | 2021.02.01 | 13 | $3.47 \pm 0.08$ |
| 59248 | 2021.02.02 | 13 | $3.57 \pm 0.12$ |
| 59264 | 2021.02.18 | 11 | $3.62 \pm 0.06$ |
| 59265 | 2021.02.19 | 20 | $5.05 \pm 0.10$ |
| 59266 | 2021.02.20 | 8 | $7.66 \pm 0.12$ |
| 59267 | 2021.02.21 | 5 | $6.91 \pm 0.06$ |
| 59270 | 2021.02.24 | 11 | $6.99 \pm 0.14$ |
| 59271 | 2021.02.25 | 6 | $6.21 \pm 0.35$ |
| 59273 | 2021.02.27 | 29 | $8.72 \pm 0.52$ |
| 59275 | 2021.03.01 | 23 | $4.29 \pm 0.33$ |
| 59276 | 2021.03.02 | 27 | $3.97 \pm 0.10$ |
| 59277 | 2021.03.03 | 26 | $3.90 \pm 0.08$ |
| 59278 | 2021.03.04 | 66 | $4.25 \pm 0.25$ |
| 59279 | 2021.03.05 | 5 | $2.59 \pm 0.02$ |
| 59281 | 2021.03.07 | 43 | $3.86 \pm 0.54$ |
| 59282 | 2021.03.08 | 34 | $4.09 \pm 0.48$ |
| 59283 | 2021.03.09 | 47 | $4.70 \pm 0.16$ |
| 59286 | 2021.03.12 | 27 | $4.32 \pm 0.41$ |
| 59287 | 2021.03.13 | 51 | $5.90 \pm 0.40$ |
| 59288 | 2021.03.14 | 48 | $6.03 \pm 0.35$ |
| 59289 | 2021.03.15 | 22 | $5.15 \pm 0.15$ |
| 59291 | 2021.03.17 | 4 | $5.39 \pm 0.14$ |
| 59292 | 2021.03.18 | 10 | $7.11 \pm 0.58$ |
| 59294 | 2021.03.19 | 7 | $6.39 \pm 0.45$ |
| 59294 | 2021.03.20 | 25 | $4.80 \pm 0.90$ |
| 59301 | 2021.03.27 | 29 | $6.71 \pm 0.41$ |
| 59302 | 2021.03.28 | 2 | $5.87 \pm 0.07$ |
| 59303 | 2021.03.29 | 25 | $4.00 \pm 0.25$ |
| 59305 | 2021.03.31 | 15 | $5.09 \pm 0.45$ |
| 59306 | 2021.04.01 | 12 | $4.84 \pm 0.20$ |
| 59307 | 2021.04.02 | 19 | $6.77 \pm 0.41$ |
| 59308 | 2021.04.03 | 3 | $6.12 \pm 0.03$ |
| 59312 | 2021.04.07 | 35 | $4.94 \pm 0.21$ |
| 59318 | 2021.04.13 | 8 | $2.73 \pm 0.18$ |
| 59319 | 2021.04.14 | 31 | $2.95 \pm 0.15$ |
| 59320 | 2021.04.15 | 3 | $4.77 \pm 0.17$ |
| 59321 | 2021.04.16 | 2 | $4.81 \pm 0.01$ |
| 59328 | 2021.04.23 | 4 | $1.44 \pm 0.01$ |
| 59332 | 2021.04.27 | 33 | $1.80 \pm 0.25$ |
| 59333 | 2021.04.28 | 22 | $1.40 \pm 0.04$ |
| 59335 | 2021.04.30 | 32 | $1.59 \pm 0.10$ |
| 59336 | 2021.05.01 | 19 | $1.42 \pm 0.07$ |
| 59337 | 2021.05.02 | 20 | $1.32 \pm 0.04$ |
| 59338 | 2021.05.03 | 14 | $1.19 \pm 0.03$ |
| 59340 | 2021.05.05 | 30 | $1.21 \pm 0.02$ |

| MJD epoch | yyyy.mm.dd | $N_{obs}$ | $R_{flux}$, mJy, $\sigma_R$, mJy |
|---|---|---|---|
| 1 | 2 | 3 | 4 |
| 59342 | 2021.05.07 | 19 | $1.57 \pm 0.03$ |
| 59347 | 2021.05.12 | 3 | $0.99 \pm 0.01$ |
| 59348 | 2021.05.13 | 21 | $1.00 \pm 0.02$ |
| 59349 | 2021.05.14 | 27 | $0.99 \pm 0.03$ |
| 59350 | 2021.05.15 | 22 | $0.95 \pm 0.04$ |
| 59351 | 2021.05.16 | 30 | $0.94 \pm 0.03$ |
| 59353 | 2021.05.18 | 9 | $0.78 \pm 0.04$ |
| 59354 | 2021.05.19 | 10 | $0.84 \pm 0.03$ |
| 59355 | 2021.05.20 | 3 | $0.83 \pm 0.01$ |
| 59365 | 2021.05.30 | 23 | $0.79 \pm 0.03$ |
| 59366 | 2021.05.31 | 7 | $0.77 \pm 0.02$ |
| 59369 | 2021.06.03 | 18 | $0.68 \pm 0.01$ |
| 59373 | 2021.06.07 | 14 | $0.83 \pm 0.01$ |
| 59374 | 2021.06.08 | 9 | $0.88 \pm 0.02$ |
| 59376 | 2021.06.10 | 3 | $1.21 \pm 0.01$ |
| 59377 | 2021.06.11 | 5 | $1.48 \pm 0.02$ |
| 59378 | 2021.06.12 | 5 | $1.24 \pm 0.01$ |
| 59379 | 2021.06.13 | 12 | $1.23 \pm 0.01$ |
| 59387 | 2021.06.21 | 3 | $1.82 \pm 0.05$ |
| 59388 | 2021.06.22 | 2 | $2.34 \pm 0.01$ |
| 59393 | 2021.06.27 | 3 | $1.09 \pm 0.07$ |
| 59398 | 2021.07.02 | 4 | $0.98 \pm 0.03$ |
| 59399 | 2021.07.03 | 3 | $0.79 \pm 0.01$ |
| 59401 | 2021.07.05 | 3 | $1.08 \pm 0.01$ |
| 59406 | 2021.07.10 | 2 | $0.98 \pm 0.03$ |
| 59407 | 2021.07.11 | 4 | $0.93 \pm 0.01$ |
| 59409 | 2021.07.13 | 4 | $1.07 \pm 0.02$ |
| 59412 | 2021.07.16 | 3 | $0.97 \pm 0.02$ |
| 59413 | 2021.07.17 | 3 | $0.93 \pm 0.03$ |
| 59414 | 2021.07.18 | 3 | $1.06 \pm 0.03$ |
| 59415 | 2021.07.19 | 6 | $1.25 \pm 0.02$ |
| 59416 | 2021.07.20 | 5 | $1.25 \pm 0.05$ |
| 59425 | 2021.07.29 | 4 | $1.29 \pm 0.02$ |
| 59427 | 2021.07.31 | 3 | $1.15 \pm 0.01$ |
| 59453 | 2021.08.26 | 3 | $2.83 \pm 0.02$ |
| 59454 | 2021.08.27 | 3 | $2.74 \pm 0.05$ |
| 59468 | 2021.09.10 | 3 | $2.91 \pm 0.06$ |
| 59478 | 2021.09.20 | 3 | $2.67 \pm 0.04$ |
| 59483 | 2021.09.25 | 3 | $1.93 \pm 0.02$ |
| 59490 | 2021.10.02 | 2 | $2.26 \pm 0.09$ |
| 59495 | 2021.10.07 | 3 | $2.60 \pm 0.02$ |
| 59496 | 2021.10.08 | 9 | $2.56 \pm 0.07$ |
| 59497 | 2021.10.09 | 11 | $2.66 \pm 0.13$ |
| 59498 | 2021.10.10 | 3 | $1.89 \pm 0.05$ |
| 59500 | 2021.10.12 | 9 | $2.11 \pm 0.09$ |
| 59504 | 2021.10.16 | 6 | $2.65 \pm 0.07$ |
| 59508 | 2021.10.20 | 10 | $2.00 \pm 0.03$ |
| 59509 | 2021.10.21 | 10 | $2.21 \pm 0.02$ |
| 59510 | 2021.10.22 | 7 | $2.34 \pm 0.02$ |
| 59511 | 2021.10.23 | 6 | $2.57 \pm 0.03$ |
| 59513 | 2021.10.25 | 4 | $4.82 \pm 0.23$ |
| 59514 | 2021.10.26 | 7 | $3.44 \pm 0.04$ |
| 59515 | 2021.10.27 | 9 | $2.73 \pm 0.02$ |
| 59516 | 2021.10.28 | 2 | $2.27 \pm 0.01$ |
| 59517 | 2021.10.29 | 7 | $2.52 \pm 0.10$ |
| 59518 | 2021.10.30 | 6 | $2.68 \pm 0.02$ |
| 59522 | 2021.11.03 | 9 | $2.40 \pm 0.03$ |
| 59523 | 2021.11.04 | 7 | $3.59 \pm 0.04$ |
| 59524 | 2021.11.05 | 9 | $2.97 \pm 0.07$ |
| 59525 | 2021.11.06 | 5 | $2.31 \pm 0.02$ |
| 59526 | 2021.11.07 | 21 | $2.18 \pm 0.05$ |
| 59527 | 2021.11.08 | 21 | $2.41 \pm 0.07$ |
| 59530 | 2021.11.11 | 20 | $2.28 \pm 0.04$ |
| 59531 | 2021.11.12 | 16 | $2.03 \pm 0.02$ |

Table A1: continued.

| MJD epoch | yyyy.mm.dd | $N_{obs}$ | $R_{flux}$, mJy, $\sigma_R$, mJy |
|---|---|---|---|
| 1 | 2 | 3 | 4 |
| 59532 | 2021.11.13 | 20 | $1.97 \pm 0.05$ |
| 59535 | 2021.11.16 | 20 | $2.48 \pm 0.07$ |
| 59545 | 2021.11.26 | 16 | $2.16 \pm 0.05$ |
| 59555 | 2021.12.06 | 3 | $3.48 \pm 0.03$ |
| 59556 | 2021.12.07 | 15 | $3.13 \pm 0.18$ |
| 59559 | 2021.12.10 | 30 | $2.60 \pm 0.15$ |
| 59560 | 2021.12.11 | 19 | $1.63 \pm 0.05$ |
| 59561 | 2021.12.12 | 24 | $2.31 \pm 0.08$ |
| 59562 | 2021.12.13 | 7 | $2.58 \pm 0.05$ |
| 59564 | 2021.12.15 | 18 | $3.55 \pm 0.46$ |
| 59576 | 2021.12.27 | 10 | $1.36 \pm 0.04$ |
| 59578 | 2021.12.29 | 15 | $1.36 \pm 0.09$ |
| 59584 | 2022.01.04 | 2 | $0.84 \pm 0.01$ |
| 59586 | 2022.01.06 | 10 | $1.02 \pm 0.01$ |
| 59587 | 2022.01.07 | 19 | $1.25 \pm 0.06$ |
| 59589 | 2022.01.09 | 13 | $2.38 \pm 0.13$ |
| 59595 | 2022.01.15 | 17 | $1.15 \pm 0.10$ |
| 59596 | 2022.01.16 | 10 | $1.16 \pm 0.02$ |
| 59605 | 2022.01.25 | 16 | $1.62 \pm 0.07$ |
| 59608 | 2022.01.28 | 16 | $1.73 \pm 0.04$ |
| 59609 | 2022.01.29 | 11 | $1.55 \pm 0.05$ |
| 59614 | 2022.02.03 | 3 | $1.28 \pm 0.02$ |
| 59616 | 2022.02.05 | 12 | $1.80 \pm 0.10$ |
| 59617 | 2022.02.06 | 14 | $2.29 \pm 0.06$ |
| 59621 | 2022.02.10 | 10 | $1.88 \pm 0.08$ |
| 59622 | 2022.02.11 | 21 | $1.90 \pm 0.10$ |
| 59626 | 2022.02.15 | 26 | $1.55 \pm 0.06$ |
| 59630 | 2022.02.19 | 3 | $1.36 \pm 0.02$ |
| 59632 | 2022.02.21 | 16 | $1.74 \pm 0.04$ |
| 59633 | 2022.02.22 | 4 | $1.38 \pm 0.01$ |
| 59635 | 2022.02.24 | 4 | $2.86 \pm 0.08$ |
| 59637 | 2022.02.26 | 3 | $2.31 \pm 0.03$ |
| 59638 | 2022.02.27 | 4 | $2.35 \pm 0.07$ |
| 59639 | 2022.02.28 | 4 | $2.05 \pm 0.14$ |
| 59652 | 2022.03.13 | 3 | $3.43 \pm 0.02$ |
| 59653 | 2022.03.14 | 5 | $2.59 \pm 0.07$ |
| 59656 | 2022.03.17 | 1 | $3.19 \pm 0.01$ |
| 59659 | 2022.03.20 | 4 | $3.10 \pm 0.07$ |
| 59660 | 2022.03.21 | 6 | $4.33 \pm 0.50$ |
| 59661 | 2022.03.22 | 6 | $3.85 \pm 0.27$ |
| 59662 | 2022.03.23 | 19 | $5.78 \pm 0.34$ |
| 59663 | 2022.03.24 | 5 | $5.08 \pm 0.15$ |
| 59664 | 2022.03.25 | 2 | $3.09 \pm 0.01$ |
| 59665 | 2022.03.26 | 11 | $3.33 \pm 0.10$ |
| 59667 | 2022.03.28 | 19 | $3.30 \pm 0.25$ |
| 59668 | 2022.03.29 | 3 | $3.59 \pm 0.05$ |
| 59669 | 2022.03.30 | 31 | $4.40 \pm 0.31$ |
| 59674 | 2022.04.04 | 4 | $2.58 \pm 0.05$ |
| 59676 | 2022.04.06 | 4 | $1.54 \pm 0.01$ |
| 59677 | 2022.04.07 | 20 | $2.00 \pm 0.06$ |
| 59679 | 2022.04.09 | 32 | $3.20 \pm 0.23$ |
| 59681 | 2022.04.11 | 15 | $8.78 \pm 0.27$ |
| 59682 | 2022.04.12 | 1 | $5.29 \pm 0.01$ |
| 59685 | 2022.04.15 | 6 | $10.92 \pm 0.31$ |
| 59686 | 2022.04.16 | 35 | $7.17 \pm 0.42$ |
| 59690 | 2022.04.20 | 9 | $9.22 \pm 0.41$ |
| 59691 | 2022.04.21 | 25 | $14.40 \pm 1.30$ |
| 59692 | 2022.04.22 | 5 | $6.70 \pm 0.15$ |
| 59693 | 2022.04.23 | 37 | $6.40 \pm 0.40$ |
| 59694 | 2022.04.24 | 29 | $4.42 \pm 0.18$ |
| 59696 | 2022.04.26 | 3 | $6.03 \pm 0.78$ |
| 59697 | 2022.04.27 | 30 | $2.85 \pm 0.15$ |
| 59698 | 2022.04.28 | 3 | $1.77 \pm 0.05$ |
| 59699 | 2022.04.29 | 31 | $2.40 \pm 0.30$ |

Table A1: continued.

| MJD epoch | yyyy.mm.dd | $N_{obs}$ | $R_{flux}$, mJy, $\sigma_R$, mJy |
|---|---|---|---|
| 1 | 2 | 3 | 4 |
| 59709 | 2022.05.09 | 2 | $8.24 \pm 0.08$ |
| 59711 | 2022.05.11 | 99 | $12.72 \pm 2.59$ |
| 59712 | 2022.05.12 | 119 | $13.32 \pm 1.61$ |
| 59713 | 2022.05.13 | 203 | $16.39 \pm 1.58$ |
| 59714 | 2022.05.14 | 32 | $6.55 \pm 0.13$ |
| 59715 | 2022.05.15 | 38 | $8.33 \pm 0.41$ |
| 59717 | 2022.05.17 | 141 | $1.90 \pm 0.17$ |
| 59718 | 2022.05.18 | 40 | $1.23 \pm 0.04$ |
| 59721 | 2022.05.21 | 7 | $1.28 \pm 0.03$ |
| 59722 | 2022.05.22 | 1 | $1.81 \pm 0.01$ |
| 59726 | 2022.05.26 | 3 | $2.79 \pm 0.08$ |
| 59727 | 2022.05.27 | 2 | $2.33 \pm 0.01$ |
| 59728 | 2022.05.28 | 112 | $6.57 \pm 0.92$ |
| 59729 | 2022.05.29 | 62 | $9.23 \pm 0.36$ |
| 59730 | 2022.05.30 | 209 | $13.37 \pm 1.90$ |
| 59731 | 2022.05.31 | 18 | $6.24 \pm 0.45$ |
| 59732 | 2022.06.01 | 26 | $5.25 \pm 0.43$ |
| 59733 | 2022.06.02 | 79 | $2.68 \pm 0.09$ |
| 59734 | 2022.06.03 | 84 | $2.55 \pm 0.10$ |
| 59736 | 2022.06.05 | 57 | $2.97 \pm 0.16$ |
| 59737 | 2022.06.06 | 48 | $8.01 \pm 1.61$ |
| 59738 | 2022.06.07 | 20 | $4.77 \pm 0.52$ |
| 59742 | 2022.06.11 | 22 | $2.11 \pm 0.13$ |
| 59744 | 2022.06.13 | 19 | $1.81 \pm 0.08$ |
| 59745 | 2022.06.14 | 10 | $1.98 \pm 0.05$ |
| 59748 | 2022.06.17 | 9 | $1.60 \pm 0.04$ |
| 59751 | 2022.06.20 | 10 | $2.06 \pm 0.06$ |
| 59758 | 2022.06.27 | 4 | $2.74 \pm 0.04$ |
| 59759 | 2022.06.28 | 4 | $3.22 \pm 0.03$ |
| 59764 | 2022.07.03 | 3 | $1.80 \pm 0.01$ |
| 59765 | 2022.07.04 | 4 | $1.80 \pm 0.06$ |
| 59766 | 2022.07.05 | 4 | $1.67 \pm 0.03$ |
| 59767 | 2022.07.06 | 4 | $1.67 \pm 0.01$ |
| 59768 | 2022.07.07 | 4 | $1.58 \pm 0.03$ |
| 59769 | 2022.07.08 | 3 | $1.83 \pm 0.03$ |
| 59770 | 2022.07.09 | 4 | $2.73 \pm 0.08$ |
| 59776 | 2022.07.15 | 4 | $4.58 \pm 0.01$ |
| 59782 | 2022.07.21 | 2 | $2.69 \pm 0.01$ |
| 59783 | 2022.07.22 | 8 | $3.24 \pm 0.04$ |
| 59784 | 2022.07.23 | 6 | $2.15 \pm 0.02$ |
| 59788 | 2022.07.27 | 3 | $2.23 \pm 0.03$ |
| 59791 | 2022.07.30 | 3 | $1.91 \pm 0.05$ |
| 59792 | 2022.07.31 | 4 | $2.35 \pm 0.01$ |
| 59797 | 2022.08.05 | 5 | $2.72 \pm 0.05$ |
| 59798 | 2022.08.06 | 10 | $2.78 \pm 0.07$ |
| 59799 | 2022.08.07 | 4 | $3.41 \pm 0.02$ |
| 59800 | 2022.08.08 | 4 | $3.15 \pm 0.08$ |
| 59805 | 2022.08.13 | 7 | $1.73 \pm 0.07$ |
| 59806 | 2022.08.14 | 4 | $2.97 \pm 0.07$ |
| 59812 | 2022.08.20 | 8 | $5.94 \pm 0.07$ |
| 59814 | 2022.08.22 | 4 | $6.87 \pm 0.06$ |
| 59815 | 2022.08.23 | 16 | $6.95 \pm 0.13$ |
| 59819 | 2022.08.27 | 9 | $2.77 \pm 0.08$ |
| 59820 | 2022.08.28 | 7 | $3.49 \pm 0.07$ |
| 59833 | 2022.09.10 | 5 | $3.72 \pm 0.03$ |
| 59848 | 2022.09.25 | 3 | $8.92 \pm 0.07$ |
| 59849 | 2022.09.26 | 11 | $10.09 \pm 0.17$ |
| 59860 | 2022.10.07 | 21 | $10.55 \pm 0.12$ |
| 59884 | 2022.10.31 | 6 | $2.43 \pm 0.04$ |
| 59889 | 2022.11.05 | 6 | $2.58 \pm 0.02$ |
| 59890 | 2022.11.06 | 6 | $5.67 \pm 0.08$ |
| 59896 | 2022.11.12 | 9 | $8.04 \pm 0.08$ |
| 59898 | 2022.11.14 | 9 | $8.98 \pm 0.06$ |
| 59899 | 2022.11.15 | 6 | $8.05 \pm 0.06$ |

Table A1: continued.

| MJD epoch | yyyy.mm.dd | $N_{obs}$ | $R_{flux}$, mJy, $\sigma_R$, mJy |
|---|---|---|---|
| 1 | 2 | 3 | 4 |
| 59912 | 2022.11.28 | 9 | $3.77 \pm 0.10$ |
| 59914 | 2022.11.30 | 8 | $3.49 \pm 0.03$ |
| 59915 | 2022.12.01 | 22 | $4.29 \pm 0.27$ |
| 59924 | 2022.12.10 | 12 | $3.36 \pm 0.19$ |
| 59928 | 2022.12.14 | 8 | $1.23 \pm 0.04$ |
| 59930 | 2022.12.16 | 7 | $0.97 \pm 0.02$ |
| 59934 | 2022.12.20 | 10 | $0.75 \pm 0.06$ |
| 59935 | 2022.12.21 | 4 | $0.65 \pm 0.04$ |
| 59937 | 2022.12.23 | 8 | $0.78 \pm 0.02$ |
| 59938 | 2022.12.24 | 11 | $0.85 \pm 0.01$ |
| 59943 | 2022.12.29 | 2 | $1.04 \pm 0.01$ |
| 59944 | 2022.12.30 | 2 | $1.59 \pm 0.01$ |
| 59945 | 2022.12.31 | 3 | $1.10 \pm 0.01$ |
| 59946 | 2023.01.01 | 4 | $1.02 \pm 0.03$ |
| 59947 | 2023.01.02 | 6 | $1.20 \pm 0.03$ |
| 59955 | 2023.01.10 | 2 | $0.48 \pm 0.01$ |
| 59956 | 2023.01.11 | 2 | $0.53 \pm 0.01$ |
| 59957 | 2023.01.12 | 2 | $0.67 \pm 0.01$ |
| 59958 | 2023.01.13 | 2 | $0.99 \pm 0.02$ |
| 59959 | 2023.01.14 | 9 | $1.21 \pm 0.01$ |
| 59960 | 2023.01.15 | 7 | $0.83 \pm 0.02$ |
| 59961 | 2023.01.16 | 17 | $0.97 \pm 0.01$ |
| 59962 | 2023.01.17 | 13 | $0.49 \pm 0.01$ |
| 59963 | 2023.01.18 | 17 | $0.85 \pm 0.05$ |
| 59964 | 2023.01.19 | 10 | $1.17 \pm 0.05$ |
| 59965 | 2023.01.20 | 12 | $0.80 \pm 0.02$ |
| 59966 | 2023.01.21 | 12 | $1.25 \pm 0.14$ |
| 59968 | 2023.01.23 | 9 | $0.69 \pm 0.02$ |
| 59969 | 2023.01.24 | 8 | $0.97 \pm 0.01$ |
| 59970 | 2023.01.25 | 6 | $0.70 \pm 0.01$ |
| 59971 | 2023.01.26 | 12 | $0.66 \pm 0.01$ |
| 59972 | 2023.01.27 | 10 | $0.93 \pm 0.01$ |
| 59975 | 2023.01.30 | 14 | $1.08 \pm 0.04$ |
| 59985 | 2023.02.09 | 5 | $1.75 \pm 0.01$ |
| 59986 | 2023.02.10 | 5 | $2.00 \pm 0.03$ |
| 59987 | 2023.02.11 | 4 | $1.60 \pm 0.02$ |
| 60002 | 2023.02.26 | 8 | $1.45 \pm 0.03$ |
| 60006 | 2023.03.02 | 3 | $1.73 \pm 0.01$ |
| 60012 | 2023.03.08 | 9 | $2.96 \pm 0.02$ |
| 60013 | 2023.03.09 | 4 | $3.35 \pm 0.03$ |
| 60018 | 2023.03.14 | 3 | $3.12 \pm 0.05$ |
| 60030 | 2023.03.26 | 105 | $11.01 \pm 1.02$ |
| 60031 | 2023.03.27 | 78 | $9.95 \pm 0.42$ |
| 60035 | 2023.03.31 | 15 | $5.37 \pm 0.15$ |
| 60036 | 2023.04.01 | 28 | $6.64 \pm 0.18$ |
| 60038 | 2023.04.03 | 28 | $3.60 \pm 0.25$ |
| 60044 | 2023.04.09 | 18 | $6.10 \pm 1.60$ |
| 60045 | 2023.04.10 | 4 | $9.56 \pm 0.04$ |
| 60050 | 2023.04.15 | 15 | $10.87 \pm 0.53$ |
| 60051 | 2023.04.16 | 12 | $8.02 \pm 0.25$ |
| 60052 | 2023.04.17 | 11 | $3.05 \pm 0.21$ |
| 60056 | 2023.04.21 | 26 | $6.74 \pm 0.48$ |
| 60057 | 2023.04.22 | 13 | $5.14 \pm 0.08$ |
| 60058 | 2023.04.23 | 9 | $5.00 \pm 0.10$ |
| 60060 | 2023.04.25 | 24 | $5.31 \pm 0.07$ |
| 60061 | 2023.04.26 | 16 | $7.25 \pm 0.25$ |
| 60062 | 2023.04.27 | 15 | $8.50 \pm 0.15$ |
| 60065 | 2023.04.30 | 9 | $4.28 \pm 0.06$ |
| 60066 | 2023.05.01 | 10 | $3.70 \pm 0.10$ |
| 60068 | 2023.05.03 | 12 | $7.14 \pm 0.13$ |
| 60069 | 2023.05.04 | 82 | $11.00 \pm 1.00$ |
| 60070 | 2023.05.05 | 32 | $5.41 \pm 0.10$ |
| 60071 | 2023.05.06 | 18 | $6.91 \pm 0.14$ |
| 60074 | 2023.05.09 | 39 | $4.05 \pm 0.55$ |

Table A1: continued.

| MJD epoch | yyyy.mm.dd | $N_{obs}$ | $R_{flux}$, mJy, $\sigma_R$, mJy |
|---|---|---|---|
| 1 | 2 | 3 | 4 |
| 60080 | 2023.05.15 | 39 | $7.25 \pm 0.50$ |
| 60082 | 2023.05.17 | 32 | $6.25 \pm 0.15$ |
| 60083 | 2023.05.18 | 17 | $6.00 \pm 0.50$ |
| 60084 | 2023.05.19 | 31 | $7.20 \pm 0.60$ |
| 60087 | 2023.05.22 | 7 | $5.60 \pm 0.55$ |
| 60089 | 2023.05.24 | 26 | $5.80 \pm 0.18$ |
| 60090 | 2023.05.25 | 18 | $5.33 \pm 0.51$ |
| 60091 | 2023.05.26 | 34 | $7.11 \pm 0.16$ |
| 60092 | 2023.05.27 | 32 | $5.36 \pm 0.16$ |
| 60094 | 2023.05.29 | 4 | $8.22 \pm 0.09$ |
| 60095 | 2023.05.30 | 70 | $15.31 \pm 0.70$ |
| 60096 | 2023.05.31 | 34 | $12.85 \pm 0.56$ |
| 60097 | 2023.06.01 | 133 | $11.90 \pm 1.20$ |
| 60098 | 2023.06.02 | 185 | $8.80 \pm 0.70$ |
| 60102 | 2023.06.06 | 53 | $3.65 \pm 0.25$ |
| 60103 | 2023.06.07 | 15 | $3.69 \pm 0.04$ |
| 60104 | 2023.06.08 | 24 | $5.68 \pm 0.19$ |
| 60108 | 2023.06.12 | 15 | $8.09 \pm 0.13$ |
| 60109 | 2023.06.13 | 18 | $10.59 \pm 0.17$ |
| 60110 | 2023.06.14 | 33 | $13.08 \pm 0.47$ |
| 60111 | 2023.06.15 | 8 | $12.50 \pm 0.16$ |
| 60112 | 2023.06.16 | 14 | $11.43 \pm 0.12$ |
| 60113 | 2023.06.17 | 16 | $7.55 \pm 0.19$ |
| 60114 | 2023.06.18 | 12 | $5.82 \pm 0.04$ |
| 60115 | 2023.06.19 | 17 | $5.98 \pm 0.34$ |
| 60117 | 2023.06.21 | 4 | $4.32 \pm 0.19$ |
| 60118 | 2023.06.22 | 3 | $3.42 \pm 0.04$ |
| 60119 | 2023.06.23 | 5 | $3.16 \pm 0.03$ |
| 60120 | 2023.06.24 | 10 | $2.70 \pm 0.08$ |
| 60121 | 2023.06.25 | 8 | $3.31 \pm 0.06$ |
| 60122 | 2023.06.26 | 7 | $3.32 \pm 0.05$ |
| 60123 | 2023.06.27 | 11 | $2.77 \pm 0.07$ |

Table A2: The RATAN-600 measurements in 1998–2023 and the RT-32 data in 2020-2023: epoch MJD (Col. 1), epoch yyyy.mm.dd (Col. 2), number of observations $N_{obs}$ (Col. 3), the flux densities at 21.7/22.3, 11.2, 7.7/8.2/8.6 GHz, 4.7/5.1, 2.3, 0.96/1.1/1.2 GHz and their errors, Jy (Cols. 4–9); name of telescope (Col. 10). The number $N_{obs}$ is indicated if information is available; for RT-32 the number of observations is given for two frequencies.

| MJD epoch | yyyy.mm.dd | $N_{obs}$ | $S_{22}, \sigma$ | $S_{11.2}, \sigma$ | $S_8, \sigma$ | $S_5, \sigma$ | $S_{2.3}, \sigma$ | $S_{1.1}, \sigma$ | Telescope |
|---|---|---|---|---|---|---|---|---|---|
| | | | | | (Jy) | | | | |
| 51079 | 1998.09.23 | – | 0.26 ± 0.04 | 0.28 ± 0.01 | 0.31 ± 0.02 | 0.33 ± 0.01 | 0.40 ± 0.08 | 0.59 ± 0.15 | RATAN-600 |
| 52248 | 2001.12.05 | – | – | 0.74 ± 0.02 | 0.78 ± 0.06 | 0.74 ± 0.02 | 0.75 ± 0.10 | – | RATAN-600 |
| 52405 | 2002.05.11 | – | 0.60 ± 0.09 | 0.75 ± 0.12 | 0.63 ± 0.06 | 0.57 ± 0.18 | – | – | RATAN-600 |
| 52580 | 2002.11.02 | – | 0.75 ± 0.05 | 0.67 ± 0.03 | 0.63 ± 0.06 | 0.50 ± 0.06 | – | – | RATAN-600 |
| 52744 | 2003.04.15 | – | 0.58 ± 0.12 | 0.55 ± 0.02 | 0.51 ± 0.04 | 0.49 ± 0.07 | – | – | RATAN-600 |
| 52803 | 2003.06.13 | – | 0.66 ± 0.08 | 0.55 ± 0.03 | 0.49 ± 0.02 | 0.36 ± 0.02 | – | – | RATAN-600 |
| 52865 | 2003.08.14 | – | 0.69 ± 0.08 | 0.54 ± 0.05 | 0.50 ± 0.04 | – | – | – | RATAN-600 |
| 52920 | 2003.10.08 | – | – | 0.61 ± 0.07 | 0.49 ± 0.08 | 0.48 ± 0.02 | 0.69 ± 0.12 | – | RATAN-600 |
| 53475 | 2005.04.15 | – | – | 1.33 ± 0.04 | 1.06 ± 0.04 | 0.91 ± 0.09 | – | – | RATAN-600 |
| 53603 | 2005.08.21 | – | 1.13 ± 0.16 | 1.35 ± 0.18 | 0.93 ± 0.06 | 1.23 ± 0.07 | – | – | RATAN-600 |
| 53888 | 2006.06.02 | – | 1.12 ± 0.08 | 1.50 ± 0.03 | 1.14 ± 0.18 | 1.16 ± 0.08 | – | – | RATAN-600 |
| 54014 | 2006.10.06 | – | – | 1.88 ± 0.06 | 1.59 ± 0.30 | – | – | 1.86 ± 0.43 | RATAN-600 |
| 54198 | 2007.04.08 | – | 1.44 ± 0.07 | 1.83 ± 0.06 | 1.76 ± 0.18 | 1.47 ± 0.09 | 1.24 ± 0.17 | 2.41 ± 0.40 | RATAN-600 |
| 54340 | 2007.08.28 | – | 1.48 ± 0.10 | 1.39 ± 0.04 | 1.39 ± 0.14 | – | 0.83 ± 0.19 | – | RATAN-600 |
| 54710 | 2008.09.01 | – | 1.71 ± 0.06 | 2.14 ± 0.05 | 2.02 ± 0.14 | 1.68 ± 0.09 | – | – | RATAN-600 |
| 55236 | 2010.02.09 | – | 1.02 ± 0.15 | 1.71 ± 0.08 | 1.66 ± 0.10 | – | – | – | RATAN-600 |
| 55425 | 2010.08.17 | – | 1.01 ± 0.12 | 1.34 ± 0.06 | 1.50 ± 0.14 | 1.48 ± 0.22 | – | – | RATAN-600 |
| 55575 | 2011.01.14 | – | 1.11 ± 0.17 | 1.13 ± 0.05 | 1.20 ± 0.04 | 1.38 ± 0.05 | – | – | RATAN-600 |
| 55607 | 2011.02.15 | – | 1.30 ± 0.13 | 1.28 ± 0.07 | 1.18 ± 0.18 | 1.13 ± 0.05 | – | – | RATAN-600 |
| 55666 | 2011.04.15 | – | – | 1.63 ± 0.07 | 1.57 ± 0.06 | 1.36 ± 0.06 | – | – | RATAN-600 |
| 55845 | 2011.10.11 | – | 0.93 ± 0.23 | 1.46 ± 0.07 | 1.35 ± 0.12 | 1.53 ± 0.06 | – | – | RATAN-600 |
| 55918 | 2011.12.23 | – | 1.05 ± 0.09 | 1.16 ± 0.04 | 1.23 ± 0.04 | 1.26 ± 0.06 | – | – | RATAN-600 |
| 55966 | 2012.02.09 | – | 0.97 ± 0.05 | 1.16 ± 0.04 | 1.13 ± 0.16 | 1.09 ± 0.03 | – | – | RATAN-600 |
| 56215 | 2012.10.15 | – | 1.35 ± 0.14 | 1.14 ± 0.05 | 0.91 ± 0.08 | 0.74 ± 0.09 | – | – | RATAN-600 |
| 56246 | 2012.11.15 | – | 1.17 ± 0.22 | 1.18 ± 0.03 | 1.10 ± 0.14 | 0.80 ± 0.08 | – | – | RATAN-600 |
| 56269 | 2012.12.08 | – | 1.42 ± 0.17 | 1.25 ± 0.05 | 1.27 ± 0.06 | – | – | – | RATAN-600 |
| 56317 | 2013.01.25 | – | 1.08 ± 0.11 | 1.23 ± 0.02 | 1.24 ± 0.08 | – | – | – | RATAN-600 |
| 56672 | 2014.01.15 | – | 1.76 ± 0.18 | 1.76 ± 0.09 | 1.80 ± 0.10 | 1.39 ± 0.03 | – | – | RATAN-600 |
| 56703 | 2014.02.15 | – | 1.91 ± 0.27 | 1.55 ± 0.07 | 1.44 ± 0.18 | 1.15 ± 0.06 | – | – | RATAN-600 |
| 57034 | 2015.01.11 | 4 | 1.33 ± 0.11 | 1.42 ± 0.08 | – | 1.52 ± 0.03 | – | – | RATAN-600 |
| 57092 | 2015.03.10 | 5 | 1.62 ± 0.13 | 1.75 ± 0.07 | – | 1.49 ± 0.05 | – | – | RATAN-600 |
| 57122 | 2015.04.09 | 4 | 1.29 ± 0.13 | 1.50 ± 0.07 | – | 1.30 ± 0.05 | – | – | RATAN-600 |
| 57153 | 2015.05.10 | 4 | 1.14 ± 0.11 | 1.03 ± 0.06 | – | 1.08 ± 0.05 | – | – | RATAN-600 |
| 57183 | 2015.06.09 | 4 | 0.90 ± 0.10 | 1.11 ± 0.08 | – | 1.11 ± 0.05 | – | – | RATAN-600 |
| 57214 | 2015.07.10 | 5 | 0.94 ± 0.10 | 0.97 ± 0.06 | – | 1.11 ± 0.05 | – | – | RATAN-600 |
| 57274 | 2015.09.08 | 2 | 1.15 ± 0.10 | 1.04 ± 0.08 | – | 0.85 ± 0.05 | – | – | RATAN-600 |
| 57305 | 2015.10.09 | 3 | 1.25 ± 0.10 | 1.08 ± 0.08 | – | 0.87 ± 0.05 | – | – | RATAN-600 |
| 57336 | 2015.11.09 | 3 | 1.36 ± 0.10 | 0.70 ± 0.05 | – | 0.67 ± 0.05 | – | – | RATAN-600 |
| 57367 | 2015.12.10 | 4 | 1.91 ± 0.15 | 1.29 ± 0.06 | – | 0.81 ± 0.05 | – | – | RATAN-600 |
| 57398 | 2016.01.10 | 5 | 2.19 ± 0.15 | 1.59 ± 0.07 | – | 0.97 ± 0.05 | – | – | RATAN-600 |
| 57429 | 2016.02.10 | 5 | 1.93 ± 0.10 | 1.63 ± 0.07 | – | 1.01 ± 0.05 | – | – | RATAN-600 |
| 57458 | 2016.03.10 | 6 | 2.07 ± 0.15 | 1.84 ± 0.06 | – | 1.15 ± 0.05 | – | – | RATAN-600 |
| 57519 | 2016.05.10 | 3 | 1.41 ± 0.10 | 1.28 ± 0.07 | – | 1.01 ± 0.05 | – | – | RATAN-600 |
| 57550 | 2016.06.10 | 4 | 1.48 ± 0.15 | 1.10 ± 0.07 | – | 1.14 ± 0.05 | – | – | RATAN-600 |
| 57580 | 2016.07.10 | 2 | 1.14 ± 0.10 | 0.93 ± 0.06 | – | 0.98 ± 0.05 | – | – | RATAN-600 |
| 57640 | 2016.09.08 | 3 | 1.26 ± 0.10 | 1.18 ± 0.07 | – | 0.91 ± 0.05 | – | – | RATAN-600 |
| 57674 | 2016.10.12 | 3 | 1.19 ± 0.10 | 1.03 ± 0.06 | – | 0.88 ± 0.05 | – | – | RATAN-600 |
| 57703 | 2016.11.10 | 6 | 0.91 ± 0.10 | 0.96 ± 0.06 | – | 0.97 ± 0.05 | – | – | RATAN-600 |
| 57733 | 2016.12.10 | 6 | 0.87 ± 0.10 | 0.79 ± 0.06 | – | 0.90 ± 0.05 | – | – | RATAN-600 |
| 57762 | 2017.01.08 | 6 | 0.83 ± 0.10 | 0.77 ± 0.06 | – | 0.69 ± 0.04 | – | – | RATAN-600 |
| 57795 | 2017.02.10 | 5 | 1.33 ± 0.10 | 0.79 ± 0.07 | – | 0.67 ± 0.03 | – | – | RATAN-600 |
| 57823 | 2017.03.10 | 6 | 1.28 ± 0.10 | 0.95 ± 0.03 | – | 0.74 ± 0.04 | – | – | RATAN-600 |
| 57854 | 2017.04.10 | 5 | 1.27 ± 0.10 | 0.99 ± 0.06 | – | 0.70 ± 0.04 | – | – | RATAN-600 |
| 57884 | 2017.05.10 | 5 | 1.22 ± 0.15 | 0.93 ± 0.06 | – | 0.71 ± 0.04 | – | – | RATAN-600 |
| 57915 | 2017.06.10 | 5 | 0.89 ± 0.10 | 0.82 ± 0.06 | – | 0.89 ± 0.05 | – | – | RATAN-600 |
| 57945 | 2017.07.10 | 3 | – | 0.84 ± 0.06 | – | 0.89 ± 0.05 | – | – | RATAN-600 |
| 58909 | 2020.03.01 | 43/– | – | – | 1.02 ± 0.06 | – | – | – | RT-32 |
| 58916 | 2020.03.08 | 62/42 | – | – | 0.95 ± 0.04 | 0.93 ± 0.08 | – | – | RT-32 |
| 58923 | 2020.03.15 | –/88 | – | – | – | 0.93 ± 0.04 | – | – | RT-32 |
| 58937 | 2020.03.29 | 67/86 | – | – | 1.22 ± 0.14 | 0.92 ± 0.03 | – | – | RT-32 |

Table A2: continued.

| MJD epoch | yyyy.mm.dd | $N_{obs}$ | $S_{22.3}, \sigma$ | $S_{11.2}, \sigma$ | $S_{7.7/8.2}, \sigma$ | $S_{4.7}, \sigma$ | $S_{2.3}, \sigma$ | $S_{1}, \sigma$ | Telescope |
| 1 | 2 | 3 | 4 | 5 | 6 | 7 | 8 | 9 | 10 |
|---|---|---|---|---|---|---|---|---|---|
| 58944 | 2020.04.05 | 61/60 | – | – | $1.02 \pm 0.08$ | $0.97 \pm 0.03$ | – | – | RT-32 |
| 58951 | 2020.04.12 | 63/82 | – | – | $1.01 \pm 0.06$ | $0.95 \pm 0.02$ | – | – | RT-32 |
| 58958 | 2020.04.19 | 62/86 | – | – | $1.12 \pm 0.06$ | $0.95 \pm 0.02$ | – | – | RT-32 |
| 58965 | 2020.04.26 | 62/44 | – | – | $1.15 \pm 0.06$ | $0.98 \pm 0.02$ | – | – | RT-32 |
| 58972 | 2020.05.03 | 47/82 | – | – | $1.12 \pm 0.06$ | $0.94 \pm 0.02$ | – | – | RT-32 |
| 58979 | 2020.05.10 | -/82 | – | – | – | $0.99 \pm 0.02$ | – | – | RT-32 |
| 58986 | 2020.05.17 | 48/- | – | – | $1.15 \pm 0.06$ | – | – | – | RT-32 |
| 58993 | 2020.05.24 | 44/88 | – | – | $1.07 \pm 0.06$ | $0.98 \pm 0.07$ | – | – | RT-32 |
| 59000 | 2020.05.31 | 62/44 | – | – | $1.12 \pm 0.06$ | $1.01 \pm 0.02$ | – | – | RT-32 |
| 59006 | 2020.06.06 | 64/20 | – | – | $1.11 \pm 0.06$ | $1.02 \pm 0.11$ | – | – | RT-32 |
| 59013 | 2020.06.13 | -/44 | – | – | – | $1.03 \pm 0.04$ | – | – | RT-32 |
| 59024 | 2020.06.24 | 65/118 | – | – | $1.17 \pm 0.04$ | $1.01 \pm 0.05$ | – | – | RT-32 |
| 59030 | 2020.06.30 | 68/112 | – | – | $1.28 \pm 0.06$ | $1.12 \pm 0.04$ | – | – | RT-32 |
| 59035 | 2020.07.05 | 45/- | – | – | $1.29 \pm 0.08$ | – | – | – | RT-32 |
| 59036 | 2020.07.06 | -/110 | – | – | – | $1.10 \pm 0.04$ | – | – | RT-32 |
| 59041 | 2020.07.11 | 64/- | – | – | $1.18 \pm 0.08$ | – | – | – | RT-32 |
| 59042 | 2020.07.12 | -/86 | – | – | – | $1.04 \pm 0.03$ | – | – | RT-32 |
| 59048 | 2020.07.18 | 44/- | – | – | $1.09 \pm 0.10$ | – | – | – | RT-32 |
| 59049 | 2020.07.19 | -/44 | – | – | – | $0.99 \pm 0.02$ | – | – | RT-32 |
| 59070 | 2020.08.09 | -/90 | – | – | – | $1.01 \pm 0.02$ | – | – | RT-32 |
| 59077 | 2020.08.16 | -/132 | – | – | – | $1.03 \pm 0.03$ | – | – | RT-32 |
| 59097 | 2020.09.05 | 46/- | – | – | $0.88 \pm 0.12$ | – | – | – | RT-32 |
| 59098 | 2020.09.06 | -/92 | – | – | – | $0.96 \pm 0.03$ | – | – | RT-32 |
| 59112 | 2020.09.20 | 43/86 | – | – | $0.92 \pm 0.06$ | $0.93 \pm 0.04$ | – | – | RT-32 |
| 59126 | 2020.10.04 | 44/- | – | – | $0.89 \pm 0.04$ | – | – | – | RT-32 |
| 59140 | 2020.10.18 | 62/90 | – | – | $0.88 \pm 0.04$ | $0.81 \pm 0.02$ | – | – | RT-32 |
| 59161 | 2020.11.08 | 46/89 | – | – | $0.89 \pm 0.04$ | $0.81 \pm 0.02$ | – | – | RT-32 |
| 59168 | 2020.11.15 | 46/82 | – | – | $0.88 \pm 0.06$ | $0.72 \pm 0.09$ | – | – | RT-32 |
| 59175 | 2020.11.22 | 44/- | – | – | $0.89 \pm 0.06$ | – | – | – | RT-32 |
| 59189 | 2020.12.06 | 36/- | – | – | $0.79 \pm 0.04$ | – | – | – | RT-32 |
| 59196 | 2020.12.13 | 45/88 | – | – | $0.93 \pm 0.06$ | $0.74 \pm 0.04$ | – | – | RT-32 |
| 59203 | 2020.12.20 | 45/84 | – | – | $0.80 \pm 0.08$ | $0.72 \pm 0.02$ | – | – | RT-32 |
| 59217 | 2021.01.03 | 30/82 | – | – | $0.88 \pm 0.04$ | $0.73 \pm 0.03$ | – | – | RT-32 |
| 59224 | 2021.01.10 | 42/90 | – | – | $0.76 \pm 0.06$ | $0.70 \pm 0.08$ | – | – | RT-32 |
| 59230 | 2021.01.16 | 43/84 | – | – | $0.91 \pm 0.06$ | $0.69 \pm 0.03$ | – | – | RT-32 |
| 59237 | 2021.01.23 | -/86 | – | – | – | $0.71 \pm 0.03$ | – | – | RT-32 |
| 59244 | 2021.01.30 | -/84 | – | – | – | $0.79 \pm 0.02$ | – | – | RT-32 |
| 59251 | 2021.02.06 | 133/- | – | – | $0.88 \pm 0.10$ | – | – | – | RT-32 |
| 59254 | 2021.02.08 | 3 | $2.04 \pm 0.17$ | $1.13 \pm 0.10$ | – | $0.74 \pm 0.04$ | – | – | RATAN-600 |
| 59259 | 2021.02.14 | 87/- | – | – | $1.10 \pm 0.08$ | – | – | – | RT-32 |
| 59266 | 2021.02.21 | 86/90 | – | – | $1.25 \pm 0.06$ | $0.93 \pm 0.03$ | – | – | RT-32 |
| 59282 | 2021.03.08 | 2 | $1.96 \pm 0.19$ | $1.39 \pm 0.10$ | – | $0.82 \pm 0.05$ | – | – | RATAN-600 |
| 59293 | 2021.03.20 | 45/90 | – | – | $1.14 \pm 0.06$ | $0.93 \pm 0.02$ | – | – | RT-32 |
| 59300 | 2021.03.27 | 42/86 | – | – | $1.21 \pm 0.08$ | $0.88 \pm 0.02$ | – | – | RT-32 |
| 59307 | 2021.04.03 | 46/- | – | – | $1.29 \pm 0.06$ | – | – | – | RT-32 |
| 59312 | 2021.04.07 | 2 | – | $1.22 \pm 0.10$ | – | $0.82 \pm 0.05$ | – | – | RATAN-600 |
| 59321 | 2021.04.17 | 44/- | – | – | $1.25 \pm 0.06$ | – | – | – | RT-32 |
| 59322 | 2021.04.18 | -/88 | – | – | – | $0.98 \pm 0.02$ | – | – | RT-32 |
| 59328 | 2021.04.24 | 21/- | – | – | $1.10 \pm 0.14$ | – | – | – | RT-32 |
| 59342 | 2021.05.08 | 45/- | – | – | $0.97 \pm 0.04$ | – | – | – | RT-32 |
| 59343 | 2021.05.08 | 2 | $1.47 \pm 0.16$ | $1.15 \pm 0.10$ | – | $0.75 \pm 0.04$ | – | – | RATAN-600 |
| 59349 | 2021.05.15 | 91/- | – | – | $0.94 \pm 0.08$ | – | – | – | RT-32 |
| 59356 | 2021.05.22 | 44/84 | – | – | $0.80 \pm 0.04$ | $0.79 \pm 0.02$ | – | – | RT-32 |
| 59375 | 2021.06.09 | 1 | – | $0.71 \pm 0.10$ | – | $0.63 \pm 0.04$ | – | – | RATAN-600 |
| 59384 | 2021.06.19 | 41/- | – | – | $0.65 \pm 0.04$ | – | – | – | RT-32 |
| 59399 | 2021.07.04 | 60/88 | – | – | $0.62 \pm 0.04$ | $0.59 \pm 0.03$ | – | – | RT-32 |
| 59405 | 2021.07.10 | 79/84 | – | – | $0.59 \pm 0.04$ | $0.55 \pm 0.02$ | – | – | RT-32 |
| 59410 | 2021.07.14 | 12 | $1.25 \pm 0.14$ | $1.03 \pm 0.10$ | – | $0.89 \pm 0.03$ | – | – | RATAN-600 |
| 59412 | 2021.07.17 | 94/- | – | – | $0.54 \pm 0.04$ | – | – | – | RT-32 |
| 59419 | 2021.07.24 | -/78 | – | – | – | $0.53 \pm 0.05$ | – | – | RT-32 |
| 59438 | 2021.08.11 | 10 | $1.17 \pm 0.10$ | $1.02 \pm 0.10$ | – | $0.81 \pm 0.05$ | – | – | RATAN-600 |
| 59474 | 2021.09.16 | 14 | $1.30 \pm 0.10$ | $1.37 \pm 0.10$ | – | $1.15 \pm 0.05$ | – | – | RATAN-600 |
| 59495 | 2021.10.07 | 25 | $1.45 \pm 0.10$ | $1.28 \pm 0.10$ | – | $1.05 \pm 0.05$ | – | – | RATAN-600 |
| 59556 | 2021.12.07 | 3 | $0.86 \pm 0.10$ | $0.88 \pm 0.10$ | – | $0.69 \pm 0.02$ | – | – | RATAN-600 |

Table A2: continued.

| MJD epoch | yyyy.mm.dd | $N_{obs}$ | $S_{22.3}, \sigma$ | $S_{11.2}, \sigma$ | $S_{7.7/8.2}, \sigma$ | $S_{4.7}, \sigma$ | $S_{2.3}, \sigma$ | $S_{1}, \sigma$ | Telescope |
|---|---|---|---|---|---|---|---|---|---|
| 1 | 2 | 3 | 4 | 5 | 6 | 7 | 8 | 9 | 10 |
| 59588 | 2022.01.08 | 2 | $0.98 \pm 0.10$ | $0.91 \pm 0.10$ | − | $0.62 \pm 0.04$ | − | − | RATAN-600 |
| 59619 | 2022.02.08 | 3 | $1.14 \pm 0.10$ | $1.11 \pm 0.10$ | − | $0.74 \pm 0.04$ | − | − | RATAN-600 |
| 59647 | 2022.03.08 | 2 | $1.58 \pm 0.10$ | $1.51 \pm 0.10$ | − | $0.79 \pm 0.05$ | − | − | RATAN-600 |
| 59678 | 2022.04.08 | 3 | $1.81 \pm 0.10$ | $1.36 \pm 0.10$ | − | $0.86 \pm 0.05$ | − | − | RATAN-600 |
| 59709 | 2022.05.09 | 2 | $1.85 \pm 0.10$ | $1.52 \pm 0.20$ | − | $0.79 \pm 0.05$ | − | − | RATAN-600 |
| 59723 | 2022.05.23 | 2 | $1.89 \pm 0.10$ | $1.46 \pm 0.10$ | − | $0.84 \pm 0.05$ | − | − | RATAN-600 |
| 59848 | 2022.09.25 | 5 | $2.27 \pm 0.18$ | $2.54 \pm 0.30$ | − | $2.24 \pm 0.05$ | − | − | RATAN-600 |
| 59891 | 2022.11.07 | 3 | $1.31 \pm 0.10$ | $1.64 \pm 0.20$ | − | $1.59 \pm 0.05$ | − | − | RATAN-600 |
| 59922 | 2022.12.08 | 3 | $1.24 \pm 0.15$ | $1.48 \pm 0.10$ | − | $1.55 \pm 0.05$ | − | − | RATAN-600 |
| 59953 | 2023.01.08 | 2 | $0.85 \pm 0.10$ | $1.44 \pm 0.10$ | − | $1.40 \pm 0.05$ | − | − | RATAN-600 |
| 59985 | 2023.02.09 | 1 | − | $0.73 \pm 0.10$ | − | $1.04 \pm 0.05$ | − | − | RATAN-600 |
| 60012 | 2023.03.08 | 2 | $1.92 \pm 0.15$ | $1.12 \pm 0.10$ | − | $0.86 \pm 0.05$ | − | − | RATAN-600 |
| 60043 | 2023.04.08 | 2 | $1.63 \pm 0.16$ | $1.57 \pm 0.20$ | − | $1.06 \pm 0.05$ | − | − | RATAN-600 |
| 60074 | 2023.05.09 | 1 | $1.36 \pm 0.14$ | $1.81 \pm 0.20$ | − | $0.86 \pm 0.05$ | − | − | RATAN-600 |
| 60104 | 2023.06.08 | 2 | − | $1.68 \pm 0.06$ | − | $0.97 \pm 0.06$ | − | − | RATAN-600 |
| 60105 | 2023.06.10 | 32/- | − | − | $1.39 \pm 0.16$ | − | − | − | RT-32 |
| 60106 | 2023.06.11 | 20/- | − | − | $1.35 \pm 0.12$ | − | − | − | RT-32 |
| 60119 | 2023.06.18 | 29/- | − | − | $1.38 \pm 0.08$ | − | − | − | RT-32 |

Table A3: The RT-22 measurements in 2009–2023: epoch MJD (Col. 1), epoch yyyy.mm.dd (Col. 2), the flux densities at 36.8 GHz and their errors, Jy (Col. 3).

| MJD epoch | yyyy.mm.dd | $S_{36.8}, \sigma$ |
| --- | --- | --- |
| 1 | 2 | 3 |
| 55196 | 2009.12.30 | $1.79 \pm 0.18$ |
| 55198 | 2010.01.01 | $1.59 \pm 0.22$ |
| 55199 | 2010.01.02 | $1.54 \pm 0.23$ |
| 55200 | 2010.01.03 | $1.39 \pm 0.21$ |
| 55202 | 2010.01.05 | $1.33 \pm 0.12$ |
| 55204 | 2010.01.07 | $0.99 \pm 0.13$ |
| 55205 | 2010.01.08 | $1.14 \pm 0.12$ |
| 55207 | 2010.01.10 | $1.74 \pm 0.11$ |
| 55208 | 2010.01.11 | $1.64 \pm 0.10$ |
| 55209 | 2010.01.12 | $1.49 \pm 0.10$ |
| 55215 | 2010.01.18 | $1.34 \pm 0.10$ |
| 55217 | 2010.01.20 | $1.29 \pm 0.11$ |
| 55218 | 2010.01.21 | $1.52 \pm 0.14$ |
| 55219 | 2010.01.22 | $1.39 \pm 0.13$ |
| 55220 | 2010.01.23 | $1.24 \pm 0.08$ |
| 55225 | 2010.01.28 | $1.48 \pm 0.07$ |
| 55228 | 2010.01.31 | $1.53 \pm 0.05$ |
| 55231 | 2010.02.03 | $1.51 \pm 0.05$ |
| 55234 | 2010.02.06 | $1.50 \pm 0.18$ |
| 55237 | 2010.02.09 | $1.47 \pm 0.04$ |
| 55238 | 2010.02.10 | $1.42 \pm 0.04$ |
| 55239 | 2010.02.11 | $1.36 \pm 0.05$ |
| 55240 | 2010.02.12 | $1.44 \pm 0.06$ |
| 55243 | 2010.02.15 | $1.33 \pm 0.07$ |
| 55247 | 2010.02.19 | $1.04 \pm 0.06$ |
| 55249 | 2010.02.21 | $0.95 \pm 0.07$ |
| 55250 | 2010.02.22 | $1.43 \pm 0.07$ |
| 55260 | 2010.03.04 | $1.20 \pm 0.16$ |
| 55261 | 2010.03.05 | $1.26 \pm 0.14$ |
| 55262 | 2010.03.06 | $1.51 \pm 0.11$ |
| 55266 | 2010.03.10 | $1.40 \pm 0.09$ |
| 55267 | 2010.03.11 | $0.97 \pm 0.19$ |
| 55270 | 2010.03.14 | $1.05 \pm 0.15$ |
| 55281 | 2010.03.25 | $0.94 \pm 0.10$ |
| 55284 | 2010.03.28 | $1.04 \pm 0.14$ |
| 55288 | 2010.04.01 | $0.95 \pm 0.10$ |
| 55289 | 2010.04.02 | $0.76 \pm 0.15$ |
| 55291 | 2010.04.04 | $0.65 \pm 0.13$ |
| 55293 | 2010.04.06 | $0.97 \pm 0.12$ |
| 55298 | 2010.04.11 | $1.10 \pm 0.10$ |
| 55308 | 2010.04.21 | $1.58 \pm 0.10$ |
| 55311 | 2010.04.24 | $1.18 \pm 0.10$ |
| 55313 | 2010.04.26 | $1.58 \pm 0.08$ |
| 55315 | 2010.04.28 | $1.16 \pm 0.13$ |
| 55333 | 2010.05.16 | $1.62 \pm 0.17$ |
| 55334 | 2010.05.17 | $1.55 \pm 0.10$ |
| 55339 | 2010.05.22 | $1.58 \pm 0.07$ |
| 55342 | 2010.05.25 | $1.52 \pm 0.05$ |
| 55343 | 2010.05.26 | $1.44 \pm 0.05$ |
| 55345 | 2010.05.28 | $1.48 \pm 0.18$ |
| 55358 | 2010.06.10 | $1.56 \pm 0.04$ |
| 55370 | 2010.06.22 | $1.46 \pm 0.04$ |
| 55371 | 2010.06.23 | $0.98 \pm 0.05$ |
| 55372 | 2010.06.24 | $1.27 \pm 0.06$ |
| 55373 | 2010.06.25 | $1.06 \pm 0.07$ |
| 55375 | 2010.06.27 | $1.35 \pm 0.06$ |
| 55378 | 2010.06.30 | $1.14 \pm 0.07$ |
| 55381 | 2010.07.03 | $1.29 \pm 0.16$ |
| 55384 | 2010.07.06 | $1.31 \pm 0.13$ |
| 55387 | 2010.07.09 | $1.55 \pm 0.11$ |
| 55392 | 2010.07.14 | $1.14 \pm 0.09$ |
| 55393 | 2010.07.15 | $1.34 \pm 0.11$ |

Table A3: continued.

| MJD epoch | yyyy.mm.dd | $S_{36.8}, \sigma$ |
| --- | --- | --- |
| 1 | 2 | 3 |
| 55398 | 2010.07.20 | $1.17 \pm 0.07$ |
| 55401 | 2010.07.23 | $1.18 \pm 0.07$ |
| 55404 | 2010.07.26 | $1.21 \pm 0.09$ |
| 55414 | 2010.08.05 | $0.55 \pm 0.10$ |
| 55417 | 2010.08.08 | $1.25 \pm 0.13$ |
| 55421 | 2010.08.12 | $1.23 \pm 0.11$ |
| 55423 | 2010.08.14 | $0.95 \pm 0.08$ |
| 55426 | 2010.08.17 | $1.19 \pm 0.07$ |
| 55435 | 2010.08.26 | $0.43 \pm 0.05$ |
| 55437 | 2010.08.28 | $0.67 \pm 0.04$ |
| 55448 | 2010.09.08 | $0.74 \pm 0.06$ |
| 55450 | 2010.09.10 | $1.17 \pm 0.07$ |
| 55452 | 2010.09.12 | $0.52 \pm 0.06$ |
| 55455 | 2010.09.15 | $0.60 \pm 0.07$ |
| 55461 | 2010.09.21 | $0.97 \pm 0.09$ |
| 55465 | 2010.09.25 | $0.85 \pm 0.10$ |
| 55469 | 2010.09.29 | $0.69 \pm 0.08$ |
| 55472 | 2010.10.02 | $0.96 \pm 0.07$ |
| 55480 | 2010.10.10 | $1.26 \pm 0.09$ |
| 55483 | 2010.10.13 | $1.39 \pm 0.10$ |
| 55484 | 2010.10.14 | $1.01 \pm 0.07$ |
| 55485 | 2010.10.15 | $1.26 \pm 0.15$ |
| 55488 | 2010.10.18 | $1.04 \pm 0.10$ |
| 55489 | 2010.10.19 | $1.23 \pm 0.14$ |
| 55491 | 2010.10.21 | $0.95 \pm 0.08$ |
| 55493 | 2010.10.23 | $1.28 \pm 0.13$ |
| 55496 | 2010.10.26 | $1.35 \pm 0.12$ |
| 55499 | 2010.10.29 | $1.16 \pm 0.11$ |
| 55500 | 2010.10.30 | $1.02 \pm 0.10$ |
| 55507 | 2010.11.06 | $1.05 \pm 0.10$ |
| 55508 | 2010.11.07 | $1.19 \pm 0.10$ |
| 55513 | 2010.11.12 | $1.21 \pm 0.11$ |
| 55517 | 2010.11.16 | $1.11 \pm 0.14$ |
| 55518 | 2010.11.17 | $0.83 \pm 0.13$ |
| 55523 | 2010.11.22 | $1.24 \pm 0.17$ |
| 55524 | 2010.11.23 | $1.05 \pm 0.10$ |
| 55525 | 2010.11.24 | $1.44 \pm 0.07$ |
| 55527 | 2010.11.26 | $1.18 \pm 0.05$ |
| 55528 | 2010.11.27 | $1.34 \pm 0.05$ |
| 55530 | 2010.11.29 | $1.35 \pm 0.18$ |
| 55537 | 2010.12.06 | $0.92 \pm 0.04$ |
| 55538 | 2010.12.07 | $1.27 \pm 0.04$ |
| 55539 | 2010.12.08 | $1.11 \pm 0.05$ |
| 55546 | 2010.12.15 | $1.26 \pm 0.05$ |
| 55548 | 2010.12.17 | $1.33 \pm 0.06$ |
| 55553 | 2010.12.22 | $1.63 \pm 0.07$ |
| 55557 | 2010.12.26 | $1.25 \pm 0.07$ |
| 55558 | 2010.12.27 | $1.55 \pm 0.16$ |
| 55563 | 2011.01.01 | $1.44 \pm 0.08$ |
| 55564 | 2011.01.02 | $1.11 \pm 0.09$ |
| 55565 | 2011.01.03 | $1.01 \pm 0.10$ |
| 55567 | 2011.01.05 | $1.12 \pm 0.11$ |
| 55568 | 2011.01.06 | $1.11 \pm 0.09$ |
| 55570 | 2011.01.08 | $1.25 \pm 0.08$ |
| 55571 | 2011.01.09 | $0.94 \pm 0.07$ |
| 55574 | 2011.01.12 | $0.74 \pm 0.09$ |
| 55578 | 2011.01.16 | $0.69 \pm 0.10$ |
| 55581 | 2011.01.19 | $1.06 \pm 0.08$ |
| 55588 | 2011.01.26 | $0.95 \pm 0.09$ |
| 55594 | 2011.02.01 | $1.17 \pm 0.15$ |
| 55596 | 2011.02.03 | $0.94 \pm 0.13$ |
| 55598 | 2011.02.05 | $1.04 \pm 0.12$ |
| 55600 | 2011.02.07 | $1.18 \pm 0.11$ |
| 55602 | 2011.02.09 | $1.22 \pm 0.10$ |

Table A3: continued.

| MJD epoch | yyyy.mm.dd | $S_{36.8}, \sigma$ |
|---|---|---|
| 1 | 2 | 3 |
| 55605 | 2011.02.12 | $1.16 \pm 0.07$ |
| 55609 | 2011.02.16 | $1.04 \pm 0.11$ |
| 55612 | 2011.02.19 | $0.84 \pm 0.14$ |
| 55616 | 2011.02.23 | $0.67 \pm 0.13$ |
| 55618 | 2011.02.25 | $0.64 \pm 0.17$ |
| 55619 | 2011.02.26 | $0.69 \pm 0.10$ |
| 55625 | 2011.03.04 | $0.74 \pm 0.07$ |
| 55626 | 2011.03.05 | $0.73 \pm 0.05$ |
| 55627 | 2011.03.06 | $0.76 \pm 0.05$ |
| 55631 | 2011.03.10 | $0.79 \pm 0.18$ |
| 55633 | 2011.03.12 | $0.84 \pm 0.04$ |
| 55634 | 2011.03.13 | $0.59 \pm 0.04$ |
| 55638 | 2011.03.17 | $0.63 \pm 0.05$ |
| 55646 | 2011.03.25 | $0.79 \pm 0.06$ |
| 55648 | 2011.03.27 | $0.96 \pm 0.07$ |
| 55650 | 2011.03.29 | $1.05 \pm 0.06$ |
| 55654 | 2011.04.02 | $1.14 \pm 0.07$ |
| 55660 | 2011.04.08 | $1.22 \pm 0.07$ |
| 55662 | 2011.04.10 | $1.24 \pm 0.16$ |
| 55663 | 2011.04.11 | $1.43 \pm 0.14$ |
| 55665 | 2011.04.13 | $1.52 \pm 0.11$ |
| 55666 | 2011.04.14 | $1.84 \pm 0.13$ |
| 55667 | 2011.04.15 | $1.46 \pm 0.13$ |
| 55669 | 2011.04.17 | $1.62 \pm 0.06$ |
| 55672 | 2011.04.20 | $1.74 \pm 0.09$ |
| 55675 | 2011.04.23 | $1.65 \pm 0.08$ |
| 55681 | 2011.04.29 | $1.74 \pm 0.17$ |
| 55688 | 2011.05.06 | $1.84 \pm 0.09$ |
| 55708 | 2011.05.26 | $1.54 \pm 0.14$ |
| 55710 | 2011.05.28 | $1.47 \pm 0.19$ |
| 55718 | 2011.06.05 | $1.43 \pm 0.15$ |
| 55721 | 2011.06.08 | $1.45 \pm 0.10$ |
| 55727 | 2011.06.14 | $1.34 \pm 0.14$ |
| 55738 | 2011.06.25 | $1.32 \pm 0.07$ |
| 55743 | 2011.06.30 | $1.12 \pm 0.12$ |
| 55746 | 2011.07.03 | $1.24 \pm 0.11$ |
| 55747 | 2011.07.04 | $0.94 \pm 0.10$ |
| 55748 | 2011.07.05 | $1.33 \pm 0.10$ |
| 55749 | 2011.07.06 | $1.29 \pm 0.10$ |
| 55753 | 2011.07.10 | $1.27 \pm 0.11$ |
| 55756 | 2011.07.13 | $1.24 \pm 0.14$ |
| 55759 | 2011.07.16 | $1.26 \pm 0.13$ |
| 55759 | 2011.07.16 | $1.28 \pm 0.17$ |
| 55766 | 2011.07.23 | $1.19 \pm 0.10$ |
| 55768 | 2011.07.25 | $1.22 \pm 0.07$ |
| 55769 | 2011.07.26 | $1.34 \pm 0.05$ |
| 55771 | 2011.07.28 | $1.44 \pm 0.05$ |
| 55775 | 2011.08.01 | $1.46 \pm 0.18$ |
| 55778 | 2011.08.04 | $1.36 \pm 0.03$ |
| 55779 | 2011.08.05 | $1.25 \pm 0.05$ |
| 55782 | 2011.08.08 | $1.44 \pm 0.06$ |
| 55784 | 2011.08.10 | $1.48 \pm 0.07$ |
| 55785 | 2011.08.11 | $1.47 \pm 0.06$ |
| 55788 | 2011.08.14 | $1.62 \pm 0.07$ |
| 55789 | 2011.08.15 | $1.46 \pm 0.07$ |
| 55791 | 2011.08.17 | $1.39 \pm 0.16$ |
| 55793 | 2011.08.19 | $0.99 \pm 0.14$ |
| 55794 | 2011.08.20 | $1.03 \pm 0.11$ |
| 55796 | 2011.08.22 | $1.50 \pm 0.13$ |
| 55798 | 2011.08.24 | $1.39 \pm 0.10$ |
| 55799 | 2011.08.25 | $1.53 \pm 0.11$ |
| 55810 | 2011.09.05 | $1.53 \pm 0.08$ |
| 55812 | 2011.09.07 | $1.69 \pm 0.16$ |
| 55813 | 2011.09.08 | $1.49 \pm 0.14$ |

Table A3: continued.

| MJD epoch | yyyy.mm.dd | $S_{36.8}$, $\sigma$ |
|---|---|---|
| 1 | 2 | 3 |
| 55817 | 2011.09.12 | $1.43 \pm 0.11$ |
| 55818 | 2011.09.13 | $1.29 \pm 0.15$ |
| 55820 | 2011.09.15 | $1.39 \pm 0.09$ |
| 55821 | 2011.09.16 | $1.29 \pm 0.07$ |
| 55834 | 2011.09.29 | $1.28 \pm 0.12$ |
| 55841 | 2011.10.06 | $1.34 \pm 0.16$ |
| 55845 | 2011.10.10 | $1.37 \pm 0.18$ |
| 55847 | 2011.10.12 | $1.19 \pm 0.08$ |
| 55848 | 2011.10.13 | $1.38 \pm 0.12$ |
| 55849 | 2011.10.14 | $1.09 \pm 0.13$ |
| 55850 | 2011.10.15 | $1.16 \pm 0.12$ |
| 55851 | 2011.10.16 | $1.43 \pm 0.15$ |
| 55852 | 2011.10.17 | $1.38 \pm 0.07$ |
| 55855 | 2011.10.20 | $1.39 \pm 0.19$ |
| 55858 | 2011.10.23 | $1.06 \pm 0.20$ |
| 55862 | 2011.10.27 | $1.29 \pm 0.09$ |
| 55865 | 2011.10.30 | $1.47 \pm 0.21$ |
| 55869 | 2011.11.03 | $1.26 \pm 0.19$ |
| 55873 | 2011.11.07 | $1.27 \pm 0.13$ |
| 55880 | 2011.11.14 | $1.24 \pm 0.10$ |
| 55883 | 2011.11.17 | $1.16 \pm 0.08$ |
| 55888 | 2011.11.22 | $1.16 \pm 0.07$ |
| 55889 | 2011.11.23 | $1.14 \pm 0.19$ |
| 55890 | 2011.11.24 | $1.18 \pm 0.14$ |
| 55894 | 2011.11.28 | $1.14 \pm 0.10$ |
| 55896 | 2011.11.30 | $1.09 \pm 0.13$ |
| 55897 | 2011.12.01 | $1.06 \pm 0.12$ |
| 55918 | 2011.12.22 | $1.05 \pm 0.13$ |
| 55923 | 2011.12.27 | $1.04 \pm 0.10$ |
| 55932 | 2012.01.05 | $0.99 \pm 0.15$ |
| 55946 | 2012.01.19 | $0.94 \pm 0.13$ |
| 55949 | 2012.01.22 | $0.92 \pm 0.19$ |
| 55950 | 2012.01.23 | $0.94 \pm 0.17$ |
| 55953 | 2012.01.26 | $1.15 \pm 0.11$ |
| 55957 | 2012.01.30 | $1.15 \pm 0.16$ |
| 55959 | 2012.02.01 | $1.12 \pm 0.12$ |
| 55965 | 2012.02.07 | $1.11 \pm 0.10$ |
| 55968 | 2012.02.10 | $1.04 \pm 0.17$ |
| 55971 | 2012.02.13 | $1.02 \pm 0.14$ |
| 55975 | 2012.02.17 | $1.05 \pm 0.10$ |
| 55978 | 2012.02.20 | $1.01 \pm 0.11$ |
| 55985 | 2012.02.27 | $1.02 \pm 0.07$ |
| 55988 | 2012.03.01 | $0.92 \pm 0.11$ |
| 55996 | 2012.03.09 | $1.04 \pm 0.09$ |
| 55997 | 2012.03.10 | $0.98 \pm 0.18$ |
| 56004 | 2012.03.17 | $1.05 \pm 0.08$ |
| 56006 | 2012.03.19 | $1.04 \pm 0.12$ |
| 56013 | 2012.03.26 | $0.98 \pm 0.13$ |
| 56025 | 2012.04.07 | $0.98 \pm 0.12$ |
| 56027 | 2012.04.09 | $0.99 \pm 0.15$ |
| 56033 | 2012.04.15 | $1.03 \pm 0.07$ |
| 56044 | 2012.04.26 | $1.07 \pm 0.19$ |
| 56082 | 2012.06.03 | $1.19 \pm 0.20$ |
| 56108 | 2012.06.29 | $1.16 \pm 0.09$ |
| 56112 | 2012.07.03 | $0.70 \pm 0.21$ |
| 56147 | 2012.08.07 | $0.87 \pm 0.19$ |
| 56153 | 2012.08.13 | $0.63 \pm 0.13$ |
| 56156 | 2012.08.16 | $0.59 \pm 0.10$ |
| 56166 | 2012.08.26 | $0.89 \pm 0.09$ |
| 56170 | 2012.08.30 | $0.87 \pm 0.11$ |
| 56173 | 2012.09.02 | $0.88 \pm 0.07$ |
| 56178 | 2012.09.07 | $0.55 \pm 0.08$ |
| 56181 | 2012.09.10 | $0.91 \pm 0.11$ |
| 56183 | 2012.09.12 | $0.91 \pm 0.13$ |

Table A3: continued.

| MJD epoch | yyyy.mm.dd | $S_{36.8}, \sigma$ |
| 1 | 2 | 3 |
|---|---|---|
| 56186 | 2012.09.15 | $0.84 \pm 0.15$ |
| 56192 | 2012.09.21 | $0.96 \pm 0.13$ |
| 56198 | 2012.09.27 | $0.98 \pm 0.10$ |
| 56200 | 2012.09.29 | $0.78 \pm 0.13$ |
| 56202 | 2012.10.01 | $1.03 \pm 0.12$ |
| 56209 | 2012.10.08 | $1.09 \pm 0.10$ |
| 56210 | 2012.10.09 | $1.10 \pm 0.14$ |
| 56215 | 2012.10.14 | $1.12 \pm 0.19$ |
| 56219 | 2012.10.18 | $1.14 \pm 0.16$ |
| 56221 | 2012.10.20 | $0.86 \pm 0.15$ |
| 56222 | 2012.10.21 | $1.15 \pm 0.13$ |
| 56226 | 2012.10.25 | $1.14 \pm 0.14$ |
| 56227 | 2012.10.26 | $1.16 \pm 0.19$ |
| 56228 | 2012.10.27 | $1.14 \pm 0.11$ |
| 56231 | 2012.10.30 | $1.13 \pm 0.13$ |
| 56235 | 2012.11.03 | $1.12 \pm 0.15$ |
| 56246 | 2012.11.14 | $1.06 \pm 0.16$ |
| 56254 | 2012.11.22 | $1.15 \pm 0.14$ |
| 56255 | 2012.11.23 | $1.15 \pm 0.15$ |
| 56256 | 2012.11.24 | $1.15 \pm 0.15$ |
| 56259 | 2012.11.27 | $1.23 \pm 0.13$ |
| 56267 | 2012.12.05 | $1.21 \pm 0.12$ |
| 56271 | 2012.12.09 | $1.22 \pm 0.11$ |
| 56279 | 2012.12.17 | $1.22 \pm 0.10$ |
| 56294 | 2013.01.01 | $1.15 \pm 0.10$ |
| 56301 | 2013.01.08 | $1.05 \pm 0.10$ |
| 56314 | 2013.01.21 | $1.09 \pm 0.11$ |
| 56316 | 2013.01.23 | $1.11 \pm 0.14$ |
| 56319 | 2013.01.26 | $0.96 \pm 0.13$ |
| 56323 | 2013.01.30 | $1.15 \pm 0.17$ |
| 56328 | 2013.02.04 | $1.46 \pm 0.10$ |
| 56331 | 2013.02.07 | $1.17 \pm 0.07$ |
| 56343 | 2013.02.19 | $1.21 \pm 0.05$ |
| 56349 | 2013.02.25 | $1.22 \pm 0.05$ |
| 56352 | 2013.02.28 | $1.17 \pm 0.18$ |
| 56356 | 2013.03.04 | $1.24 \pm 0.04$ |
| 56364 | 2013.03.12 | $1.23 \pm 0.04$ |
| 56369 | 2013.03.17 | $1.32 \pm 0.05$ |
| 56370 | 2013.03.18 | $1.06 \pm 0.06$ |
| 56373 | 2013.03.21 | $1.31 \pm 0.07$ |
| 56375 | 2013.03.23 | $1.27 \pm 0.06$ |
| 56384 | 2013.04.01 | $1.19 \pm 0.07$ |
| 56390 | 2013.04.07 | $1.14 \pm 0.07$ |
| 56399 | 2013.04.16 | $1.25 \pm 0.16$ |
| 56403 | 2013.04.20 | $1.24 \pm 0.14$ |
| 56406 | 2013.04.23 | $1.20 \pm 0.11$ |
| 56412 | 2013.04.29 | $1.16 \pm 0.13$ |
| 56427 | 2013.05.14 | $1.14 \pm 0.13$ |
| 56431 | 2013.05.18 | $1.18 \pm 0.07$ |
| 56450 | 2013.06.06 | $1.11 \pm 0.11$ |
| 56454 | 2013.06.10 | $1.03 \pm 0.12$ |
| 56471 | 2013.06.27 | $0.99 \pm 0.16$ |
| 56473 | 2013.06.29 | $0.94 \pm 0.18$ |
| 56495 | 2013.07.21 | $1.03 \pm 0.08$ |
| 56511 | 2013.08.06 | $1.06 \pm 0.12$ |
| 56517 | 2013.08.12 | $1.13 \pm 0.13$ |
| 56523 | 2013.08.18 | $1.11 \pm 0.12$ |
| 56535 | 2013.08.30 | $1.17 \pm 0.15$ |
| 56536 | 2013.08.31 | $1.14 \pm 0.07$ |
| 56539 | 2013.09.03 | $1.20 \pm 0.19$ |
| 56548 | 2013.09.12 | $1.17 \pm 0.20$ |
| 56551 | 2013.09.15 | $1.04 \pm 0.09$ |
| 56554 | 2013.09.18 | $1.22 \pm 0.17$ |
| 56559 | 2013.09.23 | $1.23 \pm 0.13$ |

Table A3: continued.

| MJD epoch | yyyy.mm.dd | $S_{36.8}$, $\sigma$ |
| 1 | 2 | 3 |
|---|---|---|
| 56562 | 2013.09.26 | $1.27 \pm 0.10$ |
| 56569 | 2013.10.03 | $1.34 \pm 0.09$ |
| 56571 | 2013.10.05 | $1.38 \pm 0.11$ |
| 56572 | 2013.10.06 | $1.42 \pm 0.07$ |
| 56574 | 2013.10.08 | $1.47 \pm 0.19$ |
| 56580 | 2013.10.14 | $1.69 \pm 0.16$ |
| 56583 | 2013.10.17 | $1.64 \pm 0.11$ |
| 56587 | 2013.10.21 | $1.66 \pm 0.12$ |
| 56594 | 2013.10.28 | $1.71 \pm 0.13$ |
| 56596 | 2013.10.30 | $1.71 \pm 0.15$ |
| 56601 | 2013.11.04 | $1.63 \pm 0.13$ |
| 56611 | 2013.11.14 | $1.73 \pm 0.11$ |
| 56613 | 2013.11.16 | $1.79 \pm 0.12$ |
| 56615 | 2013.11.18 | $1.73 \pm 0.13$ |
| 56618 | 2013.11.21 | $1.83 \pm 0.15$ |
| 56626 | 2013.11.29 | $1.79 \pm 0.16$ |
| 56630 | 2013.12.03 | $1.70 \pm 0.12$ |
| 56638 | 2013.12.11 | $1.67 \pm 0.11$ |
| 56643 | 2013.12.16 | $1.61 \pm 0.11$ |
| 56647 | 2013.12.20 | $1.66 \pm 0.13$ |
| 56658 | 2013.12.31 | $1.63 \pm 0.11$ |
| 56662 | 2014.01.04 | $1.65 \pm 0.12$ |
| 56666 | 2014.01.08 | $1.58 \pm 0.13$ |
| 56670 | 2014.01.12 | $1.59 \pm 0.08$ |
| 56671 | 2014.01.13 | $1.74 \pm 0.10$ |
| 56675 | 2014.01.17 | $1.69 \pm 0.14$ |
| 56680 | 2014.01.22 | $1.44 \pm 0.16$ |
| 56681 | 2014.01.23 | $1.34 \pm 0.20$ |
| 56682 | 2014.01.24 | $1.21 \pm 0.07$ |
| 56684 | 2014.01.26 | $1.57 \pm 0.15$ |
| 56689 | 2014.01.31 | $1.44 \pm 0.23$ |
| 56690 | 2014.02.01 | $1.31 \pm 0.20$ |
| 56692 | 2014.02.03 | $1.29 \pm 0.26$ |
| 56693 | 2014.02.04 | $1.32 \pm 0.23$ |
| 56694 | 2014.02.05 | $1.33 \pm 0.14$ |
| 56695 | 2014.02.06 | $1.19 \pm 0.11$ |
| 56699 | 2014.02.10 | $0.94 \pm 0.17$ |
| 56702 | 2014.02.13 | $0.84 \pm 0.16$ |
| 56708 | 2014.02.19 | $0.68 \pm 0.11$ |
| 56715 | 2014.02.26 | $0.75 \pm 0.06$ |
| 56719 | 2014.03.02 | $0.88 \pm 0.13$ |
| 56725 | 2014.03.08 | $0.94 \pm 0.17$ |
| 56726 | 2014.03.09 | $1.04 \pm 0.10$ |
| 56733 | 2014.03.16 | $1.14 \pm 0.12$ |
| 56736 | 2014.03.19 | $1.21 \pm 0.12$ |
| 56739 | 2014.03.22 | $1.04 \pm 0.16$ |
| 56740 | 2014.03.23 | $0.84 \pm 0.18$ |
| 56746 | 2014.03.29 | $0.69 \pm 0.10$ |
| 56750 | 2014.04.02 | $0.91 \pm 0.11$ |
| 56751 | 2014.04.03 | $1.16 \pm 0.09$ |
| 56752 | 2014.04.04 | $1.18 \pm 0.15$ |
| 56760 | 2014.04.12 | $1.20 \pm 0.09$ |
| 56761 | 2014.04.13 | $1.07 \pm 0.07$ |
| 56762 | 2014.04.14 | $1.26 \pm 0.09$ |
| 56766 | 2014.04.18 | $1.24 \pm 0.13$ |
| 56769 | 2014.04.21 | $1.18 \pm 0.14$ |
| 56769 | 2014.04.21 | $1.16 \pm 0.10$ |
| 56778 | 2014.04.30 | $1.06 \pm 0.06$ |
| 56784 | 2014.05.06 | $1.01 \pm 0.15$ |
| 56786 | 2014.05.08 | $0.99 \pm 0.15$ |
| 56787 | 2014.05.09 | $1.05 \pm 0.16$ |
| 56795 | 2014.05.17 | $1.29 \pm 0.12$ |
| 56798 | 2014.05.20 | $1.45 \pm 0.07$ |
| 56807 | 2014.05.29 | $1.59 \pm 0.10$ |

Table A3: continued.

| MJD epoch | yyyy.mm.dd | $S_{36.8}, \sigma$ |
|---|---|---|
| 1 | 2 | 3 |
| 56817 | 2014.06.08 | $1.53 \pm 0.15$ |
| 56818 | 2014.06.09 | $1.29 \pm 0.07$ |
| 56832 | 2014.06.23 | $1.18 \pm 0.11$ |
| 56842 | 2014.07.03 | $1.14 \pm 0.12$ |
| 56844 | 2014.07.05 | $1.06 \pm 0.12$ |
| 56852 | 2014.07.13 | $1.14 \pm 0.12$ |
| 56856 | 2014.07.17 | $1.53 \pm 0.09$ |
| 56860 | 2014.07.21 | $1.44 \pm 0.15$ |
| 56864 | 2014.07.25 | $1.35 \pm 0.07$ |
| 56867 | 2014.07.28 | $1.22 \pm 0.16$ |
| 56871 | 2014.08.01 | $1.17 \pm 0.09$ |
| 56881 | 2014.08.11 | $1.23 \pm 0.21$ |
| 56890 | 2014.08.20 | $1.27 \pm 0.19$ |
| 56900 | 2014.08.30 | $1.30 \pm 0.13$ |
| 56901 | 2014.08.31 | $1.34 \pm 0.10$ |
| 56902 | 2014.09.01 | $1.43 \pm 0.09$ |
| 56928 | 2014.09.27 | $1.36 \pm 0.11$ |
| 56932 | 2014.10.01 | $1.33 \pm 0.07$ |
| 56940 | 2014.10.09 | $1.46 \pm 0.09$ |
| 56943 | 2014.10.12 | $1.47 \pm 0.10$ |
| 56944 | 2014.10.13 | $1.45 \pm 0.27$ |
| 56948 | 2014.10.17 | $1.54 \pm 0.11$ |
| 56951 | 2014.10.20 | $1.64 \pm 0.14$ |
| 56954 | 2014.10.23 | $1.55 \pm 0.11$ |
| 56955 | 2014.10.24 | $1.53 \pm 0.30$ |
| 56958 | 2014.10.27 | $1.42 \pm 0.09$ |
| 56962 | 2014.10.31 | $1.44 \pm 0.12$ |
| 56967 | 2014.11.05 | $1.48 \pm 0.13$ |
| 56968 | 2014.11.06 | $1.39 \pm 0.12$ |
| 56969 | 2014.11.07 | $1.35 \pm 0.10$ |
| 56971 | 2014.11.09 | $1.24 \pm 0.16$ |
| 56977 | 2014.11.15 | $1.19 \pm 0.15$ |
| 56978 | 2014.11.16 | $1.07 \pm 0.07$ |
| 56981 | 2014.11.19 | $1.29 \pm 0.10$ |
| 56983 | 2014.11.21 | $1.46 \pm 0.10$ |
| 56988 | 2014.11.26 | $1.24 \pm 0.12$ |
| 56989 | 2014.11.27 | $1.12 \pm 0.20$ |
| 56990 | 2014.11.28 | $1.11 \pm 0.12$ |
| 56992 | 2014.11.30 | $1.12 \pm 0.13$ |
| 56995 | 2014.12.03 | $1.13 \pm 0.27$ |
| 57004 | 2014.12.12 | $1.14 \pm 0.19$ |
| 57019 | 2014.12.27 | $1.19 \pm 0.11$ |
| 57020 | 2014.12.28 | $1.09 \pm 0.13$ |
| 57021 | 2014.12.29 | $1.04 \pm 0.13$ |
| 57022 | 2014.12.30 | $1.01 \pm 0.12$ |
| 57026 | 2015.01.03 | $0.96 \pm 0.13$ |
| 57026 | 2015.01.03 | $0.95 \pm 0.15$ |
| 57027 | 2015.01.04 | $0.94 \pm 0.08$ |
| 57031 | 2015.01.08 | $1.04 \pm 0.16$ |
| 57032 | 2015.01.09 | $1.18 \pm 0.13$ |
| 57036 | 2015.01.13 | $1.24 \pm 0.17$ |
| 57047 | 2015.01.24 | $1.34 \pm 0.11$ |
| 57051 | 2015.01.28 | $1.47 \pm 0.06$ |
| 57052 | 2015.01.29 | $1.51 \pm 0.13$ |
| 57056 | 2015.02.02 | $1.55 \pm 0.07$ |
| 57058 | 2015.02.04 | $1.44 \pm 0.17$ |
| 57060 | 2015.02.06 | $1.25 \pm 0.07$ |
| 57061 | 2015.02.07 | $1.50 \pm 0.09$ |
| 57062 | 2015.02.08 | $1.39 \pm 0.15$ |
| 57063 | 2015.02.09 | $1.20 \pm 0.06$ |
| 57064 | 2015.02.10 | $1.25 \pm 0.09$ |
| 57065 | 2015.02.11 | $1.35 \pm 0.10$ |
| 57067 | 2015.02.13 | $1.59 \pm 0.08$ |
| 57068 | 2015.02.14 | $1.62 \pm 0.10$ |

Table A3: continued.

| MJD epoch | yyyy.mm.dd | $S_{36.8}, \sigma$ |
|---|---|---|
| 1 | 2 | 3 |
| 57070 | 2015.02.16 | $1.80 \pm 0.05$ |
| 57073 | 2015.02.19 | $1.77 \pm 0.10$ |
| 57074 | 2015.02.20 | $1.70 \pm 0.11$ |
| 57075 | 2015.02.21 | $1.57 \pm 0.08$ |
| 57077 | 2015.02.23 | $1.52 \pm 0.07$ |
| 57078 | 2015.02.24 | $1.29 \pm 0.12$ |
| 57079 | 2015.02.25 | $1.16 \pm 0.10$ |
| 57080 | 2015.02.26 | $1.14 \pm 0.13$ |
| 57085 | 2015.03.03 | $1.16 \pm 0.15$ |
| 57086 | 2015.03.04 | $1.24 \pm 0.06$ |
| 57087 | 2015.03.05 | $1.30 \pm 0.10$ |
| 57089 | 2015.03.07 | $1.18 \pm 0.10$ |
| 57090 | 2015.03.08 | $0.90 \pm 0.06$ |
| 57091 | 2015.03.09 | $1.05 \pm 0.09$ |
| 57092 | 2015.03.10 | $1.48 \pm 0.11$ |
| 57093 | 2015.03.11 | $1.62 \pm 0.09$ |
| 57094 | 2015.03.12 | $1.27 \pm 0.06$ |
| 57097 | 2015.03.15 | $1.40 \pm 0.10$ |
| 57102 | 2015.03.20 | $1.22 \pm 0.09$ |
| 57111 | 2015.03.29 | $0.88 \pm 0.10$ |
| 57113 | 2015.03.31 | $0.72 \pm 0.10$ |
| 57115 | 2015.04.02 | $0.79 \pm 0.11$ |
| 57116 | 2015.04.03 | $0.84 \pm 0.07$ |
| 57128 | 2015.04.15 | $1.03 \pm 0.15$ |
| 57129 | 2015.04.16 | $1.14 \pm 0.08$ |
| 57132 | 2015.04.19 | $0.87 \pm 0.08$ |
| 57134 | 2015.04.21 | $0.77 \pm 0.12$ |
| 57146 | 2015.05.03 | $0.72 \pm 0.05$ |
| 57157 | 2015.05.14 | $1.01 \pm 0.07$ |
| 57165 | 2015.05.22 | $1.00 \pm 0.07$ |
| 57166 | 2015.05.23 | $0.95 \pm 0.09$ |
| 57178 | 2015.06.04 | $1.03 \pm 0.09$ |
| 57179 | 2015.06.05 | $1.07 \pm 0.10$ |
| 57185 | 2015.06.11 | $0.75 \pm 0.11$ |
| 57199 | 2015.06.25 | $0.47 \pm 0.07$ |
| 57205 | 2015.07.01 | $0.80 \pm 0.06$ |
| 57213 | 2015.07.09 | $1.02 \pm 0.10$ |
| 57216 | 2015.07.12 | $1.12 \pm 0.09$ |
| 57222 | 2015.07.18 | $1.29 \pm 0.08$ |
| 57226 | 2015.07.22 | $1.33 \pm 0.09$ |
| 57235 | 2015.07.31 | $1.05 \pm 0.10$ |
| 57241 | 2015.08.06 | $1.06 \pm 0.11$ |
| 57257 | 2015.08.22 | $1.18 \pm 0.09$ |
| 57259 | 2015.08.24 | $1.31 \pm 0.15$ |
| 57260 | 2015.08.25 | $1.38 \pm 0.11$ |
| 57261 | 2015.08.26 | $1.52 \pm 0.07$ |
| 57269 | 2015.09.03 | $1.41 \pm 0.11$ |
| 57274 | 2015.09.08 | $1.32 \pm 0.12$ |
| 57282 | 2015.09.16 | $1.18 \pm 0.13$ |
| 57294 | 2015.09.28 | $0.93 \pm 0.08$ |
| 57304 | 2015.10.08 | $1.01 \pm 0.11$ |
| 57305 | 2015.10.09 | $1.12 \pm 0.09$ |
| 57308 | 2015.10.12 | $1.17 \pm 0.09$ |
| 57329 | 2015.11.02 | $1.53 \pm 0.08$ |
| 57336 | 2015.11.09 | $1.65 \pm 0.10$ |
| 57346 | 2015.11.19 | $1.51 \pm 0.08$ |
| 57355 | 2015.11.28 | $1.46 \pm 0.08$ |
| 57363 | 2015.12.06 | $1.51 \pm 0.07$ |
| 57368 | 2015.12.11 | $1.65 \pm 0.12$ |
| 57371 | 2015.12.14 | $1.65 \pm 0.10$ |
| 57386 | 2015.12.29 | $1.93 \pm 0.13$ |
| 57390 | 2016.01.02 | $1.91 \pm 0.11$ |
| 57398 | 2016.01.10 | $1.99 \pm 0.15$ |
| 57401 | 2016.01.13 | $1.94 \pm 0.11$ |

Table A3: continued.

| MJD epoch | yyyy.mm.dd | $S_{36.8}, \sigma$ |
|---|---|---|
| 1 | 2 | 3 |
| 57411 | 2016.01.23 | $1.78 \pm 0.11$ |
| 57417 | 2016.01.29 | $1.81 \pm 0.12$ |
| 57423 | 2016.02.04 | $1.87 \pm 0.09$ |
| 57428 | 2016.02.09 | $2.26 \pm 0.12$ |
| 57434 | 2016.02.15 | $2.05 \pm 0.10$ |
| 57453 | 2016.03.05 | $2.04 \pm 0.11$ |
| 57455 | 2016.03.07 | $1.99 \pm 0.15$ |
| 57457 | 2016.03.09 | $1.94 \pm 0.16$ |
| 57461 | 2016.03.13 | $1.96 \pm 0.10$ |
| 57480 | 2016.04.01 | $1.69 \pm 0.08$ |
| 57493 | 2016.04.14 | $1.17 \pm 0.07$ |
| 57494 | 2016.04.15 | $1.17 \pm 0.05$ |
| 57500 | 2016.04.21 | $0.96 \pm 0.04$ |
| 57511 | 2016.05.02 | $0.89 \pm 0.10$ |
| 57525 | 2016.05.16 | $0.78 \pm 0.04$ |
| 57530 | 2016.05.21 | $0.73 \pm 0.03$ |
| 57534 | 2016.05.25 | $0.78 \pm 0.13$ |
| 57543 | 2016.06.03 | $0.84 \pm 0.15$ |
| 57550 | 2016.06.10 | $1.54 \pm 0.10$ |
| 57553 | 2016.06.13 | $1.52 \pm 0.13$ |
| 57560 | 2016.06.20 | $1.48 \pm 0.15$ |
| 57571 | 2016.07.01 | $1.44 \pm 0.11$ |
| 57574 | 2016.07.04 | $1.34 \pm 0.10$ |
| 57579 | 2016.07.09 | $1.16 \pm 0.12$ |
| 57600 | 2016.07.30 | $1.10 \pm 0.17$ |
| 57603 | 2016.08.02 | $1.13 \pm 0.13$ |
| 57604 | 2016.08.03 | $1.16 \pm 0.08$ |
| 57608 | 2016.08.07 | $1.24 \pm 0.12$ |
| 57612 | 2016.08.11 | $1.13 \pm 0.13$ |
| 57626 | 2016.08.25 | $1.28 \pm 0.15$ |
| 57645 | 2016.09.13 | $1.29 \pm 0.19$ |
| 57649 | 2016.09.17 | $1.18 \pm 0.13$ |
| 57660 | 2016.09.28 | $1.21 \pm 0.11$ |
| 57665 | 2016.10.03 | $1.24 \pm 0.15$ |
| 57669 | 2016.10.07 | $1.21 \pm 0.09$ |
| 57676 | 2016.10.14 | $1.08 \pm 0.13$ |
| 57679 | 2016.10.17 | $1.06 \pm 0.11$ |
| 57684 | 2016.10.22 | $1.01 \pm 0.15$ |
| 57688 | 2016.10.26 | $0.98 \pm 0.09$ |
| 57719 | 2016.11.26 | $0.88 \pm 0.11$ |
| 57725 | 2016.12.02 | $0.86 \pm 0.19$ |
| 57729 | 2016.12.06 | $0.87 \pm 0.09$ |
| 57733 | 2016.12.10 | $0.81 \pm 0.08$ |
| 57736 | 2016.12.13 | $0.84 \pm 0.07$ |
| 57741 | 2016.12.18 | $0.87 \pm 0.10$ |
| 57744 | 2016.12.21 | $0.87 \pm 0.09$ |
| 57754 | 2016.12.31 | $0.85 \pm 0.07$ |
| 57756 | 2017.01.02 | $0.99 \pm 0.09$ |
| 57772 | 2017.01.18 | $1.14 \pm 0.11$ |
| 57774 | 2017.01.20 | $1.24 \pm 0.15$ |
| 57782 | 2017.01.28 | $1.29 \pm 0.16$ |
| 57789 | 2017.02.04 | $1.34 \pm 0.11$ |
| 57796 | 2017.02.11 | $1.39 \pm 0.15$ |
| 57800 | 2017.02.15 | $1.34 \pm 0.13$ |
| 57802 | 2017.02.17 | $1.29 \pm 0.11$ |
| 57808 | 2017.02.23 | $1.27 \pm 0.10$ |
| 57846 | 2017.04.02 | $1.24 \pm 0.15$ |
| 57851 | 2017.04.07 | $1.21 \pm 0.16$ |
| 57862 | 2017.04.18 | $1.26 \pm 0.11$ |
| 57864 | 2017.04.20 | $1.26 \pm 0.15$ |
| 57871 | 2017.04.27 | $1.24 \pm 0.13$ |
| 57878 | 2017.05.04 | $1.18 \pm 0.15$ |
| 57886 | 2017.05.12 | $1.06 \pm 0.10$ |
| 57897 | 2017.05.23 | $0.97 \pm 0.13$ |

Table A3: continued.

| MJD epoch | yyyy.mm.dd | $S_{36.8}$, $\sigma$ |
| 1 | 2 | 3 |
|---|---|---|
| 57905 | 2017.05.31 | $0.93 \pm 0.15$ |
| 57935 | 2017.06.30 | $0.77 \pm 0.11$ |
| 57939 | 2017.07.04 | $0.73 \pm 0.10$ |
| 57968 | 2017.08.02 | $0.76 \pm 0.12$ |
| 57977 | 2017.08.11 | $0.88 \pm 0.17$ |
| 57982 | 2017.08.16 | $0.78 \pm 0.13$ |
| 57990 | 2017.08.24 | $0.68 \pm 0.08$ |
| 57993 | 2017.08.27 | $0.75 \pm 0.12$ |
| 57998 | 2017.09.01 | $0.78 \pm 0.13$ |
| 58004 | 2017.09.07 | $0.81 \pm 0.15$ |
| 58011 | 2017.09.14 | $0.79 \pm 0.19$ |
| 58026 | 2017.09.29 | $0.83 \pm 0.13$ |
| 58030 | 2017.10.03 | $0.88 \pm 0.15$ |
| 58039 | 2017.10.12 | $0.93 \pm 0.16$ |
| 58048 | 2017.10.21 | $0.82 \pm 0.15$ |
| 58050 | 2017.10.23 | $0.81 \pm 0.13$ |
| 58051 | 2017.10.24 | $0.87 \pm 0.15$ |
| 58054 | 2017.10.27 | $0.88 \pm 0.13$ |
| 58059 | 2017.11.01 | $0.76 \pm 0.11$ |
| 58061 | 2017.11.03 | $0.76 \pm 0.15$ |
| 58065 | 2017.11.07 | $0.76 \pm 0.09$ |
| 58069 | 2017.11.11 | $0.78 \pm 0.11$ |
| 58071 | 2017.11.13 | $0.79 \pm 0.15$ |
| 58087 | 2017.11.29 | $0.68 \pm 0.13$ |
| 58096 | 2017.12.08 | $0.67 \pm 0.12$ |
| 58110 | 2017.12.22 | $0.63 \pm 0.11$ |
| 58114 | 2017.12.26 | $0.62 \pm 0.10$ |
| 58120 | 2018.01.01 | $0.63 \pm 0.10$ |
| 58123 | 2018.01.04 | $0.62 \pm 0.10$ |
| 58136 | 2018.01.17 | $0.62 \pm 0.11$ |
| 58141 | 2018.01.22 | $0.64 \pm 0.14$ |
| 58144 | 2018.01.25 | $0.66 \pm 0.13$ |
| 58158 | 2018.02.08 | $0.68 \pm 0.17$ |
| 58197 | 2018.03.19 | $0.73 \pm 0.03$ |
| 58198 | 2018.03.20 | $0.70 \pm 0.05$ |
| 58199 | 2018.03.21 | $0.74 \pm 0.08$ |
| 58200 | 2018.03.22 | $0.70 \pm 0.07$ |
| 58211 | 2018.04.02 | $0.72 \pm 0.08$ |
| 58221 | 2018.04.12 | $0.72 \pm 0.10$ |
| 58229 | 2018.04.20 | $0.67 \pm 0.14$ |
| 58232 | 2018.04.23 | $0.60 \pm 0.19$ |
| 58236 | 2018.04.27 | $0.54 \pm 0.04$ |
| 58277 | 2018.06.07 | $0.48 \pm 0.06$ |
| 58279 | 2018.06.09 | $0.45 \pm 0.02$ |
| 58308 | 2018.07.08 | $0.56 \pm 0.07$ |
| 58313 | 2018.07.13 | $0.58 \pm 0.09$ |
| 58390 | 2018.09.28 | $0.60 \pm 0.03$ |
| 58391 | 2018.09.29 | $0.63 \pm 0.06$ |
| 58395 | 2018.10.03 | $0.63 \pm 0.03$ |
| 58404 | 2018.10.12 | $0.60 \pm 0.03$ |
| 58409 | 2018.10.17 | $0.63 \pm 0.06$ |
| 58417 | 2018.10.25 | $0.62 \pm 0.17$ |
| 58420 | 2018.10.28 | $0.63 \pm 0.13$ |
| 58423 | 2018.10.31 | $0.61 \pm 0.08$ |
| 58433 | 2018.11.10 | $0.58 \pm 0.12$ |
| 58446 | 2018.11.23 | $0.55 \pm 0.13$ |
| 58449 | 2018.11.26 | $0.54 \pm 0.15$ |
| 58454 | 2018.12.01 | $0.56 \pm 0.10$ |
| 58456 | 2018.12.03 | $0.58 \pm 0.09$ |
| 58464 | 2018.12.11 | $0.56 \pm 0.07$ |
| 58476 | 2018.12.23 | $0.59 \pm 0.09$ |
| 58478 | 2018.12.25 | $0.56 \pm 0.10$ |
| 58481 | 2018.12.28 | $0.63 \pm 0.11$ |
| 58484 | 2018.12.31 | $0.64 \pm 0.08$ |

Table A3: continued.

| MJD epoch | yyyy.mm.dd | $S_{36.8}, \sigma$ |
|---|---|---|
| 1 | 2 | 3 |
| 58492 | 2019.01.08 | $0.64 \pm 0.10$ |
| 58499 | 2019.01.15 | $0.71 \pm 0.13$ |
| 58512 | 2019.01.28 | $0.75 \pm 0.14$ |
| 58543 | 2019.02.28 | $1.03 \pm 0.09$ |
| 58549 | 2019.03.06 | $1.06 \pm 0.11$ |
| 58562 | 2019.03.19 | $1.14 \pm 0.13$ |
| 58569 | 2019.03.26 | $1.12 \pm 0.14$ |
| 58591 | 2019.04.17 | $1.22 \pm 0.09$ |
| 58594 | 2019.04.20 | $1.29 \pm 0.11$ |
| 58600 | 2019.04.26 | $1.43 \pm 0.07$ |
| 58609 | 2019.05.05 | $1.34 \pm 0.12$ |
| 58630 | 2019.05.26 | $1.27 \pm 0.17$ |
| 58636 | 2019.06.01 | $1.25 \pm 0.13$ |
| 58644 | 2019.06.09 | $1.33 \pm 0.08$ |
| 58646 | 2019.06.11 | $1.33 \pm 0.12$ |
| 58653 | 2019.06.18 | $1.38 \pm 0.13$ |
| 58667 | 2019.07.02 | $1.34 \pm 0.15$ |
| 58681 | 2019.07.16 | $1.42 \pm 0.10$ |
| 58683 | 2019.07.18 | $1.52 \pm 0.13$ |
| 58686 | 2019.07.21 | $1.28 \pm 0.15$ |
| 58692 | 2019.07.27 | $0.94 \pm 0.11$ |
| 58693 | 2019.07.28 | $1.19 \pm 0.10$ |
| 58700 | 2019.08.04 | $1.44 \pm 0.12$ |
| 58701 | 2019.08.05 | $1.73 \pm 0.17$ |
| 58709 | 2019.08.13 | $1.59 \pm 0.13$ |
| 58711 | 2019.08.15 | $1.54 \pm 0.08$ |
| 58719 | 2019.08.23 | $1.58 \pm 0.12$ |
| 58721 | 2019.08.25 | $1.90 \pm 0.10$ |
| 58722 | 2019.08.26 | $1.74 \pm 0.19$ |
| 58736 | 2019.09.09 | $1.04 \pm 0.13$ |
| 58738 | 2019.09.11 | $1.24 \pm 0.15$ |
| 58739 | 2019.09.12 | $1.27 \pm 0.12$ |
| 58742 | 2019.09.15 | $1.27 \pm 0.10$ |
| 58743 | 2019.09.16 | $1.30 \pm 0.13$ |
| 58744 | 2019.09.17 | $1.27 \pm 0.09$ |
| 58757 | 2019.09.30 | $1.24 \pm 0.07$ |
| 58758 | 2019.10.01 | $1.27 \pm 0.16$ |
| 58759 | 2019.10.02 | $1.26 \pm 0.12$ |
| 58760 | 2019.10.03 | $1.28 \pm 0.17$ |
| 58761 | 2019.10.04 | $1.29 \pm 0.13$ |
| 58762 | 2019.10.05 | $1.27 \pm 0.08$ |
| 58766 | 2019.10.09 | $1.33 \pm 0.12$ |
| 58768 | 2019.10.11 | $1.32 \pm 0.13$ |
| 58769 | 2019.10.12 | $1.32 \pm 0.15$ |
| 58770 | 2019.10.13 | $1.30 \pm 0.11$ |
| 58771 | 2019.10.14 | $1.28 \pm 0.07$ |
| 58773 | 2019.10.16 | $1.30 \pm 0.13$ |
| 58774 | 2019.10.17 | $1.30 \pm 0.15$ |
| 58776 | 2019.10.19 | $1.28 \pm 0.07$ |
| 58777 | 2019.10.20 | $1.27 \pm 0.10$ |
| 58778 | 2019.10.21 | $1.26 \pm 0.07$ |
| 58781 | 2019.10.24 | $1.27 \pm 0.12$ |
| 58782 | 2019.10.25 | $1.26 \pm 0.13$ |
| 58785 | 2019.10.28 | $1.25 \pm 0.15$ |
| 58786 | 2019.10.29 | $1.25 \pm 0.19$ |
| 58787 | 2019.10.30 | $1.24 \pm 0.13$ |
| 58788 | 2019.10.31 | $1.25 \pm 0.15$ |
| 58790 | 2019.11.02 | $1.26 \pm 0.16$ |
| 58791 | 2019.11.03 | $1.22 \pm 0.15$ |
| 58792 | 2019.11.04 | $1.23 \pm 0.13$ |
| 58796 | 2019.11.08 | $1.44 \pm 0.15$ |
| 58799 | 2019.11.11 | $1.24 \pm 0.13$ |
| 58806 | 2019.11.18 | $1.35 \pm 0.07$ |
| 58813 | 2019.11.25 | $1.44 \pm 0.15$ |

Table A3: continued.

| MJD epoch | yyyy.mm.dd | $S_{36.8}$, $\sigma$ |
| 1 | 2 | 3 |
|---|---|---|
| 58817 | 2019.11.29 | $1.51 \pm 0.16$ |
| 58820 | 2019.12.02 | $1.58 \pm 0.10$ |
| 58823 | 2019.12.05 | $1.73 \pm 0.06$ |
| 58831 | 2019.12.13 | $1.73 \pm 0.08$ |
| 58835 | 2019.12.17 | $1.58 \pm 0.07$ |
| 58840 | 2019.12.22 | $1.70 \pm 0.13$ |
| 58843 | 2019.12.25 | $1.54 \pm 0.16$ |
| 58844 | 2019.12.26 | $1.44 \pm 0.12$ |
| 58846 | 2019.12.28 | $1.41 \pm 0.17$ |
| 58851 | 2020.01.02 | $1.49 \pm 0.13$ |
| 58852 | 2020.01.03 | $1.73 \pm 0.08$ |
| 58853 | 2020.01.04 | $1.52 \pm 0.12$ |
| 58856 | 2020.01.07 | $1.22 \pm 0.13$ |
| 58870 | 2020.01.21 | $1.14 \pm 0.15$ |
| 58898 | 2020.02.18 | $1.09 \pm 0.13$ |
| 58904 | 2020.02.24 | $1.16 \pm 0.11$ |
| 58907 | 2020.02.27 | $1.05 \pm 0.09$ |
| 58942 | 2020.04.02 | $1.09 \pm 0.06$ |
| 58946 | 2020.04.06 | $1.54 \pm 0.18$ |
| 58950 | 2020.04.10 | $1.89 \pm 0.08$ |
| 58969 | 2020.04.29 | $1.97 \pm 0.10$ |
| 58971 | 2020.05.01 | $1.84 \pm 0.14$ |
| 58976 | 2020.05.06 | $1.75 \pm 0.11$ |
| 58998 | 2020.05.28 | $1.99 \pm 0.13$ |
| 59010 | 2020.06.09 | $2.13 \pm 0.16$ |
| 59012 | 2020.06.11 | $2.05 \pm 0.11$ |
| 59027 | 2020.06.26 | $2.02 \pm 0.15$ |
| 59040 | 2020.07.09 | $1.44 \pm 0.14$ |
| 59084 | 2020.08.22 | $0.79 \pm 0.03$ |
| 59092 | 2020.08.30 | $0.77 \pm 0.03$ |
| 59103 | 2020.09.10 | $0.89 \pm 0.15$ |
| 59121 | 2020.09.28 | $1.04 \pm 0.09$ |
| 59143 | 2020.10.20 | $1.24 \pm 0.13$ |
| 59165 | 2020.11.11 | $1.19 \pm 0.19$ |
| 59183 | 2020.11.29 | $1.25 \pm 0.11$ |
| 59201 | 2020.12.17 | $1.48 \pm 0.19$ |
| 59217 | 2021.01.02 | $1.90 \pm 0.16$ |
| 59222 | 2021.01.07 | $2.29 \pm 0.12$ |
| 59226 | 2021.01.11 | $2.44 \pm 0.17$ |
| 59231 | 2021.01.16 | $2.39 \pm 0.20$ |
| 59243 | 2021.01.28 | $2.34 \pm 0.15$ |
| 59254 | 2021.02.08 | $2.14 \pm 0.16$ |
| 59281 | 2021.03.07 | $2.04 \pm 0.13$ |
| 59283 | 2021.03.09 | $1.94 \pm 0.18$ |
| 59288 | 2021.03.14 | $1.89 \pm 0.20$ |
| 59290 | 2021.03.16 | $1.86 \pm 0.21$ |
| 59319 | 2021.04.14 | $1.74 \pm 0.13$ |
| 59356 | 2021.05.21 | $1.39 \pm 0.15$ |
| 59360 | 2021.05.25 | $1.34 \pm 0.13$ |
| 59362 | 2021.05.27 | $1.29 \pm 0.10$ |
| 59366 | 2021.05.31 | $1.14 \pm 0.07$ |
| 59381 | 2021.06.15 | $1.24 \pm 0.14$ |
| 59414 | 2021.07.18 | $1.29 \pm 0.15$ |
| 59427 | 2021.07.31 | $1.24 \pm 0.16$ |
| 59473 | 2021.09.15 | $1.34 \pm 0.10$ |
| 59512 | 2021.10.24 | $0.88 \pm 0.04$ |
| 59524 | 2021.11.05 | $0.95 \pm 0.09$ |
| 59531 | 2021.11.12 | $0.80 \pm 0.02$ |
| 59532 | 2021.11.13 | $0.78 \pm 0.02$ |
| 59533 | 2021.11.14 | $0.79 \pm 0.03$ |
| 59560 | 2021.12.11 | $1.04 \pm 0.11$ |
| 59615 | 2022.02.04 | $1.22 \pm 0.05$ |
| 59618 | 2022.02.07 | $1.27 \pm 0.06$ |
| 59642 | 2022.03.03 | $1.79 \pm 0.08$ |

Table A3: continued.

| MJD epoch | yyyy.mm.dd | $S_{36.8}, \sigma$ |
|---|---|---|
| 1 | 2 | 3 |
| 59661 | 2022.03.22 | $1.81 \pm 0.09$ |
| 59667 | 2022.03.28 | $1.94 \pm 0.09$ |
| 59681 | 2022.04.11 | $2.14 \pm 0.10$ |
| 59702 | 2022.05.02 | $1.94 \pm 0.12$ |
| 59706 | 2022.05.06 | $1.89 \pm 0.16$ |
| 59717 | 2022.05.17 | $2.04 \pm 0.14$ |
| 59732 | 2022.06.01 | $3.04 \pm 0.15$ |
| 59762 | 2022.07.01 | $4.00 \pm 0.19$ |
| 59766 | 2022.07.05 | $4.09 \pm 0.21$ |
| 59775 | 2022.07.14 | $3.22 \pm 0.16$ |
| 59784 | 2022.07.23 | $3.50 \pm 0.17$ |
| 59801 | 2022.08.09 | $3.28 \pm 0.16$ |
| 59802 | 2022.08.10 | $3.16 \pm 0.15$ |
| 59805 | 2022.08.13 | $3.81 \pm 0.19$ |
| 59816 | 2022.08.24 | $3.14 \pm 0.23$ |
| 59832 | 2022.09.09 | $3.00 \pm 0.15$ |
| 59842 | 2022.09.19 | $2.24 \pm 0.20$ |
| 59850 | 2022.09.27 | $2.75 \pm 0.13$ |
| 59858 | 2022.10.05 | $3.15 \pm 0.23$ |
| 59860 | 2022.10.07 | $3.68 \pm 0.11$ |
| 59865 | 2022.10.12 | $4.65 \pm 0.17$ |
| 59872 | 2022.10.19 | $4.53 \pm 0.22$ |
| 59876 | 2022.10.23 | $3.09 \pm 0.15$ |
| 59879 | 2022.10.26 | $2.79 \pm 0.13$ |
| 59883 | 2022.10.30 | $2.37 \pm 0.12$ |
| 59886 | 2022.11.02 | $2.34 \pm 0.09$ |
| 59894 | 2022.11.10 | $2.07 \pm 0.10$ |
| 59895 | 2022.11.11 | $2.16 \pm 0.10$ |
| 59896 | 2022.11.12 | $2.39 \pm 0.11$ |
| 59904 | 2022.11.20 | $2.90 \pm 0.14$ |
| 59911 | 2022.11.27 | $2.64 \pm 0.13$ |
| 59913 | 2022.11.29 | $2.14 \pm 0.17$ |
| 59917 | 2022.12.03 | $2.05 \pm 0.10$ |
| 59921 | 2022.12.07 | $1.62 \pm 0.07$ |
| 59929 | 2022.12.15 | $1.21 \pm 0.05$ |
| 59942 | 2022.12.28 | $1.03 \pm 0.05$ |
| 59958 | 2023.01.13 | $1.00 \pm 0.18$ |
| 59976 | 2023.01.31 | $0.98 \pm 0.04$ |
| 59984 | 2023.02.08 | $1.03 \pm 0.04$ |
| 59987 | 2023.02.11 | $1.08 \pm 0.05$ |
| 59990 | 2023.02.14 | $1.31 \pm 0.06$ |
| 59992 | 2023.02.16 | $1.39 \pm 0.07$ |
| 59994 | 2023.02.18 | $1.32 \pm 0.06$ |
| 60003 | 2023.02.27 | $1.54 \pm 0.07$ |
| 60004 | 2023.02.28 | $1.49 \pm 0.07$ |
| 60007 | 2023.03.03 | $1.61 \pm 0.16$ |
| 60014 | 2023.03.10 | $1.92 \pm 0.14$ |
| 60021 | 2023.03.17 | $2.29 \pm 0.11$ |
| 60031 | 2023.03.27 | $2.42 \pm 0.13$ |
| 60053 | 2023.04.18 | $2.81 \pm 0.13$ |
| 60057 | 2023.04.22 | $2.91 \pm 0.14$ |

This paper has been typeset from a TeX/LaTeX file prepared by the author.